\documentclass[twocolumn]{emulateapj}
\usepackage{graphicx}
\usepackage{amssymb}
\usepackage{amsmath}
\usepackage{subfigure}
\usepackage{multirow}
\usepackage{threeparttable}
\usepackage{array}
\usepackage{natbib}
\usepackage{xcolor}
\usepackage{subfigure}
\usepackage{hyperref}

\hypersetup{
    colorlinks=true,
    linkcolor=red,
    filecolor=magenta,
    urlcolor=cyan,
}

\def\kms{~\rm km~s^{-1}}
\def\cmsq{~\rm cm^{-2}}
\def\cc{~\rm cm^{-3}}

\defcitealias{Qu:2019ab}{QB19}

\begin{document}
\title{The Warm Gas in the MW: A kinematical Model}
\author{Zhijie Qu$^1$, Joel N. Bregman$^1$, Edmund Hodges-Kluck$^{2, 3}$, Jiang-Tao Li$^1$, and Ryan Lindley$^1$}
\affil{$^1$ Department of Astronomy, University of Michigan, Ann Arbor, MI 48109, USA}
\affil{$^2$ Department of Astronomy, University of Maryland, College Park, MD 20742, USA}
\affil{$^3$ NASA/GSFC, Code 662, Greenbelt, MD 20771, USA}
\email{quzhijie@umich.edu}

\begin{abstract}
We develop a kinematical model for the Milky Way \ion{Si}{4}-bearing gas to determine its density distribution and kinematics.
This model is constrained by a column density line shape sample extracted from the {\it HST}/COS archival data, which contains 186 AGN sight lines.
We find that the \ion{Si}{4} ion density distribution is dominated by an extended disk along the $z$-direction (above or below the midplane), i.e., $n(z)=n_0\exp(-(z/z_0)^{0.82})$, where $z_0$ is the scale height of $6.3_{-1.5}^{+1.6}$ kpc (northern hemisphere) and $3.6_{-0.9}^{+1.0}$ kpc (southern hemisphere).
The density distribution of the disk in the radial direction shows a sharp edge at $15-20$ kpc given by, $n(r_{\rm XY})=n_0\exp(-(r_{\rm XY}/r_0)^{3.36})$, where $r_0 \approx 12.5\pm0.6$ kpc.
The difference of density distributions over $r_{\rm XY}$ and $z$ directions indicates that the warm gas traced by \ion{Si}{4} is mainly associated with disk processes (e.g., feedback or cycling gas) rather than accretion.
We estimate the mass of the warm gas (within 50 kpc) is $\log (M(50 {\rm kpc})/M_\odot)\approx8.1$ (assuming $Z\approx0.5Z_\odot$), and a $3\sigma$ upper limit of $\log (M(250 {\rm kpc})/M_\odot)\approx9.1$ (excluding the Magellanic system).
Kinematically, the warm gas disk is nearly co-rotating with the stellar disk at $v_{\rm rot}=215\pm3\kms$, which lags the midplane rotation by about $8\kms\rm~kpc^{-1}$ (within 5 kpc).
Meanwhile, we note that the warm gas in the northern hemisphere has significant accretion with $v_{\rm acc}$ of $69\pm 7\kms$ at 10 kpc (an accretion rate of $-0.60_{-0.13}^{+0.11}~M_\odot\rm~yr^{-1}$), while in the southern hemisphere, there is no measurable accretion, with an upper limit of $0.4~M_\odot\rm~yr^{-1}$.
\end{abstract}
\maketitle

\section{Introduction}
As part of the galaxy baryon cycle, the multi-phase gas exists in both the gaseous disk (interstellar medium; ISM; \citealt{Dickey:1990aa, Cox:2005aa}) and the gaseous halo (circumgalactic medium; CGM; \citealt{Putman:2012aa, Tumlinson:2017aa}).
The gaseous disk is roughly cospatial with the stellar disk and provides fuels for current star formation.
The existence of the gaseous halo not only supplies the gaseous disk for continuous star formation but also gathers the feedback materials from stellar evolution.
The gas exchange between the gaseous disk and the gaseous halo involves fundamental processes in galaxy formation and evolution: the gas assembly and the galactic feedback, which are still highly uncertain.

The warm-hot gas ($\log T \approx 5$) in galaxies is a unique tracer for accretion and feedback processes, because it is at the peak of the radiative cooling curve, which leads to short cooling timescale ($\log \tau \lesssim 10$ Myr; e.g., \citealt{Oppenheimer:2013aa, Gnat:2017aa}).
The existence of this gas is unstable, so it needs to be refreshed by accretion (e.g., accretion shocks \citealt{McQuinn:2018aa, Qu:2018aa, Stern:2018aa}) and feedback processes (e.g., galactic fountain and galactic wind; \citealt{Shapiro:1976aa, Bregman:1980aa, Thompson:2016aa}).

The warm-hot gas is commonly observed in both external galaxies and the Milky Way (MW).
For external galaxies, the extended warm-hot gas is detected in multi-wavelength emissions \citep{Howk:2000aa, Rand:2008aa, Li:2014aa, Hodges-Kluck:2016aa, Boettcher:2016aa}.
The detection approaches utilizing warm gas emission has a limitation of low emissivity at large radii ($\gtrsim 20$ kpc).
However, the low-density warm-hot gas at large radii could be detected as absorption lines against the continua of background AGN/stellar objects \citep{Stocke:2013aa, Werk:2013aa, Lehner:2015aa, Johnson:2015aa, Bowen:2016aa, Tumlinson:2017aa, Burchett:2019aa}.
For the warm-hot gas, the most popular intermediate-to-high ionization state ions are in the UV band, such as \ion{Si}{4}, \ion{C}{4}, and \ion{O}{6}, with a limiting column density of $\log N \approx 13$ at $S/N=10$.
The major limitation of the absorption line studies is that the bright background UV targets (AGN or UV-bright star for local galaxies) are rare to have a large sample (more than 10 sight lines) for individual galaxies \citep{Lehner:2015aa, Bowen:2016aa, Zheng:2017aa, Qu:2019aa}.

The only exception is the MW, which has hundreds of sight lines toward AGN and stars within the MW halo observed in past decades \citep{Jenkins:1978aa, Cowie:1979aa, Bruhweiler:1980aa, Savage:1981aa, de-Boer:1983aa, de-Kool:1985aa, Sembach:1992aa, Sembach:1994aa, Shull:1994aa, Sembach:1997aa, Savage:2001aa, Savage:2003aa, Sembach:2003aa, Fox:2004aa, Bowen:2008aa, Savage:2009aa, Lehner:2011aa, Wakker:2012aa, Fox:2014aa, Fox:2015aa, Bordoloi:2017aa, Karim:2018aa, Werk:2019aa, Zheng:2019aa}.
These studies suggested that the MW has a thick warm gas disk close to the stellar disk (e.g., \citealt{Savage:2009aa}) and a massive warm gas halo (e.g., \citealt{Zheng:2019aa}).
However, the radial density distribution of the warm gas is still poorly known, because the column density integrated over the sight line cannot determine the density distribution directly.

Here, we propose a new method to extract the density distribution by considering warm gas kinematics.
The basis of this method is that, given a bulk velocity field, different radial density distributions will lead to significantly different absorption line shapes.
This is because gas close to the Sun will have small projected velocities, while distant gas will have large velocity shifts.
Then, different velocities (in the absorption line shape) could be converted to distances of the gas.
Combining with column densities (amount of gas) at different velocities, the density distribution of the warm gas could be derived.

The issue for this method is that the kinematics of the MW warm gas is not completely understood, although it is important to understand the Galaxy evolution (i.e., the continuous star formation; \citealt{Lehner:2011aa}).
Previous studies suggested that the warm gas in the MW shows signatures from both galaxy rotation \citep{Wakker:2012aa} and gas inflow ($-100 \lesssim v_{\rm LSR} \lesssim 0 \kms$; \citealt{Lehner:2011aa, Zheng:2019aa}).
This is consistent with both the \ion{H}{1} disk (and the \ion{H}{1} halo; \citealt{Dickey:1990aa, Marasco:2011aa}) measured from 21 cm line mapping, and hot gas traced by X-ray absorption features \citep{Hodges-Kluck:2016aa}.
For both \ion{H}{1} and X-ray observations, kinematical models have been applied to reproduce the observed features (e.g., line centroids or line widths), and extract kinematics information.
The \ion{H}{1} halo (up to $z$ height $\approx 2-3$ kpc) are co-rotating with disk with a rotation velocity of $v_{\rm rot} =220 \kms$ and a vertical velocity gradient (disk-halo lagging) of ${\rm d}v_{\rm rot}/{\rm d} z =-15\pm 4 \kms~\rm kpc^{-1}$ \citep{Marasco:2011aa}.
Also, the \ion{H}{1} halo has a significant inflow with a velocity along the radial direction of $30_{-5}^{+7} \kms$ and a vertical velocity of $20_{-7}^{+5} \kms$.
Similarly, the hot halo is also co-rotating with the disk at $v_{\rm rot} = 183\pm41\kms$, while the hot halo does not have detected inflow, outflow, or lagging features due to the limitation of the X-ray instrument \citep{Hodges-Kluck:2016aa}.

However, no such kinematical model has been applied to reproduce the warm gas absorption features.
Previous studies only modeled the (column) density distribution without kinematics, hence the density distribution at large radii cannot be obtained (\citealt{Savage:2009aa, Zheng:2019aa}; \citealt{Qu:2019ab}; hereafter \citetalias{Qu:2019ab}).
To make up this gap, we build up a kinematical model, which contains free parameters for both the warm gas density distribution, the gas kinematics (e.g., rotation, inflow. or outflow), and the gas properties (e.g., the broadening velocity).
In this kinematical model, the absorption features are predicted to exist in the velocity range of $\approx -200$ to $200\kms$, which is mainly determined by galaxy rotation.
To constrain this kinematical model, we extract a \ion{Si}{4} differential column density line shape sample based on the {\it Hubble Space Telescope}/Cosmic Origins Spectrograph ({\it HST}/COS; \citealt{Green:2012aa}) archival data, and obtain the best parameters to optimize the likelihood of reproducing all \ion{Si}{4} line shapes.

In this paper, Section \ref{data_sample} summarizes the employed sight lines (mainly extracted from the {\it Hubble} Spectroscopic Legacy Archive; HSLA; \citealt{Peeples:2017aa}) and introduces the data reduction of individual sight lines.
In Section \ref{previous_models}, we introduce the previous models of the MW warm gas (column) density distribution (no kinematics in models), which is the basis of the new kinematical model in this work.
In the new model, we assume that the warm gases are clouds or layers rather than a continuous distribution, and assumptions of the cloud-like feature are introduced in Section \ref{cloud_model}.
The kinematical models are described in Section \ref{kmodel}, which includes the density distribution (Section \ref{density_field}), the kinematics (Section \ref{velocity_field}), calculation of the differential column density line shape (Section \ref{model_prediction}), and the Bayesian model to optimize parameters (Section \ref{Bayesian_frame}).
Section \ref{fitting_results} describes the fitting results from the kinematical model, where we extract the warm gas density distribution, the rotation velocity, the radial velocity, and properties of single warm gas clouds.
We discuss the results in Section \ref{discussion}, such as the origin of the warm gas, the implications of kinematics, and the warm gas mass and accretion rate.
We summarize key conclusions in Section \ref{conclusion}.

\section{Sample and Data Reduction}
\label{data_sample}
We consider both the stellar sample and the AGN sample in our analyses.
The CGM at large radii is only detected against the AGN continuum ($r \gtrsim 10$ kpc; $r$ is the distance to the Galactic center; GC), and the large scale variation of the disk is also dominated by the AGN sight lines \citepalias{Qu:2019ab}.
Due to the high sensitivity, the {\it HST}/COS obtains hundreds of AGN sight lines, which could be employed to extract absorption line shapes at high signal-to-noise ratios ($S/N > 10$).
Using the line shape, one could extract both the MW gas density distribution and kinematics (details in Section 5).
The stellar sample is employed to better constrain the midplane gas properties (hence the disk properties). 

In this study, we focus on the intermediate ionization state ion \ion{Si}{4}, which has doublet lines at $1393.8\rm~\AA$ and $1402.8\rm~\AA$.
The doublet could be used to exclude contamination and check saturation.
Therefore, \ion{Si}{4} is a good choice to extract the differential column density line shape.
\ion{C}{4} is another important ion of interests, which has doublet lines at $1548.2\rm~\AA$ and $1550.8\rm~\AA$.
With a higher element abundance, the \ion{C}{4} absorption column density are typically $\approx 0.5$ dex stronger than the \ion{Si}{4} column density, which helps to extract weak features.
However, the stronger absorption leads to more serious saturation issues for \ion{C}{4} at the peak of the differential column density line profile (about the half of sight lines have flattened peaks due to saturation).
The flattened peaks will significantly affect the model constraints on the gas distribution (i.e., the higher peak around $v=0\kms$ means more gas close to the Solar system).
The method is beyond the scope of this paper to extract the column density line profile from the modestly saturated lines, so we do not analyze \ion{C}{4} in this paper.

We construct the \ion{Si}{4} line shape sample for AGN sight lines based on the HST Spectroscopic Legacy Archive (HSLA; \citealt{Peeples:2017aa}).
For the stellar sight lines, we only use the column density measurements (without line shapes) for \ion{Si}{4} in the literature \citep{Savage:2001aa, Savage:2009aa, Lehner:2011aa}, which is extracted using observations obtained by the {\it International Ultraviolet Explorer} ({\it IUE}) and {\it HST}/Space Telescope Imaging Spectrograph ({\it HST}/STIS).
There are 186 AGN sight lines with differential column density line profiles and 88 stellar sight lines with column density measurements.

\subsection{Stellar sight lines}
\label{stellar_sl}
The \ion{Si}{4} stellar samples are adopted from \citet{Savage:2009aa} and \citet{Lehner:2011ab}.
\citet{Savage:2009aa} mainly summarized column density measurements from {\it IUE} observations \citep{Savage:2001aa}.
The \citet{Savage:2009aa} sample includes five transitional ions (\ion{Al}{3}, \ion{Si}{3}, \ion{Si}{4}, \ion{C}{4}, and \ion{O}{6}) for 109 MW stellar sight lines, 6 Large Magellanic Clouds/Small Magellanic Clouds (LMC/SMC) stellar sight lines, and 25 AGN sight lines.
In our analysis, we only use the MW stellar sightlines, since COS gives a better AGN sample.
These {\it IUE} observations have typical spectral resolutions of $\approx 20\kms$ and $S/N \gtrsim 5$.
The \citet{Lehner:2011ab} sample is composed of the {\it HST}/STIS observations, which typically have higher S/N than the IUE sample, so we adopted STIS measurements for overlapping sight lines.
There are 14 sight lines observed by both {\it IUE} and STIS, among which 12 sight lines are consistent within $2\sigma$.
Two sight lines have lower limits in the {\it IUE} sample, while the STIS sample has measurements lower than these lower limits.
This indicates that the {\it IUE} observation may overestimate some continuum levels, which  is limited by the $S/N$.

To test the possible systematic uncertainty of the {\it IUE} sample, we built two models in the fitting process (Section \ref{fitting_results}).
One model uses the combination of both {\it IUE} and STIS samples, while another one only uses the STIS sample.
These two samples give similar results (within $1 \sigma$), which indicates that the IUE sample is consistent with the STIS sample (more details in Section \ref{results}).
Therefore, we still use the combination of {\it IUE} and STIS samples for following analyses.

\citet{Savage:2009aa} showed that \ion{H}{2} regions have a significant contribution to the \ion{Si}{4} column density, so we omit the sight lines that have known foreground \ion{H}{2} regions. 
The final sample has 65 \ion{Si}{4} column density measurements, 11 lower limits and 12 upper limits, among which 27 are from STIS.

We do not use the COS archival stellar sight lines in the following analyses.
To constrain the midplane gas properties, we need sight lines that have suitable distances ($\approx 1- 10$ kpc) and low Galactic latitude ($b \lesssim 20^\circ$).
However, there are few useful sight lines in the COS archival data.
The COS instrument was used to obtain spectra in hundreds of stellar sight lines, with 354 of them having $S/N>10$, but most sight lines do not have suitable distances.
More than two-thirds of these stellar targets are nearby white dwarfs with distances of $\lesssim 0.1$ kpc.
These sight lines mostly have non-detection for \ion{Si}{4} due to small path-lengths.
Among the remaining $\approx 100$ targets, about half are LMC/SMC targets \citep{Roman-Duval:2019aa}, similar to the AGN (the column densities are sensitive to gas within $\approx 50 \rm~kpc$), but affected by the LMC/SMC ISM (at $v\approx 200-300\kms$).
There are 8 stellar sight lines in M33 \citep{Zheng:2017aa}, which have the same role as AGN sight lines nearby.

The remaining sight lines ($\approx 50$) have distances of $\approx 1- 10$ kpc.
However, about half of these targets have strong stellar features (i.e., strong stellar winds, emission lines, and structured continua), which make the extraction of absorption features unreliable.
Finally, there are only $\approx 20$ sight lines close to the disk and with well-behaved continua.
These targets are mainly UV-bright stars in globular clusters (i.e., blue horizontal branch stars; \citealt{Werk:2019aa}) at high $b$ and $|z|$-height (above or below the disk), which is opposite to our purpose to constrain the midplane density of \ion{Si}{4}.
Therefore, the archival COS data cannot improve our fitting significantly, and we do not include the line shapes from COS for stellar sight lines in this study.

\subsection{AGN sight lines}
For AGN sight lines, we limit the sample to $S/N > 10$, which is higher than the threshold of the stellar sample ($S/N \gtrsim 5$).
This is because the line shapes in the AGN sample is required to constrain the kinematical model, while for the stellar sample, we only use the column density measurements.
\ion{Si}{4} features normally have a velocity width of $\approx 100 \kms$ ($\approx 6$ resolution elements for both {\it IUE} and COS; the STIS sample has a higher resolution, but the {\it IUE} sample is dominant).
Therefore, the uncertainty of line shape (per resolution element) for AGN sight lines at $S/N = 10$ should be comparable to the total uncertainty of the integrated column density for the stellar sample at $S/N = 5$.

We extract the line shape sample based on the HSLA database, which provides a uniformly-reduced scientific-level database \citep{Peeples:2017aa}.
As a quick summary, the HSLA database archives all of the COS public data, extracts one-dimension spectra for individual exposures, and coadds all exposures for one target to generate the final coadded spectrum.
The output spectra have wavelength bins of $9.97\times 10^{-3}\rm~ \AA$ for the grating G130M and $12.23\times 10^{-3} \rm~ \AA$ for the grating G160M.
Based on the first data release of the HSLA, \citet{Zheng:2019aa} constructed the COS-GAL sample, an AGN sample for the MW absorption features, which includes \ion{Si}{4} column density and line centroid measurements within $|v_{\rm LSR}| \leq 100\kms$.
Here, we construct an updated \ion{Si}{4} sample for three reasons.
First, the HSLA database has the second release, which includes hundreds of new AGN sight lines, which also contains tens of $S/N>10$ sight lines.
Second, \citet{Zheng:2019aa} employed an arbitrary velocity criterion of 100 $\kms$ to truncate the measurements, which excludes some wings of high-velocity clouds (HVCs), which could affect our model constraints.
Third, we need to combine the two lines of the \ion{Si}{4} doublet to reduce the noise for the differential column density line shape.

\begin{figure*}
\begin{center}
\includegraphics[width=0.49\textwidth]{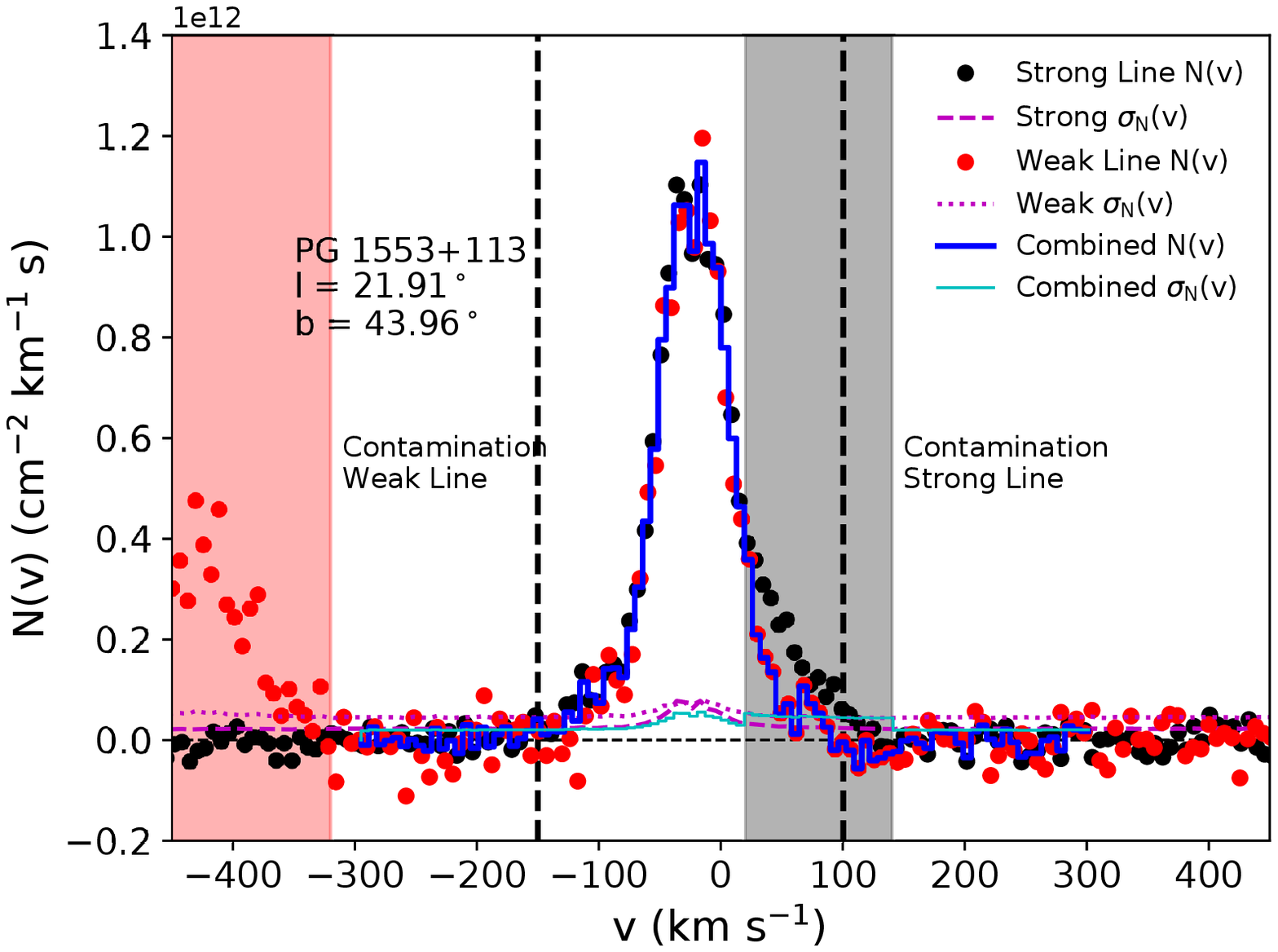}
\includegraphics[width=0.49\textwidth]{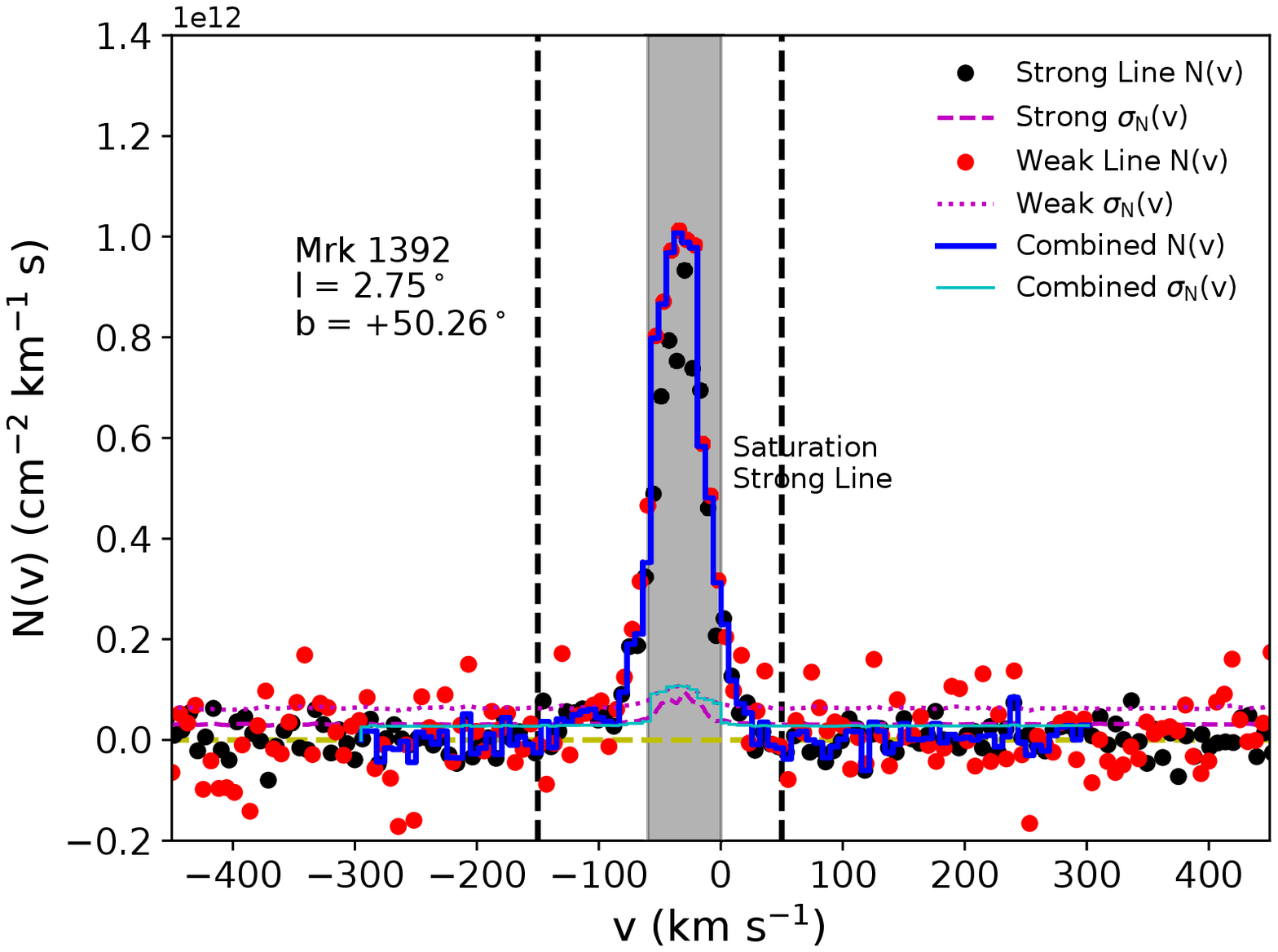}
\end{center}
\caption{Two example sight lines of PG 1553+113 (left panel) and Mrk 1392 (right panel) showing combination of the \ion{Si}{4} doublet to obtain the column density line shape (the blue line). The black and red dots are the apparent column density of the strong ($1393.8\rm~\AA$) and the weak ($1402.8\rm~\AA$) lines, respectively. The shadowed regions are blocked out in the combination of the doublet due to contamination or saturation.}
\label{example_sightline}
\end{figure*}

\ion{Si}{4} has the doublet at $1400\rm~\AA$, so we focus on the G130M spectrum that covers the wavelength range of $1100 -1450\rm~ \AA$.
In the second data release of the HSLA, there are 802 AGN sight lines, among which 402 sight lines have spectra using the grating G130M.

Our construction of the \ion{Si}{4} line shape sample has two parts.
In the first part, we determine the continuum near $1393\rm~\AA$ and $1402\rm~\AA$ in a $5.6 \rm~\AA$ interval ($-600 \kms$ to $600 \kms$), and select the sight lines with $S/N > 10$ per resolution element (6 original pixels).
To determine the continuum, we use an iteration method, which masks out absorption features.
First, we mask out pixels lower than the mean value of the flux by 1.5 times the error and obtain the initial guess of the continuum using the spline fitting for each interval.
The factor of 1.5 is used to avoid the absorption features that might affect the fitting of the continuum.
With the initial continuum, we mask out the pixels with flux lower than this continuum by 1.5 error and estimate the new continuum.
This step is iterated until the continuum converges, which means the masked out region is the same for two successive continuum fittings.
With the final continuum, the $S/N$ values are calculated for the two intervals of $1393\rm ~\AA$ and $1402\rm ~\AA$.
Normally, these two regions have similar S/N values, except that an AGN broad emission line occurs in the \ion{Si}{4} region.
We select all sight lines with $S/N > 10$ for either interval ($1393\rm~\AA$ or $1402\rm~\AA$).
The 186 selected sight lines are summarized in Table \ref{sample}, and there are 9 sight lines with relatively low $S/N < 10$ for one interval.
Because we will combine two lines to obtain the final differential column density line shape, the combined $S/N$ is always higher than 10.

In the second part, we combine the differential column density line shape and calculate the integrated column density.
The line shape is calculated based on the apparent optical depth method (AODM; \citealt{Savage:1991aa}), which converts the absorption depth into the apparent optical depth, hence the differential column density $N(v)$ of the velocity ($v$):
\begin{equation}
N(v) =\frac{m_{\rm e}c}{\pi e^2 f \lambda} \ln \frac{I_0(v)}{I_{\rm obs}(v)},
\end{equation}
where $I_0(v)$ and $I_{\rm obs}(v)$ are the continuum flux and the observed flux.
To reduce the uncertainty per data point in the line shape, the spectra are rebinned by 3 pixels (half of the resolution element; $6.4\kms$).
Then the differential column densities are calculated for both strong and weak lines.
By comparing the shapes of the strong line and weak line, we mask out the contamination regions in either line.
We mainly calculate the line shape in the velocity range of $-300\kms$ to $300\kms$, which is the velocity region that could be accounted for in our model (Section \ref{kmodel}) and continua of about $100 \kms$ in both sides.
The coadded regions could be varied if necessary to include HVCs with extremely high velocities of $|v| >300\kms$.
Fig. \ref{example_sightline} shows two examples of the sight lines toward PG 1553+113 and Mrk 1392.
The coadded column density is calculated using column density errors as weights, then the uncertainty of the coadded column density is $1/(1/\sigma_{N_{\rm s}}^2 + 1/\sigma_{N_{\rm w}}^2)^{1/2}$, where ``s" and ``w" denote the strong and the weak lines.

\begin{table*}
\begin{center}
\caption{The Column Density Measurements of the Selected \ion{Si}{4} Sample}
\label{sample}
\begin{tabular}{lrrrrrrrrrr}
\hline
\hline
Sightline & $l$ & $b$  & $S/N$ & $S/N$ & $v_{\rm min}$ & $v_{\rm max}$ & $\log N$ & $\sigma_{\log N}$ & $v_{\rm c}$ & $\sigma_{v_{\rm c}}$ \\
& deg. & deg. & Strong & Weak & $\kms$ & $\kms$ & dex & dex & $\kms$ & $\kms$ \\
(1) & (2) & (3) & (4) & (5) & (6) & (7) & (8) & (9) & (10) & (11)\\
\hline
        Mrk 1392 & $   2.8$ & $  50.3$ & $ 24$ & $ 23$ & $-150$ & $  50$ & $13.75$ & $0.02$ & $ -37.1$ & $  1.2$ \\
LQAC 209+017 004 & $   2.9$ & $  71.8$ & $ 25$ & $ 25$ & $-140$ & $  30$ & $13.70$ & $0.01$ & $ -42.8$ & $  1.4$ \\
     PG 1352+183 & $   4.4$ & $  72.9$ & $ 37$ & $ 39$ & $-180$ & $  60$ & $13.65$ & $0.01$ & $ -45.0$ & $  1.1$ \\
        RBS 1768 & $   4.5$ & $ -48.5$ & $ 20$ & $ 20$ & $-220$ & $ -60$ & $13.17$ & $0.03$ & $-141.7$ & $  3.4$ \\
                 &          &          &       &       & $ -60$ & $ 120$ & $13.58$ & $0.02$ & $  29.7$ & $  1.6$ \\
LQAC 350-034 001 & $   5.5$ & $ -69.4$ & $ 17$ & $ 18$ & $-130$ & $ -70$ & $12.68$ & $0.07$ & $-102.2$ & $  2.6$ \\
                 &          &          &       &       & $ -70$ & $  50$ & $13.23$ & $0.03$ & $ -15.7$ & $  2.3$ \\
\hline
\end{tabular}
\end{center}
Columns: (1) Target name; (2) Galactic longitude; (3) Galactic latitude; (4) $S/N$ of the strong line continuum; (5) $S/N$ of the weak line continuum; (6) Lower bound of absorption component; (7) Upper bound of absorption component; (8) Total column density of a component; (9) Column density uncertainty; (10) Line centroid of a component; (11) Line centroid uncertainty.\\
The entire version of this table is in the online version.
\end{table*}

\begin{figure*}
\begin{center}
\includegraphics[width=0.49\textwidth]{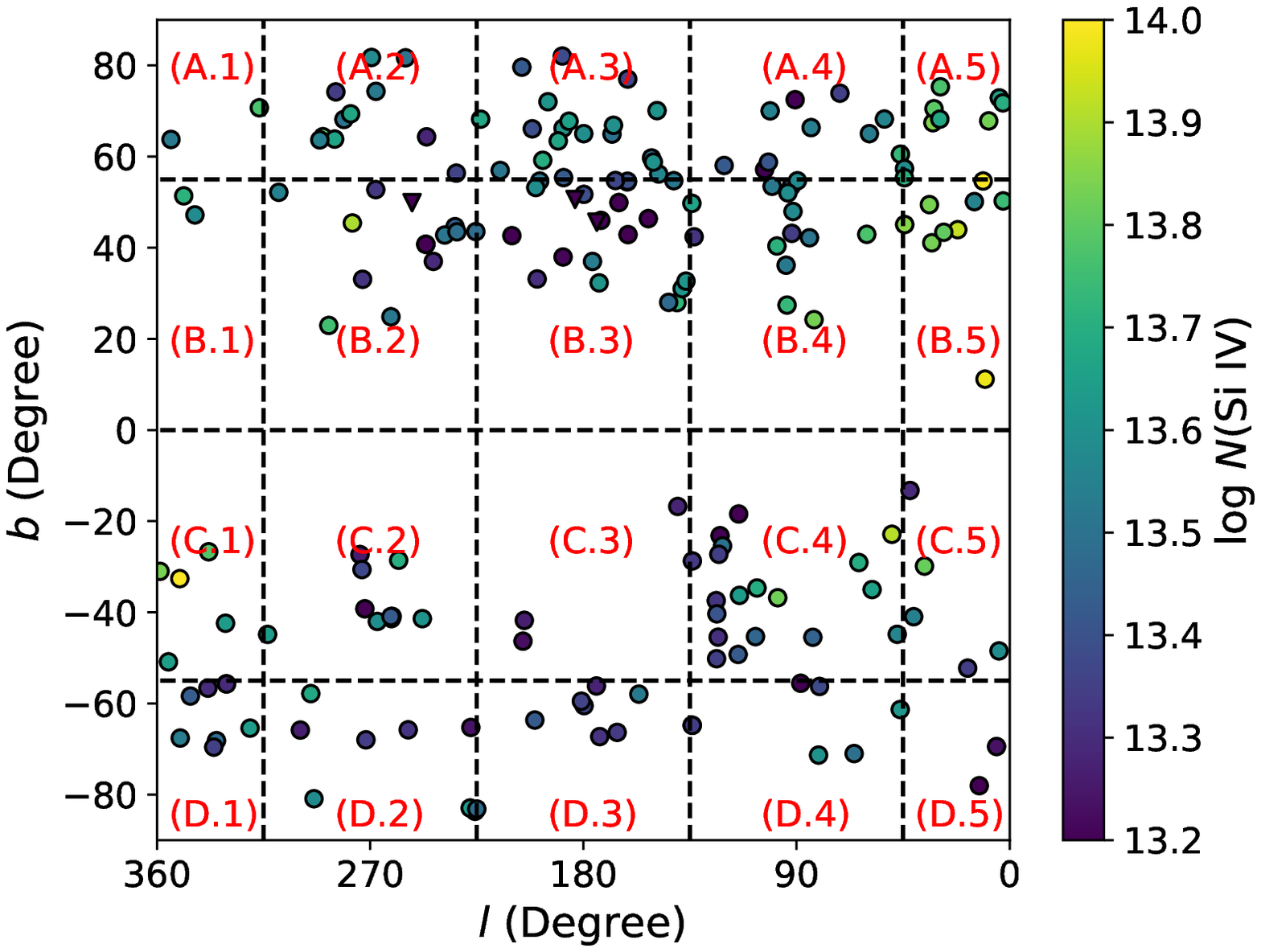}
\includegraphics[width=0.49\textwidth]{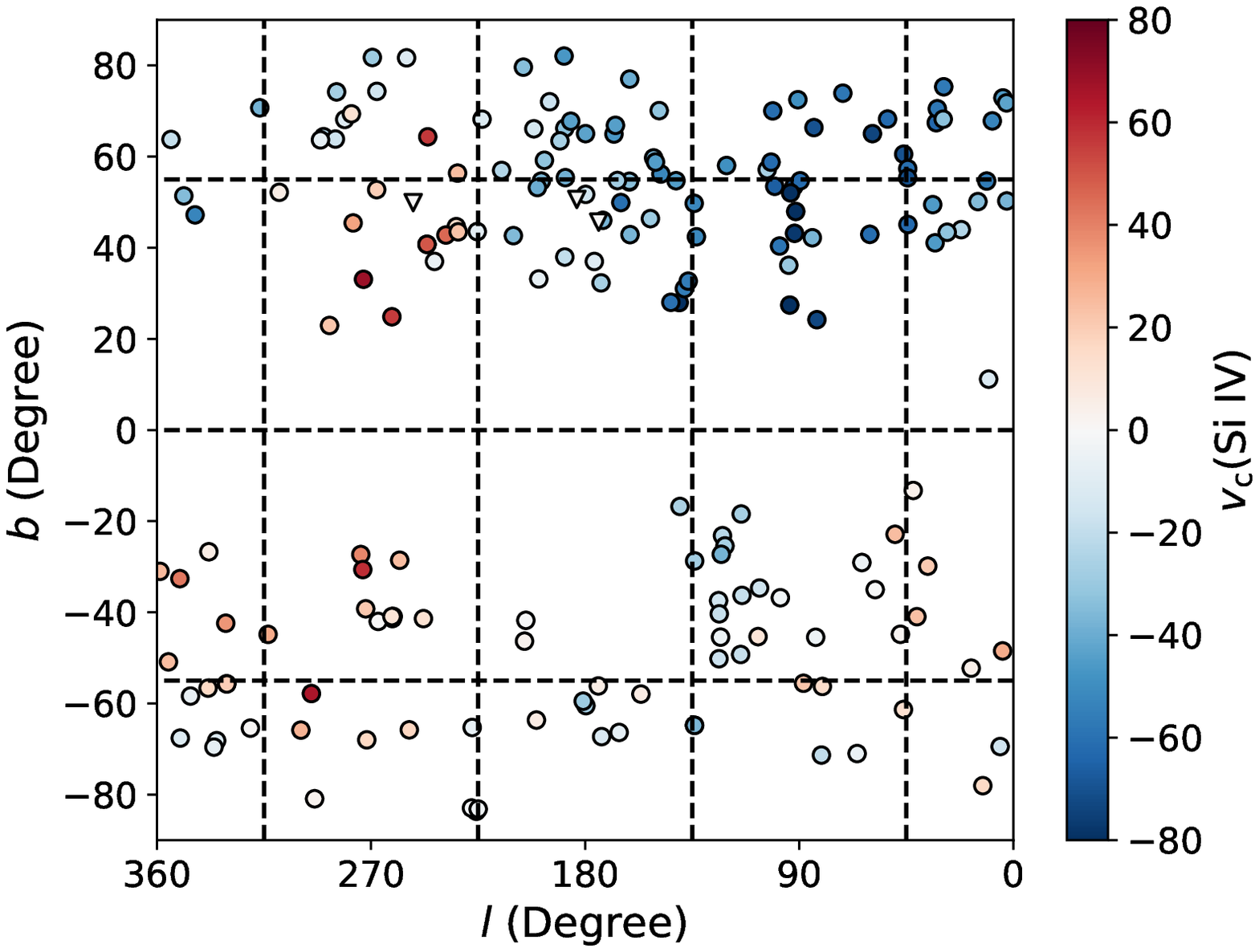}
\end{center}
\caption{The column density (left panel) and the line centroid (right panel) of the MW \ion{Si}{4} line shape sample. The entire sky is divided into 20 regions based on the Galactic longitude and Galactic latitude grids. For each region, we stack the column density line shape to obtain an average line shape in Fig. \ref{combine_ls}.}
\label{N_v_dist}
\end{figure*}

Because we will model the line shape, it is not necessary to decompose the absorption features into individual components for the following analyses.
However, for the common use for the community, we decompose the components based on separated peaks (Table \ref{sample}).
These components can be divided into two classes roughly based on the line centroids ($v_{\rm c}$): the MW disk with low-velocity CGM ($v_{\rm c} \lesssim 150 \kms$) and HVC ($v_{\rm c} \gtrsim 150 \kms$).
Using the coadded line shape, the total column density and line centroid for each component are calculated by integrals:
\begin{equation}
\begin{array}{l}
N = \int_{v_{\rm min}}^{v_{\rm max}} N(v) {\rm d} v \\
~\\
v_{\rm c} = \int_{v_{\rm min}}^{v_{\rm max}} N(v) v {\rm d} v / \int_{v_{\rm min}}^{v_{\rm max}} N(v) {\rm d} v,
\end{array}
 \end{equation}
where $v_{\rm min}$ and $v_{\rm max}$ are the minimum and maximum velocity for each component.
The column densities and line centroids for individual components are summarized in Table \ref{sample}, and plotted in Fig. \ref{N_v_dist}.
The HSLA spectrum is in the heliocentric frame, so the reported velocities are also in the heliocentric frame.
We do not convert this velocity into the LSR frame, and the motion of the Solar system is modeled in the kinematical model (Section \ref{velocity_field}).

There are three special issues need to note for the data reduction:

1. {\it Wrong continuum}.
Most ($>70\%$) of the sight lines use the continua generated in the first part of our method, using the automatically iterative method.
However, the automatic continuum may deviate from the true continuum significantly, due to the AGN broad emission line (i.e., sharp features), and high S/N (i.e., the continuum fitting progress catching the wing of absorption features).
Therefore, we inspect every automatic continuum fit, and do the continuum fitting by hand when necessary, where we select the continuum regions by hand and apply the spline fitting.

2. {\it Saturation of the strong line}.
Due to the higher oscillator $f$ factor, the strong line of the \ion{Si}{4} doublet is twice stronger than the weak line, which might be affected by saturation.
The saturation is shown as the feature that the weak line has a higher AODM column density than the strong line.
However, the higher column density of the weak line does not necessarily mean saturation occurred, because it can also be due to contamination in the weak line.
Therefore, we use Voigt fitting to test whether the weak line and the strong line are matched (i.e., two lines can be modeled by one Voigt model).
If the two lines are matched, we need to check whether the weak line is significantly affected by saturation.
We suggest that the weak line is not significantly affected by saturation if the AODM column density of the weak line is within $2 \sigma$ of the summation of the fitting components.
Then, we use the peak of the weak line column density shape as the peak of the combined line shape instead of the combination of both the strong and weak lines (Mrk 1392 in Fig. \ref{example_sightline}).
In practice, all of the saturation features are weak saturation features, where we could extract the peak shape of the column density line shape from the weak line, although the strong line is partially saturated.

\begin{figure*}
\begin{center}
\includegraphics[width=0.49\textwidth]{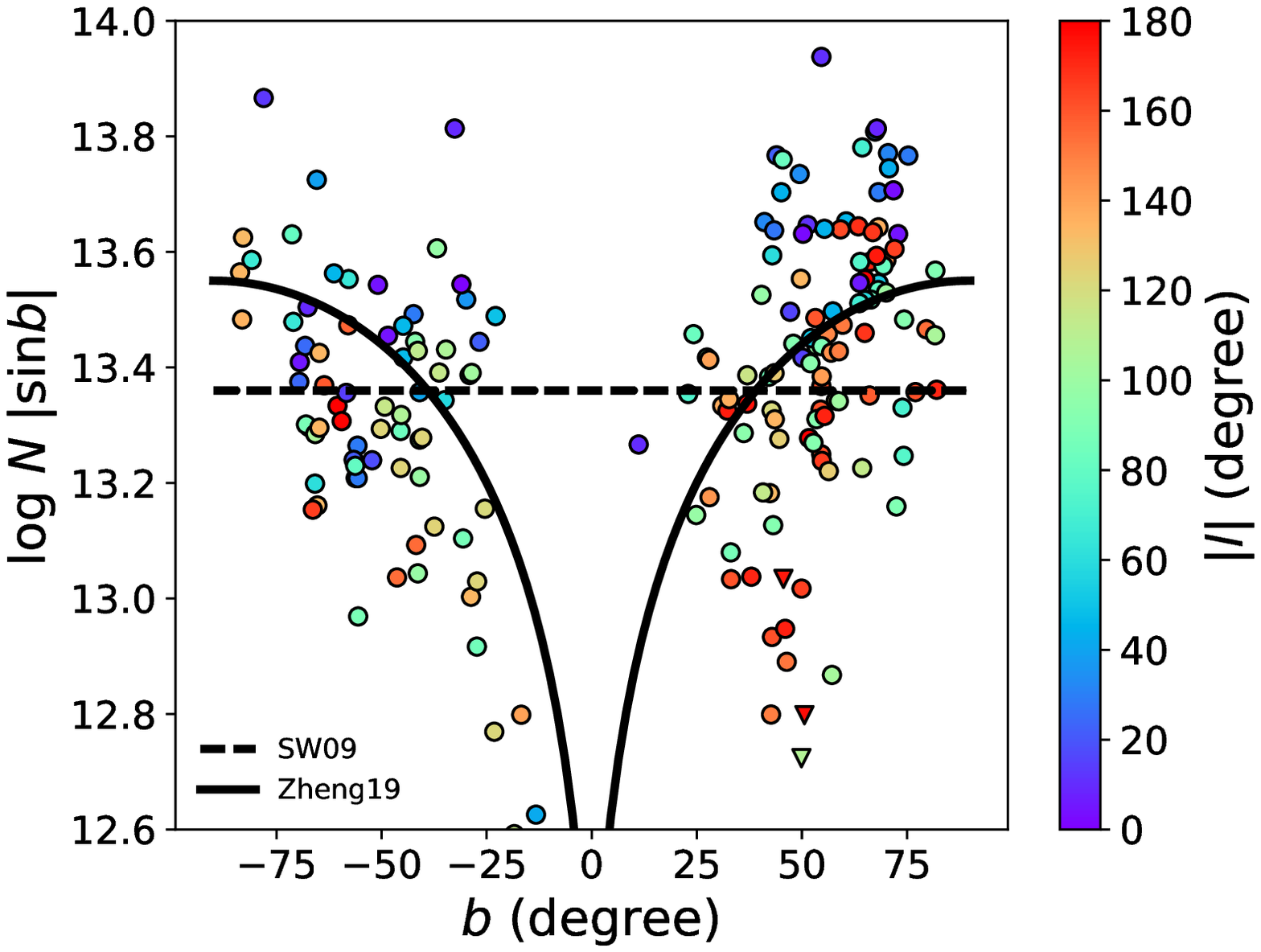}
\includegraphics[width=0.49\textwidth]{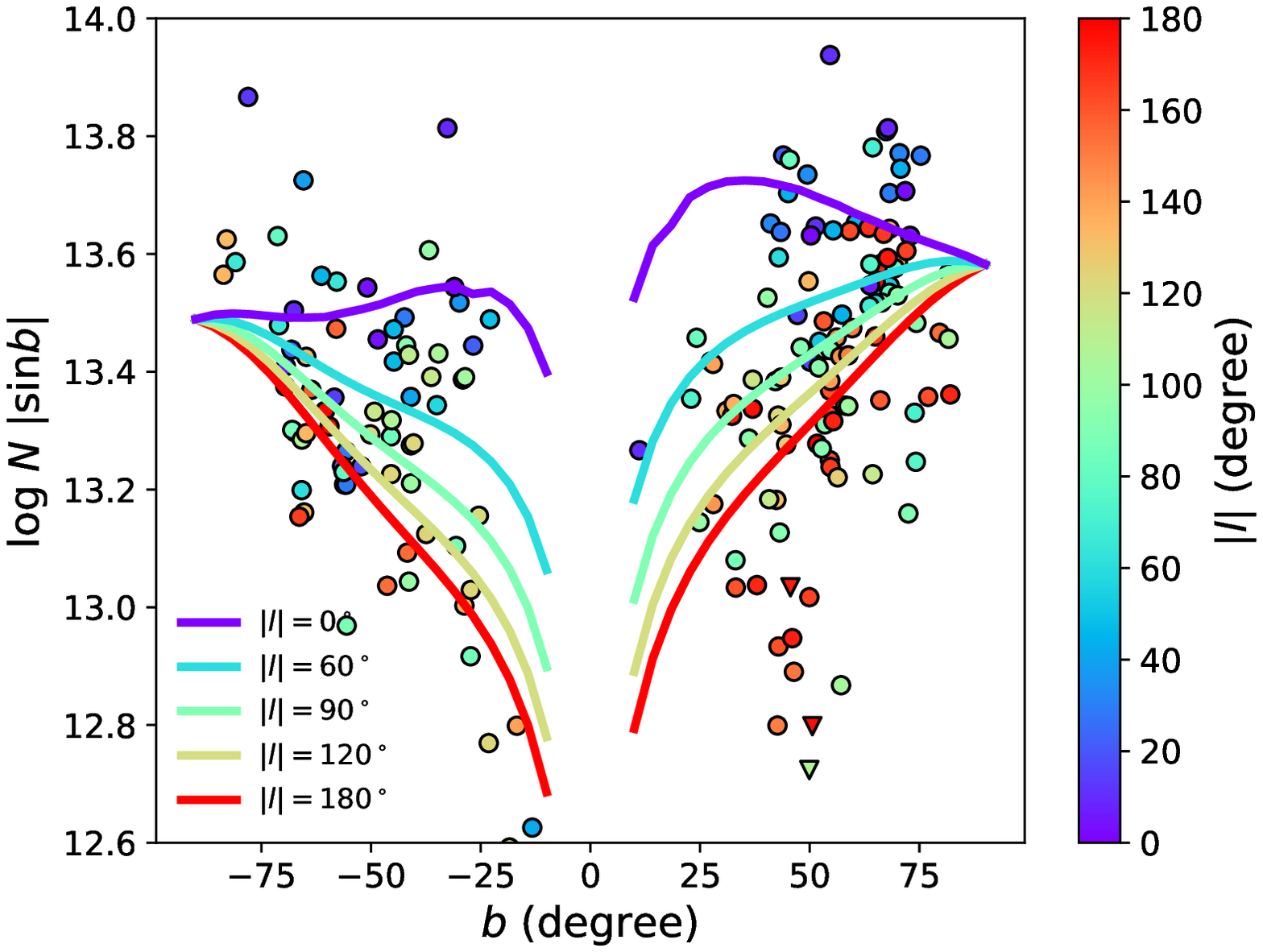}
\end{center}
\caption{Three previous models are compared to the new line shape sample: the \citet{Savage:2009aa} model (left panel), the \citet{Zheng:2019aa} model (left panel), and the \citetalias{Qu:2019ab} model (right panel). The circles are the measurements of column densities, while triangles are upper limits. The \citet{Savage:2009aa} model and the \citet{Zheng:2019aa} model are the original ones in literature, while the \citetalias{Qu:2019ab} model is the new one fitted to the new sample, which is consistent with the original model in \citetalias{Qu:2019ab}. With the new sample, it is more clear that the projected \ion{Si}{4} column density has dependences on both Galactic longitude and Galactic latitude, which implies the necessity of the 2D disk-CGM model.}
\label{qu2019}
\end{figure*}

3. {\it Badly blended features}.
If the two lines of the doublet are both blended with contamination lines, the true line shape cannot be extracted using the AODM approach.
There are 12 sight lines with the badly blended features, and we omit these sight lines from our sample (not in Table \ref{sample}).

One additional issue is name consistency.
The HSLA archive uses the target name in the {\it HST} proposals, among which some are nonstandard, and can be challenging to follow.
Therefore, we check the database SIMBad to extract more formal target names.
AGN are normally first detected in radio, optical, and X-ray surveys, so we choose names from these surveys.
Because this work is in the UV band, we prefer the UV name first (e.g., the Mrk survey), then the optical,  X-ray, and radio names,
The high-frequency surveys are HE, Mrk, PG, and RBS, as shown in Table \ref{sample}.

\section{Previous Models without kinematics}
\label{previous_models}
For the MW warm gas, various models have been proposed to explain the measured column density in both stellar and AGN sight lines (\citealt{Savage:2009aa}; \citetalias{Qu:2019ab}; \citealt{Zheng:2019aa}).
In Fig. \ref{qu2019}, the comparison between these three models are shown, and the observation data are from Table \ref{sample}.
The detailed comparisons between these models are in \citet{Zheng:2019aa} and \citetalias{Qu:2019ab}.
Here we just describe these models briefly.

Based on the stellar dominated sample ($\approx 100$ stellar sight lines and $\approx 20$ AGN sight lines), a thick warm gas disk is suggested \citep{Savage:2009aa}.
Previous studies employed a plane-parallel slab model, where $n(z) = n_0 \exp(-|z|/z_0)$, and $z_0$ is the scale height.
For the intermediate to high ionization state ions (e.g., \ion{Si}{4}, \ion{C}{4}, and \ion{O}{6}), the scale heights are about $2-4$ kpc \citep{Savage:2003aa, Bowen:2008aa, Savage:2009aa}.
This is a model with a one-dimensional (1D) variation over the $z$ height (above or below the disk), and the expected column density has a dependence of the column density on Galactic latitude ($N_{\rm slab}(b) = N_{\rm slab, 0}/\sin b$) for AGN sightlines, showing $\log N_{\rm slab, 0} = 13.36$ for \ion{Si}{4} \citep{Savage:2009aa}.
However, \citet{Zheng:2019aa} found the expectation in the slab disk model shows conflicts with their AGN sample (the COS-GAL \ion{Si}{4} sample).
At low Galactic latitudes, AGN sight lines have lower column density than the prediction of the slab disk model.
To solve this problem, \citet{Zheng:2019aa} proposed a two-component model, which contains both the 1D slab disk and an isotropic CGM component.
Applying this model to the AGN sample, they obtained a massive CGM model ($\log N_{\rm CGM} \approx 13.5$) with a relatively small disk ($\log N_{\rm slab, 0}\approx 12.11$).
This solution improves the fitting on the AGN sample (Fig. \ref{qu2019}), however, it is inconsistent with the thick disk supported by the stellar sample.
Therefore, there is tension between the slab disk model (dominated by the stellar sample) and the two-component disk-CGM model (determined by the AGN sample).

To relieve the tension between these two models, \citetalias{Qu:2019ab} proposed the two-dimensional (2D) disk-CGM model by introducing the radial profile of the disk into the two-component disk-CGM model.
Instead of the 1D variation of the disk $n(z) = n_0 \exp(-|z|/z_0)$, the 2D disk has a density distribution given by $n(z) = n_0 \exp(-|z|/z_0)\exp(-|r_{\rm XY}|/r_0)$, where $r_{\rm XY}$ is the radius in the Galactic XY plane (the disk midplane) and $r_0$ is the scale length.
Compared to the previous two models, the column density not only depends on Galactic latitude but also Galactic longitude (i.e., sight lines toward the GC $|l|=0$ have higher measured column densities; Fig. \ref{qu2019}).
Based on the 2D disk-CGM model, \citetalias{Qu:2019ab} suggested that the stellar sample and the AGN sample are not mutually incompatible, but can be fit with one model.

We applied the \citetalias{Qu:2019ab} model to the new AGN sample (Table \ref{sample}) and the \citet{Savage:2009aa} stellar sample (the same one used in \citetalias{Qu:2019ab}).
The AGN sample only uses components within $|v_{\rm c}| < 150 \kms$, which could be modeled by our kinematical model (Section \ref{kmodel}).
The fitting suggested similar results as \citetalias{Qu:2019ab} (notations taken from \citetalias{Qu:2019ab}, and \citetalias{Qu:2019ab} values in brackets): the disk scale length $r_0 = 6.62\pm0.86$ kpc ($5.9\pm1.1$ kpc); the disk scale height in northern hemisphere $z_0^{\rm N} = 3.86\pm 0.89$ kpc ($3.5\pm0.5$ kpc); the disk scale height in northern hemisphere $z_0^{\rm N} = 3.43\pm 0.61$ kpc ($2.3\pm0.4$ kpc); the CGM component perpendicular to the disk direction $\log N_{\rm nd}^{\rm CGM} = 12.96\pm0.31$ ($13.26\pm 0.08$);  the CGM component along the disk direction $\log N_{\rm mp}^{\rm CGM} = 10.70\pm1.90$ ($12.43\pm 0.42$). 
This similarity of the fitting results suggests that the absorption column densities between $100\kms<|v|\lesssim 150\kms$ have a minor effect on the large scale structure in the column density-only model.
In the following analyses, we will introduce the kinematics into the 2D disk-CGM model and adopt conclusions from \citetalias{Qu:2019ab} as the basis of the fiducial model.
Namely, we assume the north-south difference is mainly due to the scale height of the disk and will fix the midplane disk density and CGM density to be the same for the northern and southern hemispheres (details in Section \ref{overview}).

\section{The Cloud Path-Length Density}
\label{cloud_model}
The observations of the MW absorption features reveal that the warm gaseous components are clumpy rather than smoothly and uniformly distributed in the gaseous disk and halo \citep{Savage:1997aa, Savage:2003aa, Lehner:2003aa, Wakker:2003aa, Zsargo:2003aa, Bowen:2008aa, Savage:2009aa, Wakker:2012aa, Zheng:2019aa}.
The clumpy nature of the warm gas could be modeled by the patchiness parameter method.
The patchiness parameter is an additional uncertainty to lower the reduced $\chi$ to 1, when one wants to compare the observation with the model (\citealt{Savage:2009aa} and reference therein).
For intermediate-to-high ionization state ions (e.g., \ion{Si}{4}, \ion{C}{4}, and \ion{O}{6}), the typical values of the patchiness parameter are about $0.2-0.4$ dex for stellar-dominated samples, and $0.1-0.2$ dex for AGN-dominated samples.

This intrinsic scatter could have physical meanings.
To account for the intrinsic scatter, we assume the cloud nature of the warm gas, which is suggested by simulations \citep{Kwak:2010aa, Hummels:2018aa, Liang:2018aa, Shelton:2018aa, Ji:2019aa}.
In our modeling, the warm gas is assumed to be separated clouds with a typical column density of $\log N_{\rm sg}$ (``sg" denotes single).
Along a given sight line ($l$, $b$, $d$), the predicted number of clouds in the model is the $\mathcal{N}_{\rm c}$, which is an integral of the path-length density of clouds ($X$).
Then, the predicted column density is $N_{\rm sg} \mathcal{N}_{\rm c}$.
Because the number of clouds $\mathcal{N}_{\rm c}$ follows the Poisson distribution, the number of clouds has an intrinsic uncertainty of $\mathcal{N}_{\rm c}^{1/2}$, hence the column density uncertainty is $N_{\rm sg} \mathcal{N}_{\rm c}^{1/2}$.
Then, the uncertainty due to the number of cloud variation is $\mathcal{N}_{\rm c}^{-1/2} \log_{10} e$ dex.

\begin{table*}
\begin{center}
\caption{Parameters in the kinematical Model}
\label{params}
\begin{tabular}{cl}
\hline
\hline
Syms & Description \\
 \hline
$X_0^{\rm disk}$ $^a$ & The path-length density of clouds for the disk component at the GC. \\
$r_0$ \& $\alpha_{\rm rxy}$& The scale length and the index parameter of the disk density distribution along the radial direction: \\
& $n(r_{\rm XY}) = \exp (-(r_{\rm XY}/r_0)^{\alpha_{\rm rxy}})$. \\
$z_0$ $^a$ \& $\alpha_{\rm z}$ & The scale height and the index parameter of the disk density distribution along the $z$ direction: $n(z) = \exp (-(z/z_0)^{\alpha_{\rm z}})$. \\
$R_{\rm disk}$ $^a$ & The size of the disk without radial velocities, in the units of the scale height or scale length. \\
$\sigma_{\rm p}$ & The additional patchiness parameter for the stellar sample. \\
$X_0^{\rm mp}$ $^a$ & The path-length density of cloud for the CGM component at the core region along the midplane of the disk. \\
$X_0^{\rm nd}$ $^a$ & The path-length density of cloud for the CGM component at the core region along the normal direction of the disk. \\
$r_{\rm c}$ & The core radius of the $\beta$-model for the CGM component. \\
$\beta$ & The slope in the $\beta$-model for the CGM density distribution. \\
$v_{\rm rot}$ & The rotation velocity on the flat part of the rotation curve in the midplane. \\
$v_{\rm rad, 10}$ $^a$ & The radial velocity at 10 kpc assuming constant accretion $\dot{M}$.\\
$v_{\rm abs}$ & The intrinsic broadening velocity of the \ion{Si}{4} absorption, $=0.707~ b$ factor. \\
$v_{\rm rand}$ & The random motion of the cloud along the sight line direction. \\
$N_{\rm sg}$ & The \ion{Si}{4} column density of single cloud. \\
\hline

\end{tabular}
\end{center}
$^a$ For these parameters, they may have NS asymmetry. Then, ``N" or ``S" denote values for the northern or southern hemispheres, respectively.
\end{table*}

If this uncertainty is caught by the patchiness parameter, one could use the patchiness parameter to estimate the typical column density for the warm gas.
In Section \ref{previous_models}, we applied the \citetalias{Qu:2019ab} model to the new AGN sample, and obtain a set of new parameters.
Using the new parameters, we estimate the patchiness parameters (fitting the residuals with a Gaussian function) are 0.162 dex for the AGN sample and 0.360 dex for the stellar measurements.
For the AGN sample, $\sigma_{\mathcal{N}_{\rm c}}/\mathcal{N}_{\rm c} = \sigma_N / N = 0.162 / \log_{10} e = \mathcal{N}_{\rm c}^{-1/2}$, then we know the average number of clouds is $\mathcal{N}_{\rm c} \approx 7.2$.
The median column density of the AGN sample is $\log N = 13.53$, then the estimated typical column density of single cloud is $\log N_{\rm sg} = 12.67$.
Similarly, the average number of clouds is $\mathcal{N}_{\rm c} \approx 1.5$ for the stellar sample, and the typical column density of single cloud is $\log N_{\rm sg} = 13.00$ (with a median stellar sample column density of $\log N = 13.15$).

The typical column density of a single cloud of the stellar sample is much larger than the AGN sample by a factor of $0.3-0.4$ dex.
This difference has two interpretations on whether this difference is real.
First, if this difference is real, then there is a physical difference between the gas on the warm gas disk and the gas beyond the disk (i.e., CGM), which indicates the warm cloud in the CGM has a smaller typical column density than the disk.
This might be caused by the pressure difference in the disk (high pressure; $\approx 10^3\rm~K~ cm^{-2}$) and in the CGM (low pressure; $\approx 1-10^2\rm~K~cm^{-2}$).
Assuming that clouds in the disk and the CGM have a similar total mass and temperature (determined by ionization state), the column density has a dependence on the pressure as $N \propto P^{2/3}$, which leads to lower typical column density in the CGM.

Another possibility for the physical difference is that the ISM might be more structured than the CGM, which is affected by stellar activity, such as from spiral arms.
Based on \ion{O}{6} observations, \citet{Bowen:2008aa} suggested that the scatter of the \ion{O}{6} column density does not depend on the distance, which indicates that the scatter caused by the ISM structures dominates the statistical scatter due to a number of clouds.
In this case, the column density of a cloud should be a distribution rather than a fixed column density.
Here, we emphasize that it is also unclear whether the CGM could be approximated by a fixed column density, so the cloud model of the warm gas is still an assumption.
However, it is certain that there are significant differences between the stellar sample (for disk) and the AGN sample (mainly for the CGM).

Second, the difference between the disk and the CGM clouds may not be real, because there are several other observational uncertainties for the stellar sample.
For example, the distance to a star may have significant uncertainties.
The reported distance uncertainties in \citet{Savage:2009aa} are about $\approx 0.1$ dex.
The distances in the \citet{Savage:2009aa} sample are from \citet{Bowen:2008aa}, which are mainly spectroscopic distances (i.e., obtaining absolute magnitude based on the spectral type and estimating the distance by comparing with apparent magnitude; see their Appendix B for more details).
We believe that this is the best way to obtain a large uniformly-reduced distance sample, but the uncertainty of this method can be large \citep{Shull:2019aa}, which contributes to the difference between the stellar sample and the AGN sample.
Also, the stellar continua sometimes contain intrinsic features (e.g., stellar winds, and absorption close to the star).
Especially in the relatively low $S/N$ ($\approx 5$) spectrum, the stellar features may be difficult to identify, which could introduce additional uncertainty in the column density measurements of interstellar absorption.
This explanation is supported by the result that the additional patchiness parameter of the stellar sample is sightly reduced when we only use the STIS sample (Section \ref{results}).
However, the STIS-only fitting model still has a none-zero additional patchiness parameter ($\sigma_{\rm p} = 0.15\pm 0.09$).
This may be due to the systematic uncertainty of the STIS sample, but it is highly likely that the ISM is physically different from the CGM.

The reason for the difference between ISM and CGM is beyond the scope of this paper, since the structure of the ISM is a huge topic.
Here we just introduce an additional patchiness parameter to the stellar sample ($\sigma_{\rm p}$).
This patchiness parameter may be due to the stellar intrinsic features, or physical difference between the warm gas in the disk and the CGM gas.
Then, the cloud assumption is only applicable to the CGM observation to model the intrinsic scatter, which cannot be applied to ISM, because the additional patchiness parameter $\sigma_{\rm p}$ significantly affects the scatter of column density measurements in stellar sight lines (Section \ref{results}).

In the following analyses, we use the path-length density ($X$) and $\log N_{\rm sg}$ for a single cloud to replace the density distribution.
This implementation of the density formulation not only predicts the total column density, but also predicts model uncertainty for each sight line due to the variation of the total number of clouds along the sight line ($\mathcal{N}_{\rm c}$).
For the AGN sample, the model uncertainty will be used to reproduce the intrinsic scatter (previously, the patchiness parameter).
For the stellar sight line, the predicted column density uncertainty will be combined with the additional patchiness parameter (more details in Section \ref{model_prediction}).

\section{The 2D disk-CGM Model with kinematics}
\label{kmodel}
In addition to the \citetalias{Qu:2019ab} model, we consider the kinematics in the new model, which includes two major parts: the ion density distribution and the bulk velocity field.
The density distribution is divided into two components as the disk and the CGM phenomenologically (Section \ref{density_field}).
Here, we note that this decomposition does not imply that these two components are physically de-associated with each other.
Instead, the combination of both components is to approximate the real warm gas distribution.
For the velocity field, there are also two components, the rotation velocity and the radial velocity (Section \ref{velocity_field}).
With all of these assumptions, we calculate the differential column density distribution to compare with the observed column density line shape sample (Section \ref{model_prediction}).
The Bayesian framework employed to estimate the best parameters is introduced in Section \ref{Bayesian_frame}.

\begin{figure*}
\begin{center}
\includegraphics[width=0.99\textwidth]{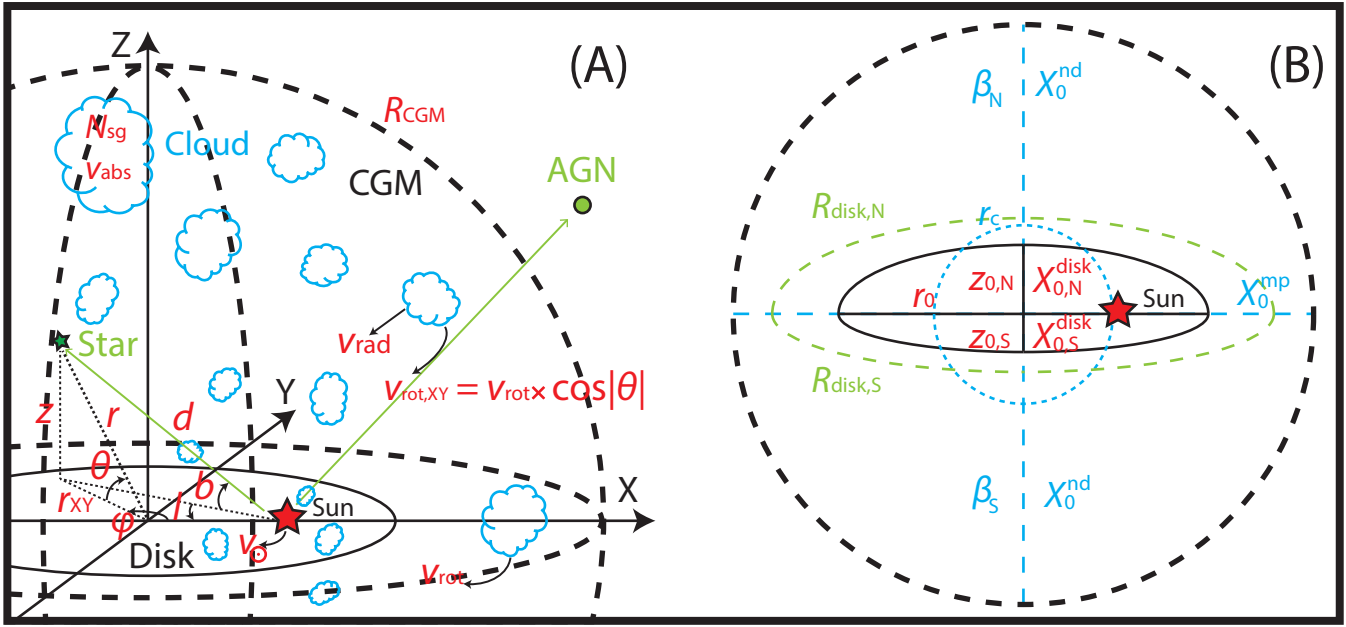}
\end{center}
\caption{An illustration of the kinematical model (not scaled). Left panel: the coordinate parameters ($x$, $y$, $z$, $\phi$, and $\theta$) and parameters to describe the velocity field ($v_{\rm rot}$, $v_{\rm rad}$, and $v_\odot$). Right panel: the parameters that describe the ion density distribution: the disk component ($r_0$, $z_0$, and $X_0^{\rm disk}$) and the CGM component ($r_{\rm c}$, $\beta$, $X_0^{\rm mp}$, and $X_0^{\rm nd}$).}
\label{cartoon}
\end{figure*}

The assumptions and implementation of this model are described in the following subsections, while the varied parameters in the kinematical model are summarized in Table \ref{params}.

\subsection{The Density Model}
\label{density_field}

Before introducing the \ion{Si}{4} density distribution for both disk and CGM components, we note that the model has the origin at the Galactic center (GC) rather than the Solar system.
Therefore, we need to convert a given position in the Galactic coordinate system ($l$, $b$, and $d$) to the Galactic XYZ coordinates ($x$, $y$, and $z$; Fig. \ref{cartoon}):
\begin{eqnarray}
        x &=& 8.5 {\rm~kpc} - d \cos l \cos b, \notag \\
        y &=& - d \sin l \cos b, \notag \\
        z &=& d\sin b.
\end{eqnarray}
Here, we assume the Solar system is at $r_\odot = 8.5$ kpc \citep{Ghez:2008aa}.
The coordinates $x$ and $y$ are in the disk midplane, while $z$ is the height above or below the disk midplane.
Then, the distance to the GC ($r$), the XY plane distance to the GC ($r_{\rm XY}$), and the corresponding spherical angles ($\phi$ and $\theta$; Fig. \ref{cartoon}) are
\begin{eqnarray}
r &=& (x^2 + y^2 +z^2)^{1/2}, \notag \\
r_{\rm XY} &=& (x^2 + y^2)^{1/2}, \notag \\
\sin \phi &=& \frac{y}{r_{\rm XY}},~ \cos \phi = \frac{x}{r_{\rm XY}},~ \sin \theta = \frac{z}{r}.
\end{eqnarray}
These parameters will be used in the definition of the ion density distribution and the velocity field.

Following the \citetalias{Qu:2019ab} model, the total cloud path length density is the summation of the disk and the CGM components.
The total cloud path length density is 
\begin{equation}
\label{eqn_X}
X(l, b, q) = X^{\rm disk}(r_{\rm XY}, z) + X^{\rm CGM}(r, \theta),
\end{equation}
where $X^{\rm disk}$ and $X^{\rm CGM}$ are the path density of clouds for the disk and the CGM components, respectively.

The disk density distribution is 2D ($r_{\rm XY}$ and $z$) and axial symmetric with the axis as the disk normal line through the GC.
The 2D disk density distribution includes both radial ($r_{\rm XY}$) and vertical ($z$) profiles, which are independent between each other.
Empirically, both radial and vertical profiles are exponential.
Then, the cloud path length density of the disk component has a format of:
\begin{equation}
X^{\rm disk}(r_{\rm XY}, z) = X_0^{\rm disk}\exp(-\frac{|r_{\rm XY}|}{r_0}) \exp(-\frac{|z|}{z_0}),
\end{equation}
where $X^{\rm disk}$ is the path-length density for the disk component, and $r_0$ and $z_0$ are the scale length and the scale height of the disk.
As discussed in \citetalias{Qu:2019ab}, the radial and vertical profiles might have other formats (e.g., the Gaussian function).
Therefore, we parameterize this variation by introducing two index parameters ($\alpha_{\rm rxy}$ and $\alpha_{\rm z}$):
\begin{equation}
X^{\rm disk}(r_{\rm XY}, z) = X_0^{\rm disk}\exp(-(\frac{|r_{\rm XY}|}{r_0})^{\alpha_{\rm rxy}} -(\frac{|z|}{z_0})^{\alpha_{\rm z}}).
\end{equation}
Larger $\alpha_{\rm rxy}$ or $\alpha_{\rm z}$ values lead to sharper edges of the ion distribution for the radial or the $z$-height directions.
If $\alpha_{\rm rxy}$ or $\alpha_{\rm z}$ are 1, the density distributions are exponential, while if $\alpha_{\rm rxy}$ or $\alpha_{\rm z}$ are 2, the density distributions are Gaussian.

Previously, the CGM component is modeled as the column density distribution (isotropic in \citealt{Zheng:2019aa} or dependent on Galactic latitude in \citetalias{Qu:2019ab}) rather the density distribution.
This is the limitation of models that only consider the integrated column density (without line shape; i.e., kinematics) sample.
In the new kinematical model, we could constrain the CGM density distribution with the velocity field.
Here, we assume the $\beta$-model for the CGM density distribution, which is commonly adopted in X-ray investigations to study the quasi-hydrostatic equilibrium hot gas for both the MW and external galaxies \citep{Bogdan:2013aa, Miller:2013aa, Li:2017aa, Li:2017ac, Li:2018aa}.
\begin{equation}
X^{\rm CGM}(r, \theta) = X_0^{\rm CGM}(\theta) \times (1+(\frac{r}{r_{\rm c}})^2)^{-3\beta/2},
\end{equation}
where $X_0^{\rm CGM}(\theta)$ is the core density, $r_{\rm c}$ is the core radius, and $-3\beta$ is about the power law index at large radii.
$X_0^{\rm CGM}(\theta)$ has a dependence on $\theta$, which accounts for the anisotropy of the MW CGM as introduced in \citetalias{Qu:2019ab}.

\citetalias{Qu:2019ab} found that the MW warm CGM column density distribution (traced by both \ion{Si}{4} and \ion{O}{6}) is anisotropic, which is not a constant column density over the entire sky. 
This is also found in external galaxies for the cool gas ($\approx10^3\rm~K$; traced by \ion{Mg}{2} and \ion{Fe}{2}) in the CGM \citep{Bordoloi:2011aa, Lan:2018aa, Martin:2019aa}.
These studies showed that the \ion{Mg}{2} or \ion{Fe}{2} column density depends on the azimuthal angle (the projected angle related to the minor axis).
In \citetalias{Qu:2019ab},  we considered this variation by assuming a column density distribution as a function of Galactic latitude.
Here, we improve on this assumption by introducing a dependence on $\theta$ ($X_0^{\rm CGM}(\theta)$) instead of Galactic latitude $b$, where $\theta$ is the angle measured from the midplane with the origin at the GC (see the cartoon in Fig. \ref{cartoon}):
\begin{equation}
\log X_0^{\rm CGM}(\theta) = \log X_0^{\rm mp} \cos^2 \theta + \log X_0^{\rm nd} \sin^2 \theta,
\end{equation}
where $ X_0^{\rm mp}$ is the cloud path-length density along the radial direction in the disk midplane (``mp"), and $ X_0^{\rm nd}$ is the cloud path-length density along the $z$ direction (the normal direction; ``nd") perpendicular to the disk.
This is similar to the format in \citetalias{Qu:2019ab} ($\log N_{\rm CGM}(b) = \sqrt{\log^2 N_{\rm mp} \cos^2 b+ \log^2 N_{\rm nd} \sin^2 b}$).
Here, we omit the square root in the new function, because $\log X$ could be negative, while  $\log N \approx 12-13$ is always positive.
These two formats show similar variations (with differences $< 10\%$) when the difference of $\log N$ or log$X$ in two directions are within $~1-2$ (i.e., $|\log X_{\rm mp}-\log X_{\rm nd}|<2$ or $|\log N_{\rm mp} - \log N_{\rm nd}| < 2$).

The $\beta$-model converges to the power law model at large radii ($r >> r_{\rm c}$).
Typically, the $r_{\rm c}$ of external galaxies are not resolved by current X-ray instrument, so it is $\lesssim 10$ kpc \citep{Bogdan:2013ab, Li:2018aa}.
For the MW, $r_{\rm c}$ is measured to be $2.5\pm 0.2 \rm~ kpc$ using \ion{O}{7} and \ion{O}{8} X-ray emission lines \citep{Li:2017aa}.
However, in our \ion{Si}{4} absorption line study, this parameter cannot be constrained, because few sightlines pass through the GC at distances $<2.5$ kpc.
Therefore, we fix $r_{\rm c}$ to 2.5 kpc in the following analyses, which will not affect the power-law approximation at large radii.

The $\beta$-model should have a maximum radius ($R_{\rm max}$), because the total number of ions does not converge when $\beta < 0.7$, which is pretty likely ($\beta=0.5$ for hot gas in the MW; \citealt{Li:2017aa}).
Here, $R_{\rm max}$ is fixed to 250 kpc (about the virial radius of the MW).
Our model is not sensitive to this parameter as discussed in Section \ref{gas_dist}.

\subsection{The Velocity field}
\label{velocity_field}
The bulk velocity field has two major contributors: the Galactic rotation and the radial motion (outflow or inflow).
For a given position ($r$, $\theta$, and $\phi$; Fig. \ref{cartoon}), the total velocity is the combination of both rotation and radial velocities:
\begin{equation}
\vec{v}_{\rm bulk} (r, \theta, \phi) = \vec{v}_{\rm rot} (r, \theta, \phi) + \vec{v}_{\rm rad} (r, \theta, \phi).
\end{equation}

The rotation velocity is approximated by a linear part ($r_{\rm XY} \leq 0.5$ kpc) and a flat part ($r_{\rm XY}> 0.5$ kpc; \citealt{Kalberla:2008aa}).
At the midplane, the velocity of the flat part is a free parameter in our model ($v_{\rm rot}$):
\begin{eqnarray}
v_{\rm rot} (r_{\rm XY}) &=& (r_{\rm XY}/{0.5~\rm kpc}) v_{\rm rot},~ r_{\rm XY} \leq 0.5 \rm ~kpc \notag \\
& = & v_{\rm rot},~ r_{\rm XY} > 0.5 \rm ~kpc.
\end{eqnarray}
When it is above (the northern hemisphere) or below (the southern hemisphere) the midplane, we consider a rotating cylinder (i.e., no $z$-component velocity).
Then the value of rotation velocity is
\begin{equation}
v_{\rm rot} (r, \theta) = v_{\rm rot}(r_{\rm XY}) \cos \theta,
\end{equation}
and the three-dimensional (3D) rotation velocity vector is
\begin{equation}
\vec{v}_{\rm rot} (r, \theta, \phi) = (-\sin \phi, \cos \phi, 0) \cdot v_{\rm rot}(r, \theta). 
\end{equation}

The radial velocity is introduced to account for the possible inflow or outflow of the warm gas.
The warm gas disk is expected to have no radial velocity, which is similar to the \ion{H}{1} disk \citep{Marasco:2011aa}.
Therefore, we assume that there is a boundary for the radial velocity, and the boundary follows the shape of the isodensity line of the disk component (as stated in Section \ref{density_field}):
\begin{equation}
(\frac{|r_{\rm XY}|}{r_0})^{\alpha_{\rm rxy}} +(\frac{|z|}{z_0})^{\alpha_{\rm z}} = R_{\rm disk},
\end{equation}
where $R_{\rm disk}$ is a dimensionless parameter to describe the size the disk without the radial velocity.
At this boundary the cloud path density of the disk component is a constant of $X_0^{\rm disk} \exp(-R_{\rm disk})$.

We assume that the radial velocity is spherically symmetric, so it only has a dependence on the distance to the GC ($r$).
For the radial velocity dependence on the radius, we assume a constant accretion/outflow rate at different radii for the CGM.
\begin{equation}
4\pi r^2 v_{\rm rad}(r) N_{\rm sg} X^{\rm CGM}(r) = \dot{M} = const.,
\end{equation}
where $N_{\rm sg} X^{\rm CGM}(r)$ is equivalent to the ion density of the CGM component, and $\dot{M}$ is the mass loading rate (outflow) or the accretion rate (inflow) of the CGM component.
For simplicity, we use the radial velocity at 10 kpc as the characteristic radial velocity.
Then, the radial velocity dependence on the radius is
\begin{equation}
v_{\rm rad}(r) = \frac{10^2 X^{\rm CGM}(10)}{r^2 X^{\rm CGM}(r)} v_{\rm rad, 10},
\end{equation}
where $v_{\rm rad, 10}$ is the free parameter in our model (Table \ref{params}).
Positive value of $v_{\rm rad, 10}$ indicates outflow, while negative $v_{\rm rad, 10}$ means gas accretion.
Then, the 3D radial velocity is
\begin{equation}
\vec{v}_{\rm rad} (r, \theta, \phi) = (\frac{x}{r}, \frac{y}{r}, \frac{z}{r}) \cdot v_{\rm rad}(r).
\end{equation}

Because the calculated velocity of the line shape sample is in the heliocentric frame, we need to consider the Solar motion in the velocity calculation.
We have two ways to include the Solar motion.
One is fixing the Solar motion to $\vec{v}_{\odot} = (-9, -232, 7) \kms$ \citep{Delhaye:1965aa}, while another choice is varying the velocities of the Solar motion as free parameters.
Based on our tests, the \ion{Si}{4} sample is insufficient to constrain the Solar motion, because of the sample size and the complicated variation of the warm gas distributions and kinematics (affected by stellar activity).
Therefore, we fix the Solar motion as $(-9, -232, 7) \kms$ in the fiducial motion in Section \ref{overview}.
The warm gas velocity relative to the Solar motion is:
\begin{equation}
\vec{v}_{\rm bulk, rel} = \vec{v}_{\rm rot} + \vec{v}_{\rm rad} - \vec{v}_{\odot}. \notag
\end{equation}
Finally, the projected velocity is (because we define the Galactic radial velocity in Section \ref{velocity_field}, we use the ``projected velocity" to represent the normal radial velocity):
\begin{eqnarray}
\label{vproj_eqn}
v_{\rm proj}(l, b, q) = \vec{v}_{\rm bulk, rel}(l, b, q) \cdot \vec{\mathbf{n}},
\end{eqnarray}
where $\vec{\mathbf{n}}=(-\cos l\cos b, -\sin l \cos b, \sin b)$ is the direction vector toward $l$ and $b$.

In addition to the bulk velocity field, a single cloud may have a random motion.
For this variation, we assume the random motion follows a normal distribution with a mean value of $0\kms$ and a standard deviation of $v_{\rm rand}$, which is a free parameter in our model (Table \ref{params}). 
The inclusion of $v_{\rm rand}$ in the kinematical model is described in Section \ref{model_prediction} and Section \ref{Bayesian_frame}.

\subsection{The Model Prediction}
\label{model_prediction}
Based on the assumptions of the density distribution and the bulk velocity field, we could predict the observed column density (stellar) and differential column density line shape (AGN) for each sight line.
As stated in Section \ref{cloud_model}, we also consider the associated model uncertainty due to the number variation of clouds in each sight line.

For each stellar sight line ($l$, $b$, and $d$), we calculate the total number of clouds ($\mathcal{N}_{\rm c}$) by integrating $X$ (the cloud path length density) over the path length to the target.
Then, the predicted column density of stellar sight line is expected to be $N_{\rm st} = \mathcal{N}_{\rm c} N_{\rm sg}$, where $N_{\rm sg}$ is the column density for individual clouds.
To consider the model uncertainty, we assume the distribution of the cloud number follows a Poisson distribution.
Therefore, the cloud number uncertainty is $\mathcal{N}_{\rm c}^{1/2}$, and the model uncertainty of the column density is $\sigma_{N} = \mathcal{N}_{\rm c}^{1/2} N_{\rm sg}$ in the linear scale.
In the logarithm scale, the model uncertainty is $\sigma_{N} = \mathcal{N}_{\rm c}^{1/2} \log_{10} e$ dex.
Besides the model-predicted uncertainty, the stellar sample also suffers from an additional uncertainty (patchiness parameter; $\sigma_{\rm p}$) as discussed in Section \ref{cloud_model}.
Then, the total uncertainty of one stellar sight line is $\sigma_{\rm st}^2 = \sigma_{N}^2 +\sigma_{\rm p}^2$.

For the AGN sample, we need to consider the line shape by introducing the velocity field (Section \ref{velocity_field}).
For each AGN sight line ($l$, $b$), there are five steps to obtain the model prediction of the line shape.
(1) We calculate the max path length ($d_{\rm max}$) within the MW halo ($R_{\rm max}=250$ kpc), which is the distance between the Sun and the maximum radius of the sphere along a sight line ($l$ and $b$).
(2) The entire path length is divided into bins with a width of $\Delta q =0.1$ kpc.
Because the cloud path length density at 250 kpc is extremely low, we ignore that the last bin that may not be 0.1 kpc.
(3) For one bin at a distance of $q$, we calculate the cloud path-length density ($X(l, b, q)$) in the bin using Eq. (\ref{eqn_X}) and the corresponding projected (observed radial) velocity $v_{\rm proj}(l, b, q)$ using Eq. (\ref{vproj_eqn}).
Then, the cloud number in the path length bin is $\Delta \mathcal{N}_{\rm c}(l, b, q) = X(l, b, q) \Delta q$.
This indicates that there are $\Delta \mathcal{N}_{\rm c}$ clouds at the projected velocity of $v_{\rm proj}$.
(4) After the calculation of all path length bins, we rebin the cloud number based on the projected velocity.
The adopted projected velocity bin is the observed velocity bin for each AGN sight line.
Therefore, we could know the cloud number in each observed velocity bin ($\mathcal{N}_{\rm c}(v)$).
(5) The cloud number in each velocity bin is converted to the column density ($N_{\rm c}(v) = (\mathcal{N}_{\rm c}(v) N_{\rm sg}$).

\begin{figure*}
\begin{center}
\includegraphics[width=0.49\textwidth]{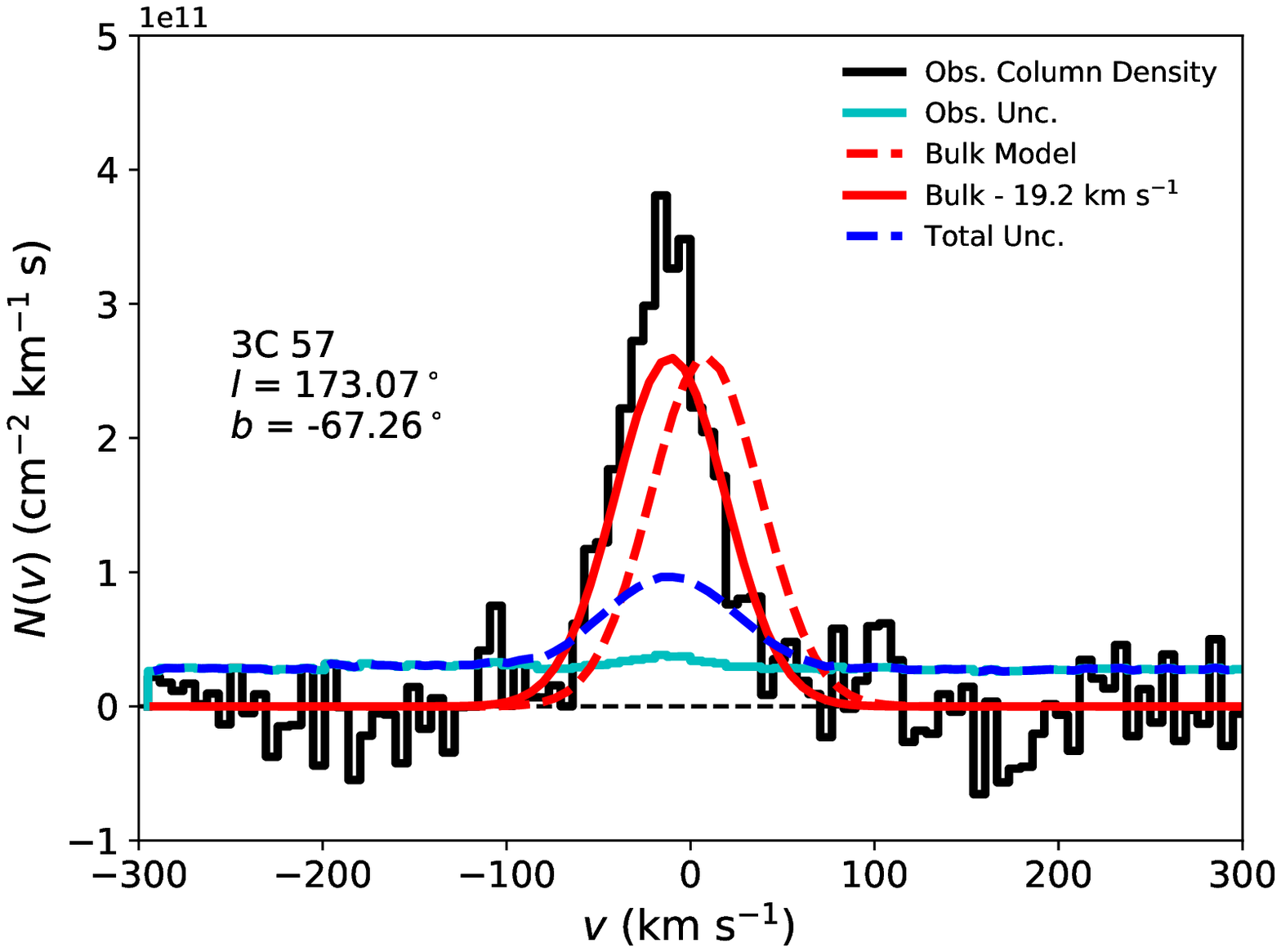}
\includegraphics[width=0.49\textwidth]{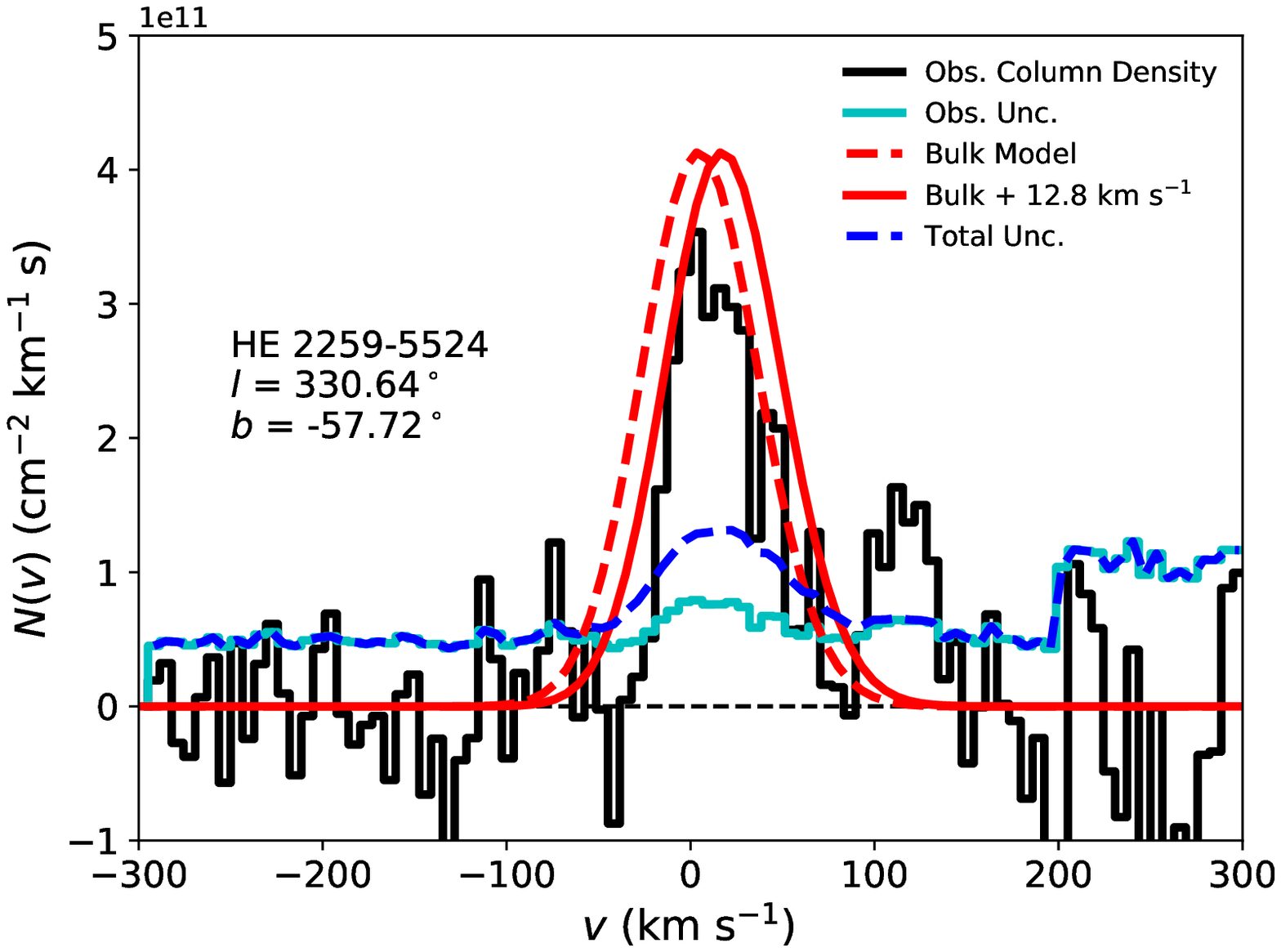}
\end{center}
\caption{Two examples sight lines of 3C 57 (left) and HE 2259-5524 (right) showing the velocity shift along the sight line direction. The red dashed lines are the predicted column density line shape without the random motion (i.e., only with the bulk velocity field). The red solid lines are the model considering the random motion along the sight line direction determined from the power spectrum of the cross-correlation. The blue dashed lines are the total uncertainty combining both the observation uncertainty (cyan lines) and the model uncertainty.}
\label{vshift_example}
\end{figure*}

Similar to the stellar sample, we also calculate the corresponding model uncertainty for the line shape.
For each bin, the uncertainty of the cloud number is $\mathcal{N}_{\rm c}(v)^{1/2}$, based on the Poisson statistic assumption.
Then, the naive column density uncertainty is $\mathcal{N}_{\rm c}(v)^{1/2}  N_{\rm sg}$ for individual velocity bins.

However, there is one issue when applying this naive column density uncertainty to the column density line shape.
By varying $\log N_{\rm sg}$, we expect to capture the intrinsic scatter of the integrated column density as stated in Section \ref{cloud_model}.
Nevertheless, this $\log N_{\rm sg}$ value for the integrated column density overestimates the uncertainty in each bin of the column density line shape.

Here, we use the $\chi^2$ framework to show the basic idea, although we implement the Bayesian frame in the following analyses (Section \ref{Bayesian_frame}).
For one sight line, the total observed column density is $N$, the model column density is $M$, and the model uncertainty is $E$.
With a suitable $\log N_{\rm sg}$, one could always obtain a reduced $\chi^2 = (N-M)^2/E^2 = 1$.
The situation is different for the line shape calculation.
Assuming there are $m$ bins for the spectrum, each bin has an average column density of $N/m$, a model column density of $M/m$, and a model uncertainty of $E/m^{1/2}$, based on the Poisson statistics.
Then, for each bin in the spectrum, the corresponding $\chi^2 = 1/m$, which is unacceptable, although the adopted $\log N_{\rm sg}$ makes the total column density sample have reduced $\chi^2 = 1$.
We make the model self consistent by using the line width (in number of spectral bins) to reduce the naive model uncertainty for each bin (i.e., $\mathcal{N}_{\rm c}(v)^{1/2}  N_{\rm sg} / m^{1/2}$).
Here, for each sight line the value of $m$ is not the number of bins for the entire spectrum, because the spectrum also contains the continuum, which should not contribute to $m$.
Finally, the value of $m$ is fixed to be the number of spectral bins that have model column densities $> 10^{10}\rm \cmsq$ (much lower than the detection limits in our sample).
Using this criterion, we exclude the continuum region, which would lower the significance of the model.

By now, the model column density and the uncertainty are both intrinsic values, which have not been affected by several broadening processes (e.g., thermal or turbulence broadening).
These broadening processes cannot be distinguished from observations, so we introduce the absorption broadening velocity $v_{\rm abs}$ to model them together.
Using $v_{\rm abs}$, we generate a Gaussian kernel and convolve this kernel to the intrinsic column density model and uncertainty.
The convolved model and uncertainty are the model for AGN sight lines, considering the density distribution and the bulk velocity field.

As introduced in Section \ref{velocity_field}, warm gas clouds could also have random motions besides the bulk velocity.
To determine the random motion component, we extract the power spectrum from the cross-correlation of the convolved model (without the random motion) and the observation for every sight line.
Then, the peak position of the power spectrum corresponds to the required velocity shift ($v_{\rm shift}$) to account for the random motion.
We show two examples (3C 57 and HE 2259-5524) in Fig. \ref{vshift_example}, where the input model is the preferred model in Section \ref{overview}.
The bulk velocity model of 3C 57 shows a similar shape as the observed line shape, but the line centroid is shifted.
The bulk velocity model with a velocity shift of $-19.2\kms$ reproduces the line shape better, which is the final model to compare with the data.
For HE 2259-5524, the line shape is not reproduced as well as 3C 57, but the velocity shift also shows up by the peak of the column density line shape.
Here, we emphasize that the velocity shift is part of the kinematical model to obtain the final model-predicted line shape for AGN sight lines as well as the ion density distribution and the bulk velocity field.

Finally, for the stellar sample, we have the model predicted column density and the total uncertainty of the column density (the combination of both the model uncertainty and the stellar patchiness parameter).
For each AGN sight line, we have the predicted column density line shape, associated uncertainty line shape, and the velocity shift for the random motion.

\subsection{The Bayesian Analysis}
\label{Bayesian_frame}
We use the Bayesian framework (together with the Markov Chain Monte Carlo method; MCMC) to compare the model prediction and the observation sample and obtain the best parameters of the kinematical model \citep{Zheng:2019aa}.
Compared to the $\chi^2$ minimization method, the Bayesian method could obtain the entire posterior distribution rather than the most likely parameter (with the minimum $\chi^2$).
The posterior distribution is necessary when there are unconstrained parameters in the model, which is the case here (Section \ref{overview}).
The MCMC simulation is run by using the {\it emcee} package \citep{Foreman-Mackey:2013aa}.

In the Bayesian approach, we need to optimize the likelihood function ($p_{\rm tot}(\theta|D)$), where $D$ and $\theta$ are the observation data and the model parameters, respectively.
The total likelihood function contains two major parts for both stellar ($p_{\rm st}$) and AGN ($p_{\rm AGN}$) samples:
\begin{equation}
p_{\rm tot}(\theta|D) = p_{\rm st}(\theta|D) \times p_{\rm AGN}(\theta|D). \notag
\end{equation}
Based on Bayes' theorem, the likelihood depends on
\begin{equation}
p(\theta|D) = \frac{p(D|\theta) p(\theta)}{p(D)}, \notag
\end{equation}
where $p(D|\theta)$ is the possibility for a model with parameters of $\theta$ to reproduce the data $D$, $p(\theta)$ is the prior knowledge for the parameters, and $p(D)$ is the possibility to have the observation data.
In our implementation of the Bayesian simulation, $p(\theta)$ is assumed to be uniform in the given parameter space, so it is a constant.
The boundary of the parameter space is determined to allow each parameter to have sufficient variation (except for parameters only have limits) as shown in Fig. \ref{params_corner}.
$p(D)$ is a complex term, but this term is the same for all model, so it can be ignored in practice.
Finally, $p(D|\theta)$ is determined by comparing the model prediction and the observation.
In the following analyses, $\ln p$ stands for $\ln p(D|\theta)$ or $\ln p(\theta|D)$.
The difference between these two values is constant, which does not affect the results.

The stellar sample only provides a column density, so the likelihood function only has one term for the total column density.
Following \citet{Zheng:2019aa}, the logarithmic value of the column density likelihood (for the stellar sample in our work) is 
\begin{eqnarray}
\ln p{\rm st} = -\sum_{k = 1}^{75} (\ln \sigma_{{\rm t}, k}  + \frac{1}{2} \frac{(\log N_{k} - \log N_{{\rm m}, k})^2}{\sigma^2_{{\rm t}, k}}),
\end{eqnarray}
where $\sigma_{{\rm t}, k}^2 = \sigma_{\log N, k}^2 + \sigma_{\log N_{\rm m}, k}^2 + \sigma_{\rm p}^2$ is the total uncertainty of the $k$-th stellar sight line, and $\rm m$ denotes for the model prediction.
$N_{k}$ is the $k$-th observed column density, while $N_{{\rm m}, k}$ and $\sigma_{\log N_{\rm m}, k}$ are the model predicted column density and uncertainty.

For the AGN sight line, there are three components for the total likelihood ($p_{\rm AGN}$): the line shape ($p_{\rm ls}$), the random motion along the sight line ($p_{\rm rand}$), and the total column density ($p_{\rm cd}$):
\begin{equation}
\ln p_{\rm AGN} = \sum_{k=1}^{186} (\ln p_{{\rm ls}, k} + \ln p_{{\rm rand}, k} + \ln p_{{\rm cd}, k}).
\end{equation}
The total likelihood of the AGN sample is the summation for the 186 sight lines.
For each sight line, the first term is the line shape, where we compare the model-predicted line shape (Section \ref{model_prediction}) and the observed column density line shape (Section \ref{data_sample}).
We note that the model prediction includes the velocity shift along the sight line ($v_{\rm shift}$).
Then for each bin, we could calculate the likelihood, and the total likelihood is the summation of the logarithmic likelihoods of individual bins:
\begin{equation}
\ln p_{\rm ls} = - \frac{1}{m'} \sum_{j = 1}^{m} (\ln \sigma_{{\rm t}, j}  + \frac{1}{2} \frac{(N_{j} - N_{{\rm m}, j})^2}{\sigma^2_{{\rm t}, j}}),
\end{equation}
where $\sigma_{{\rm t}, j}^2 = \sigma_{N, j}^2 + \sigma_{N_{\rm m}, j}^2$, $m$ is the total number of bins in the spectrum, and $j$ denotes the $j$-th bin in the column density line shape.
Compared to Eq. (19), there are three differences.
First, the calculation here is in the linear scale, while Eq. (19) is in the logarithm scale.
This is because the total column density is always a large number ($\log N \gtrsim 12$), and the distribution is better approximated by a lognormal distribution.
However, the differential column density in each bin of the line shape could be negative (continuum region or low significance features), which is also included in the modeling.
For these bins, the logarithm scale is not appropriate, so we use a linear scale.
Second, $\sigma_{{\rm t}, k}$ does not have an additional patchiness parameter as discussed in Section \ref{cloud_model}.
Third, we normalize the total logarithm likelihood by a factor of $1/m'$, where $m'$ is defined as the number of bins with observed column density $> 10^{10} \cmsq$ rather than the total number of bins in the entire spectrum.
Similar to the $\chi^2$ arguments, each bin in the line shape has a reduced $\chi^2 = 1$, then AGN sight lines have $m$ times more weights than stellar sight lines if no normalization is applied.
By adopting this normalization, all sight lines (both stellar and AGN) have similar weights.
Here, we use $m'$ instead of $m$ to do the normalization because the continuum region in the line shape is constant for different models, which will not affect the fitting results.
The choice of $10^{10} \cmsq$ criterion is to exclude the continuum region that dilutes the model significance.

The second term ($\ln p_{\rm rand}$) is the likelihood for the velocity shift ($v_{\rm shift}$) along the AGN sight line.
We assume that $v_{\rm shift}$ follows a normal distribution with a mean of zero and a standard deviation of $v_{\rm rand}$:
\begin{equation}
\ln p_{\rm rand} = - \ln v_{\rm rand} - \frac{1}{2} \frac{v_{\rm shift}^2}{v_{\rm rand}^2}.
\end{equation}
The last term ($\ln p_{\rm cd}$) is for the total column density derivation for AGN sight lines.
This term is similar to the stellar sample, but without the patchiness parameter:
\begin{equation}
\ln p_{\rm cd} = - \ln \sigma_{\rm t} - \frac{1}{2} \frac{(\log N - \log N_{\rm m})^2}{\sigma_{\rm t}^2},
\end{equation}
where $\sigma_{\rm t}^2 = \sigma_{\log N}^2 + \sigma_{\log N_{\rm m}}^2$ is the total uncertainty of the column density.
These two terms ($\ln p_{\rm rand}$ and $\ln p_{\rm cd}$) do not need the normalization as the line shape term ($\ln p_{\rm ls}$), because these two terms should be applied to each pixel (e.g., every pixel should be moved by $v_{\rm shift}$; Fig. \ref{vshift_example}).
Therefore, the calculation is already normalized.

\section{Fitting Results and Implications}
\label{fitting_results}

\subsection{Overview of the Fiducial Model}
\label{overview}
In Section \ref{kmodel}, we introduce the basic assumptions of the kinematical model.
In practice, there are additional improvements that could be considered, such as the NS asymmetry.
There is a systematic NS asymmetry for the \ion{Si}{4} and \ion{O}{6} column density distribution for northern and southern hemisphere (measured in the AGN sample; \citealt{Savage:2003aa, Zheng:2019aa}).
Based on the column density-only sample (and model), \citetalias{Qu:2019ab} suggested that this NS asymmetry is mainly caused by the difference of the exponential disk density distribution in two hemispheres (i.e., the scale height).
In this subsection, we introduce the our fiducial model and tests that we used to finalize it, while a more detailed discussion are on the fitting results (Section \ref{results}), the density distribution (Section \ref{gas_dist}), the gas kinematics (Section \ref{gas_kinematics}), and the gas properties (Section \ref{gas_property}).

With the new kinematical model, we can learn more about the variation for the NS asymmetry: the ion density distribution, the warm gas kinematics, and the gas properties.
For the density distribution, the disk component has five possible variations: the disk normalization (the GC density; $\log X_0^{\rm disk}$); the scale length ($r_0$); the scale height ($z_0$); and two index parameter for disk expansion ($\alpha_{\rm rxy}$ and $\alpha_{\rm z}$).
The CGM component has another four parameters: the CGM normalization (the density at the core; $\log X_{\rm mp}$ and $\log X_{\rm nd}$), the core radius ($r_{\rm c}$), and the $\beta$ factor.

\citetalias{Qu:2019ab} excluded the NS variation of the disk normalization and the CGM normalization based on tests, so we adopt this conclusion in the kinematical model.
\citetalias{Qu:2019ab} suggested the scale length is the same, and we adopt this assumption for two reasons.
First, the scale length is related to the stellar disk, which is supposed to be the same for both hemispheres.
Also, the difference of the scale length is at low Galactic latitude, but there are very few AGN sight lines at low Galactic latitude ($|b| < 20$; 4/186) to distinguish it.
We assume that $\alpha_{\rm rxy}$ of both hemispheres are the same for similar reasons.
For the CGM component, the core radius of the $\beta$-model is fixed to 2.5 kpc (obtained from the X-ray emission modeling), because the warm gas absorption line sample cannot be used to measure the gas within $\lesssim 2$ kpc of the GC.
Finally, there are only three parameters left: the scale height, the $\alpha_{\rm z}$ index parameter and the $\beta$ factor.
As stated in Section \ref{density_field}, we are using two components (disk and CGM, with different profiles) to approximate the real warm gas distribution.
Then, the index parameters and $\beta$ are degenerate with each other.
As will shown in Section \ref{gas_dist}, $\alpha_{\rm z}$ and $\beta$ both determine the gas slope at $20-50$ kpc along the $z$-direction, so we only keep the NS variation for one of these parameters.
Here we choose $\beta$ in following modelings.
In our test, the only varying the $\beta$ leads to a difference of $\approx 5$ for the Bayesian information criterion (BIC; moderately significant), so we retain this in the fiducial model.
Adding variation of $\alpha_{\rm z}$ does not improve the model (BIC $<2$).

For the kinematics, the rotation velocity is assumed to be the same for both hemispheres, because the midplane velocity should be continuous.
The radial velocity is found to be different between two hemispheres because the same radial velocity model leads to a BIC difference of $> 10$ (very significant).
Therefore, the NS asymmetric radial velocity is included in the fiducial model.
Additionally, we also consider the radial velocity boundary difference for both hemispheres ($R_{\rm disk}$).
This parameter mainly determines the amount of gas with a significant radial velocity, while the radial velocity determines the velocity shift of the gas.

The gas property may be different in both hemispheres, such as the column density per cloud, the absorption width, and the random velocity of absorption features along the sight line.
Based on our test, these parameters are roughly same with BIC $\lesssim 3$, which indicates that the model with more parameters (NS asymmetry for these parameters) is not a better model than the simple model (with the same parameter for both hemispheres).
For the absorption broadening velocity, the variation among individual sight lines is larger than the difference between the two hemispheres.
Therefore, our model only obtains the average broadening velocity.
The random velocity of absorption features shows smaller scale variation, rather the NS asymmetry, as discussed in Section \ref{gas_property}.
Therefore, we do not include the NS asymmetry for these three parameters.

\begin{figure*}
\begin{center}
\includegraphics[width=0.99\textwidth]{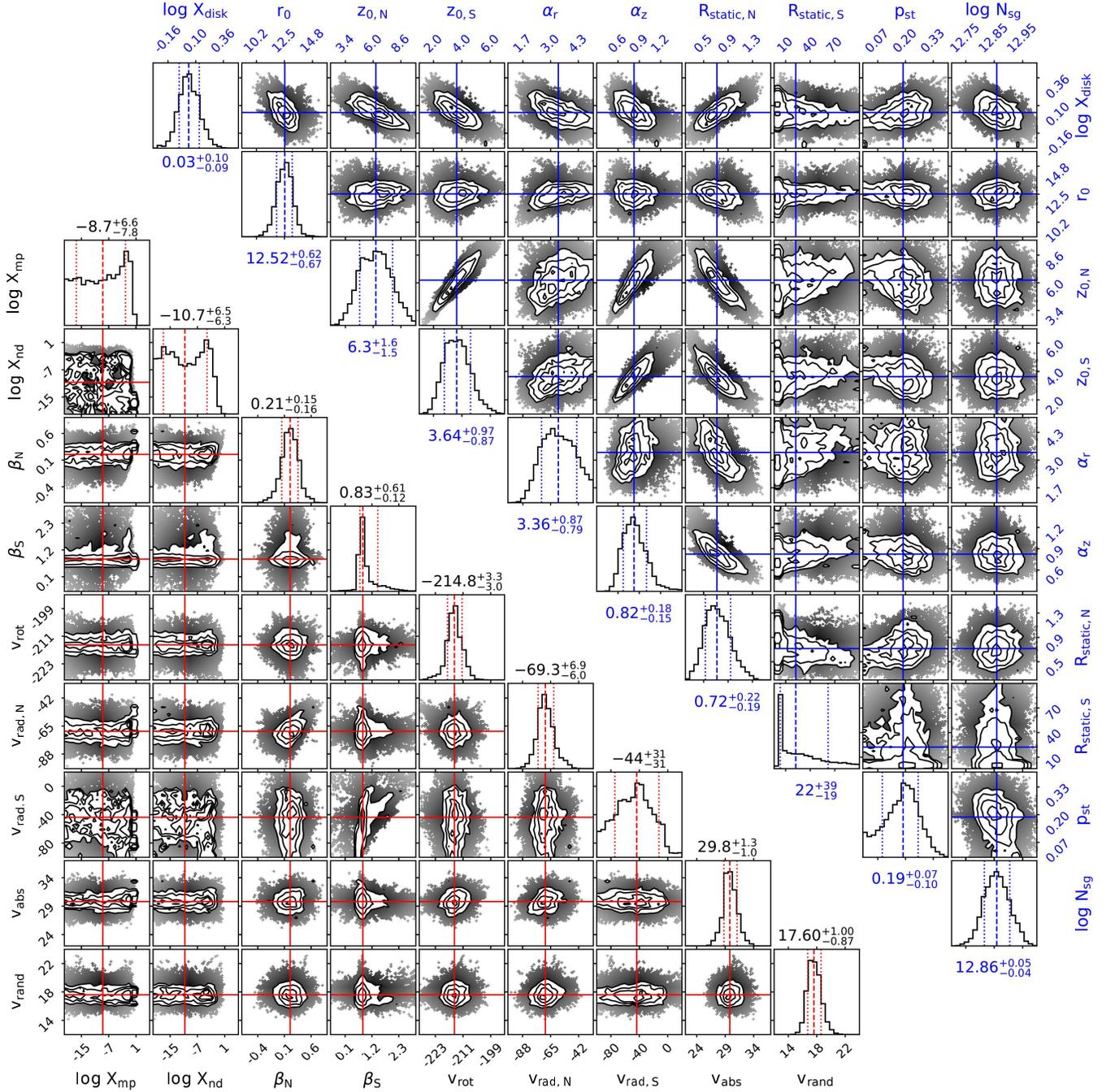}
\end{center}
\caption{The posterior distribution of the parameters in the fiducial model. In this model, the NS asymmetry are modeled by four parameters: the disk scale height ($z_0$), the boundary of the radial velocity ($R_{\rm disk}$), the $\beta$ factor, and the radial velocity ($v_{\rm rad, 10}$). Most parameters are constrained, except for 4/19 parameters: the boundary of the radial velocity in the southern hemisphere ($R_{\rm disk, S} > 3.4$), the radial velocity in the southern hemisphere ($v_{\rm rad, 10, S} < -12 \kms$), the CGM core density ($\log X_0^{\rm mp} < -2.1$ and $\log X_0^{\rm nd} < -4.2$).}
\label{params_corner}
\end{figure*}

Finally, the fiducial model has 19 parameters (Fig. \ref{params_corner} and Table \ref{params}).

\subsection{Fitting Results}
\label{results}

For the MCMC fitting to our Bayesian model, the parameter space is sampled with 100 walkers and each walker has 20000 steps to fully thermalize the posterior distribution.
As stated in Section \ref{stellar_sl}, we fit the fiducial model with two different sets of observation samples.
The preferred fitting model uses both {\it IUE} and STIS stellar sample, while another model only uses the STIS sample.
The COS AGN sample is the same for these two fitting models.
The fitting results are consistent between two models within $1\sigma$ uncertainty (expect for the additional patchiness parameter of the stellar sample).
Typically, the STIS-only model has larger uncertainties for some parameters (e.g., cloud number density of the disk component), because fewer stellar sight lines are considered in this model.
The patchiness parameter of the stellar sample is slightly smaller for the STIS-only model ($0.15\pm0.09$) compared to the combined-sample model ($0.19_{-0.10}^{+0.07}$).
This indicates that the IUE sample has a larger systematic uncertainty, but this uncertainty is modeled by the patchiness parameter, which does not affect other parameters significantly.
Therefore, we suggest that there is no physical difference between these two models.
In the following analyses, we only discuss the results from the combined-sample model.

The fitting results are presented in Fig. \ref{params_corner}, where most parameters are constrained, except for four of them.
The 1 $\sigma$ limits for these four unconstrained parameters are the disk boundary for the radial velocity in the south $R_{\rm disk, S}>3.4$, the radial velocity in the south $v_{\rm rad, S}<-12\kms$, the CGM component in the midplane $\log X_0^{\rm mp}<-2.1$) and the CGM component cloud path-length density perpendicular to the disk ($\log X_0^{\rm nd} < -4.2$).

\begin{figure*}
\begin{center}
\includegraphics[width=0.49\textwidth]{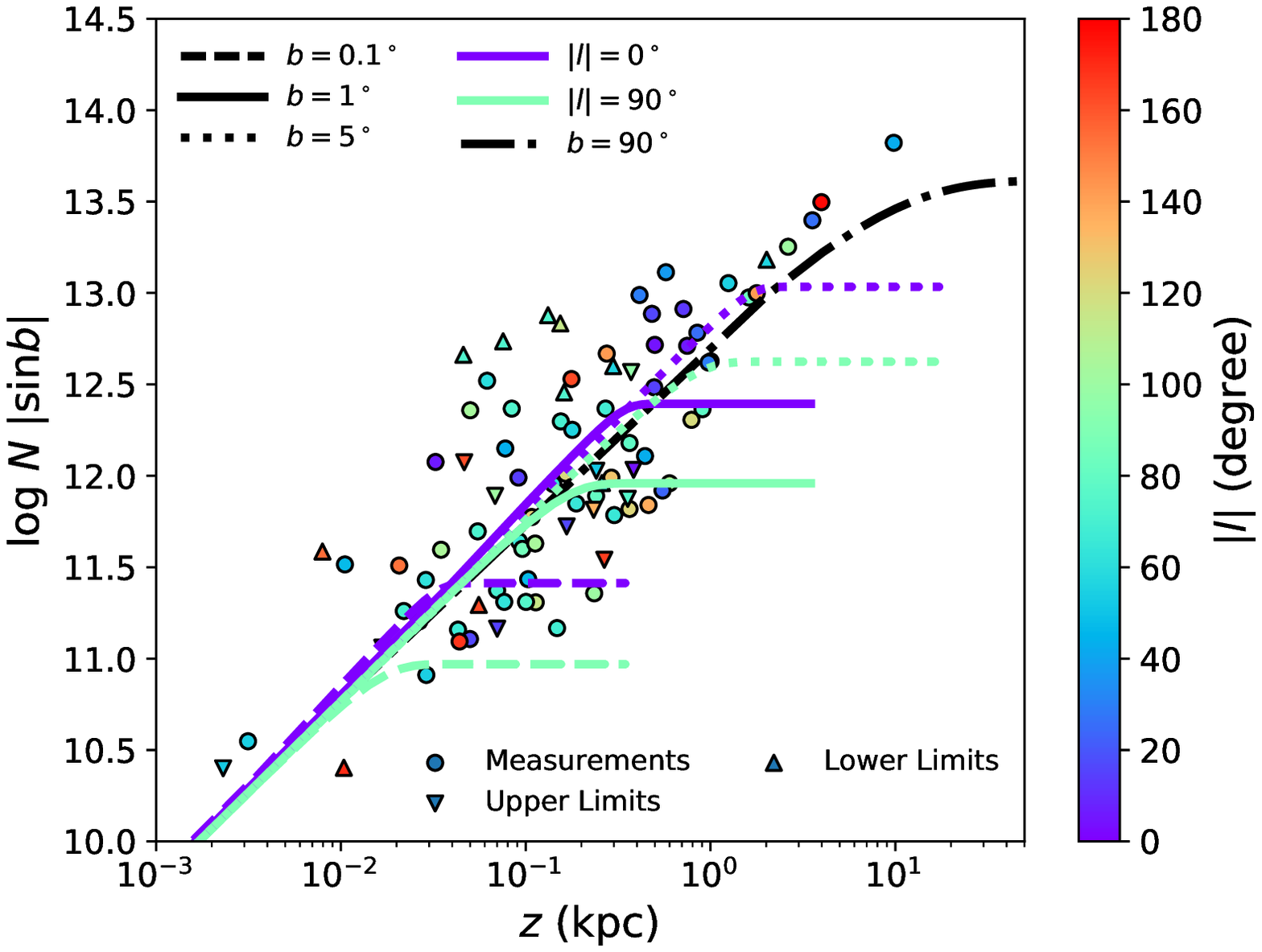}
\includegraphics[width=0.49\textwidth]{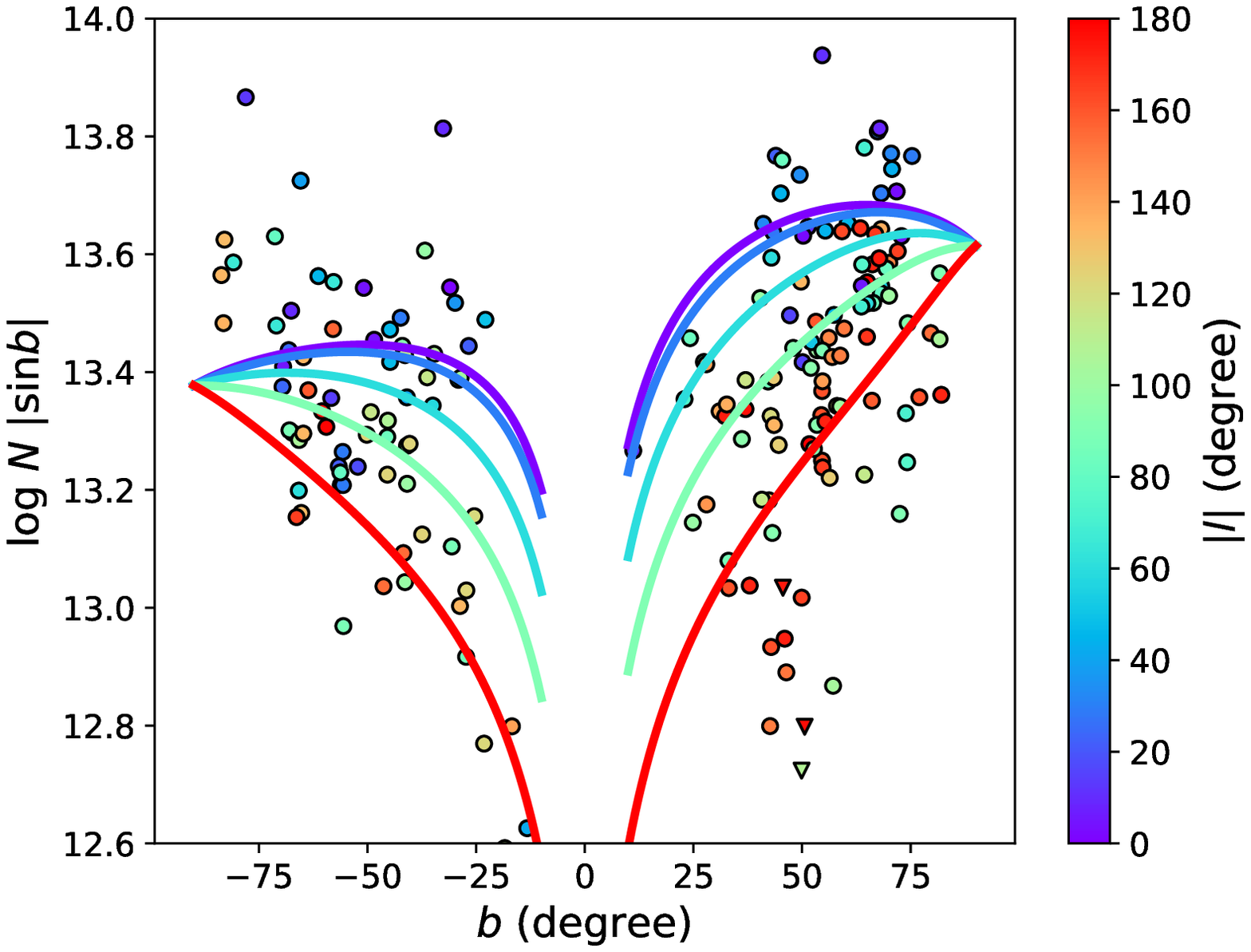}
\end{center}
\caption{The comparison between the fiducial model and the observed column density of the stellar sample (left) and the AGN sample (right). The behavior of the kinematical model is similar to the \citetalias{Qu:2019ab} because we use the 2D disk-CGM model as the basis of the kinematical model. In the left panel, we only plot the model of the northern hemisphere for simplicity. The southern hemisphere model is slightly lower than the northern hemisphere due to the NS asymmetry.}
\label{stellar_agn}
\end{figure*}

These unconstrained parameters ($R_{\rm disk, S}$, $v_{\rm rad, S}$, $\log X_0^{\rm mp}$, and $\log X_0^{\rm nd}$) has two physical implications.
First, there is an insufficient number of absorbing clouds in the southern hemisphere at $10-20$ kpc, which are needed to determine the radial velocity.
This also leads to the unconstrained radial velocity and boundary (Section \ref{gas_kinematics}).
This is consistent with the large $\beta$ value in the south, which confirms less gas at large radii (excluding the LMC/SMC).
Second, the upper limits of the CGM component path-length densities suggest that the gas distribution is dominated by the disk component (Section \ref{gas_dist}).
We note that the disk component defined in Section \ref{density_field} is more extended in the $z$-direction than the previously adopted exponential disk (e.g., \citealt{Savage:2009aa}).
The implication of all parameters is described and discussed in details in following subsections on various topics: the warm gas distribution, the warm gas kinematics, and the warm gas property.
Here, we compare the model prediction to the observation.

In Fig. \ref{stellar_agn}, we plot the column density distribution and the model predictions for both the stellar sample and the AGN sample.
The behaviors of the models are similar to the observations and the \citetalias{Qu:2019ab} model (details are described in \citetalias{Qu:2019ab}).
We find that sight lines with high Galactic latitudes ($|b|$) and small Galactic longitudes ($|l|$) have higher projected column density ($\log N \sin |b|$) for stellar sight lines at the same $z$ heights.
For the AGN sample, the kinematical model captures the feature that the projected column density is higher for sight lines toward the GC ($|l|=0^\circ$) compared to the anti-GC ($|l|=180^\circ$) direction.
For sight lines toward the anti-GC, the projected column density shows a rapid decrease toward low Galactic latitudes.
Then, the inclusion of the kinematics in the model does not break the self-consistency with the previous column density-only model (\citetalias{Qu:2019ab}; more comparisons are in Section \ref{comparison}).

\begin{figure*}
\begin{center}
\includegraphics[width=0.99\textwidth]{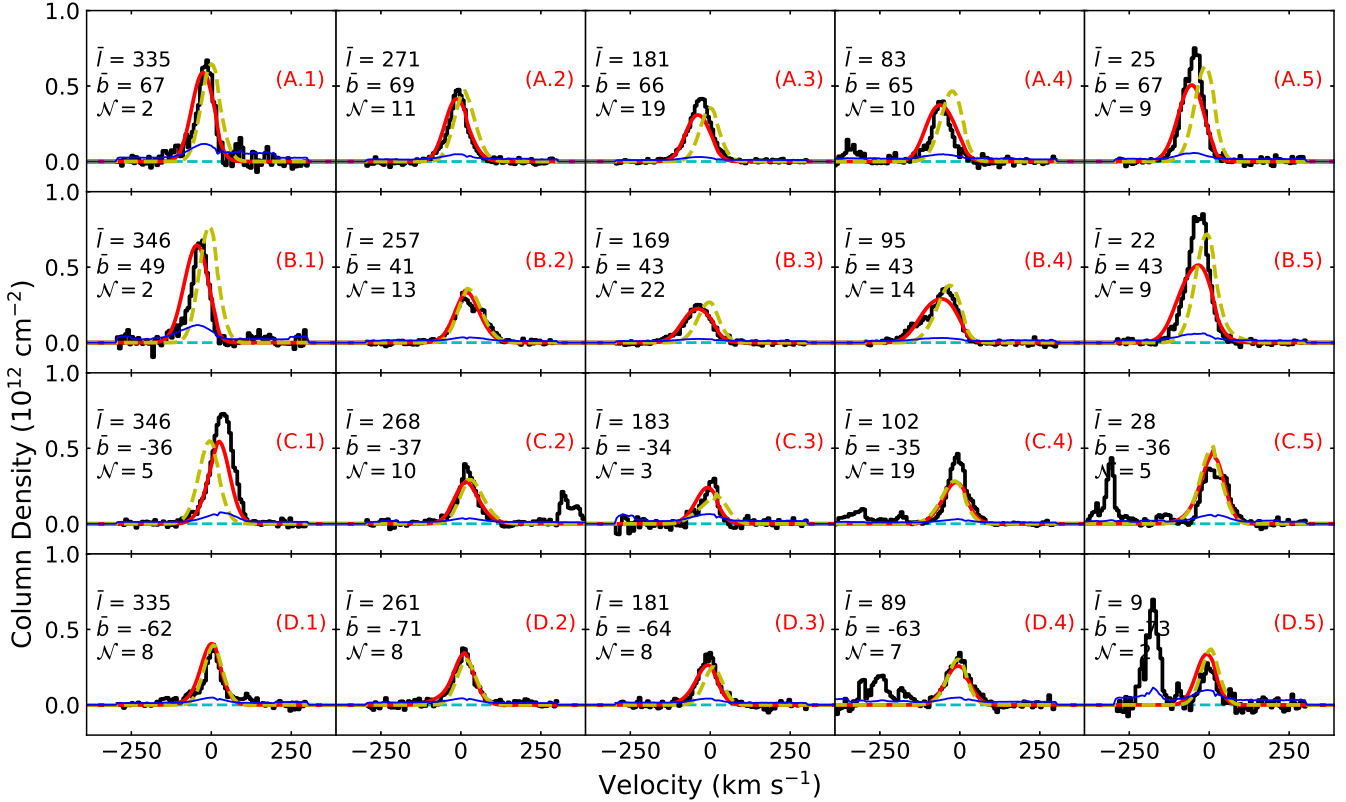}
\end{center}
\caption{The comparison between the stacked line shape of observation and the kinematical model prediction: the sky is divided into 20 regions (the grids in Fig. \ref{N_v_dist}). The observed line shapes (black lines) are consistent with the model predictions (red lines) within the uncertainty (blue lines) for most regions. The dashed yellow lines are the model without the radial velocity component (other parameters are the same as the fiducial model). In the northern hemisphere, the yellow lines have systematically positive shifts compared to the observation and the fiducial model, which is the evidence of systematic inflow. The southern hemisphere does not show these systematic shifts, although the region C.1 shows an outflow feature (positive shifts). In some regions (e.g., B.5), the absorption broadening velocity is smaller than the average broadening velocity in the fiducial model, which leads to broader features. Some regions show features that cannot be accounted for by the kinematical model, which is suggested to have other origins: C.2 might be affected the MS \citep{Fox:2014aa}, while C.4 and nearby regions might be an HVC population associated with the LG \citep{Bouma:2019aa}.}
\label{combine_ls}
\end{figure*}

To test the role of the warm gas kinematics, we compare the stacked line shapes based on the AGN sample and the model predictions.
The kinematical model predicts that the line shape has a dependence on Galactic longitude and Galactic latitude.
Therefore, we divide the entire sky into 20 regions, which roughly have a similar number of sight lines ($\sim 10$) in each region.
Galactic longitude have grids of $0^\circ-45^\circ$, $45^\circ-135^\circ$, $135^\circ-225^\circ$, $225^\circ-315^\circ$, and $315^\circ-360^\circ$, while the Galactic latitude grids are $-90^\circ$ to $-55^\circ$, $-55^\circ$ to $0^\circ$,  $0^\circ$ to $55^\circ$, and $55^\circ$ to $90^\circ$ (grids in Fig. \ref{N_v_dist}).
In each region, we stack the differential column density line shape by obtaining the mean value for all sight lines as shown in Fig. \ref{combine_ls}, where we also plot the fiducial model.
The fiducial model reproduces the major absorption features around $-150$ to $150 \kms$ including some intermediate-velocity clouds (IVCs) and HVCs in specific regions (e.g., sight lines in region B.4 have HVCs at $< -100 \kms$).

However, there are still some HVC features are not accounted for in the kinematical model, such as negative HVCs at $\approx -300$ to $-150 \kms$ in regions C.4, C.5, D.4, and D.5, and the positive HVC at $\approx 300 \kms$ in region C.2.
The negative HVCs could be a population of clouds associated with the Local Group (LG; \citealt{Bouma:2019aa}), while the positive HVCs are likely to be associated with the LMC/SMC (discussed further in Section \ref{IHVC}.

\begin{figure*}
\begin{center}
\includegraphics[height=0.22\textheight]{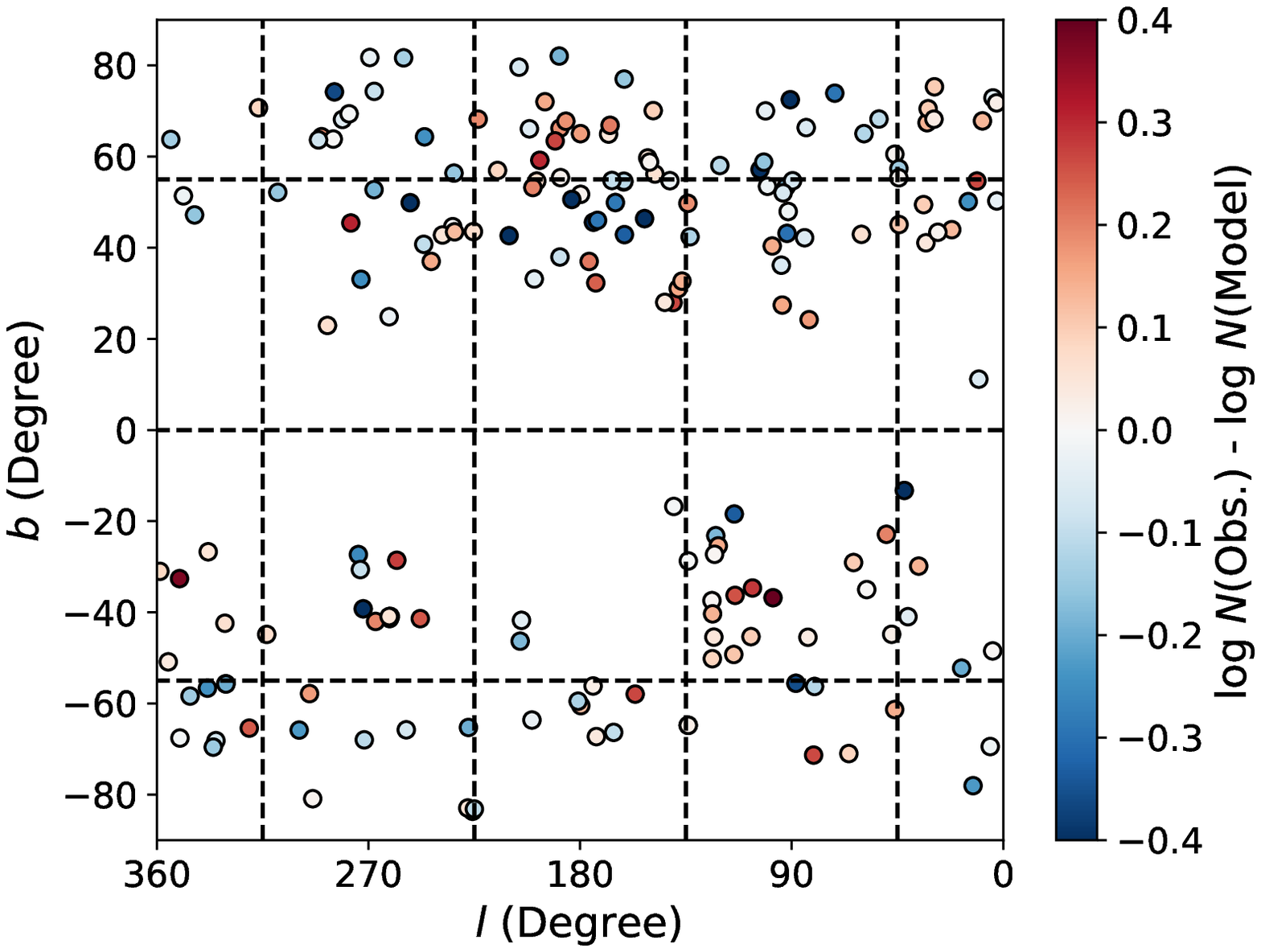}
\includegraphics[height=0.22\textheight]{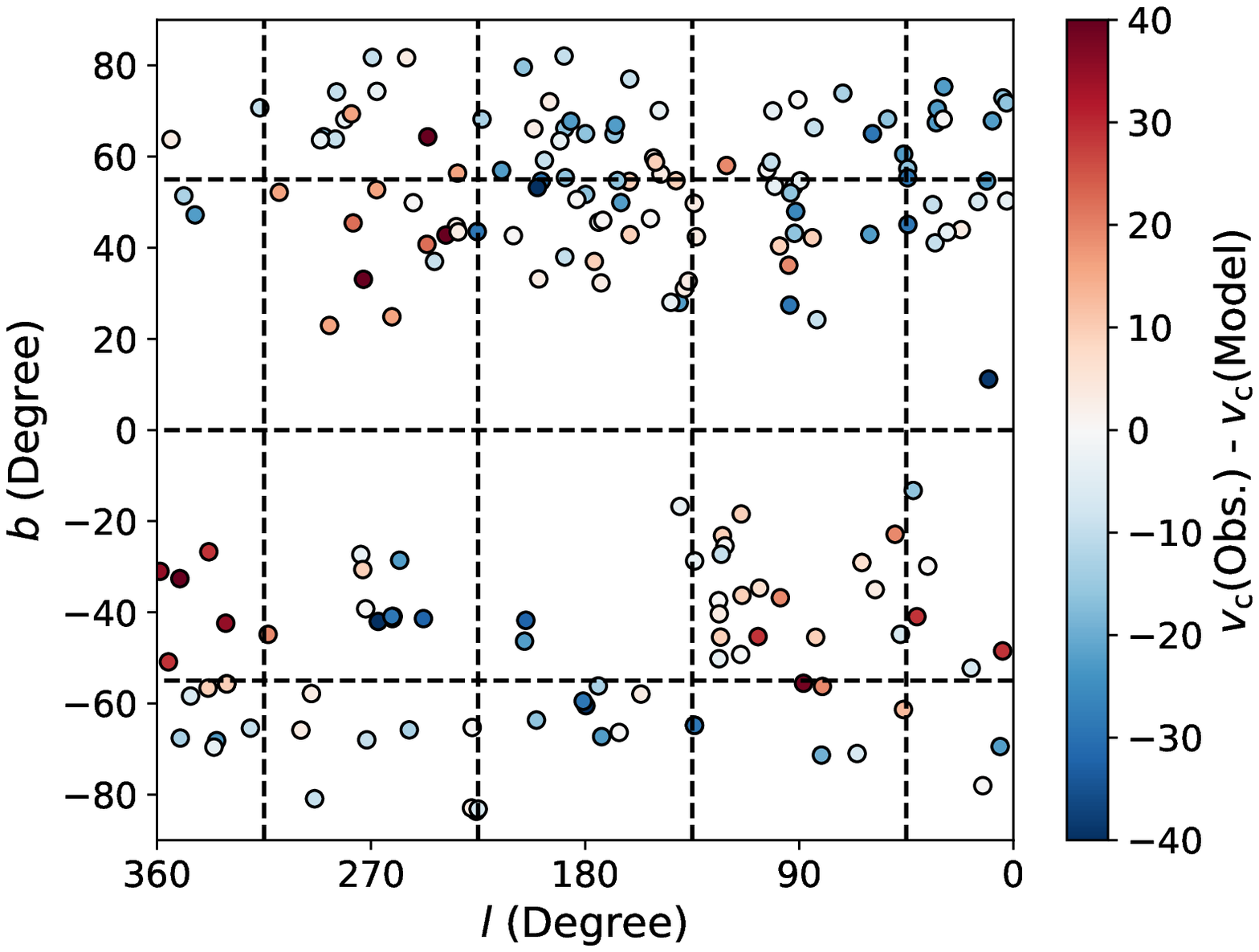}
\includegraphics[height=0.22\textheight]{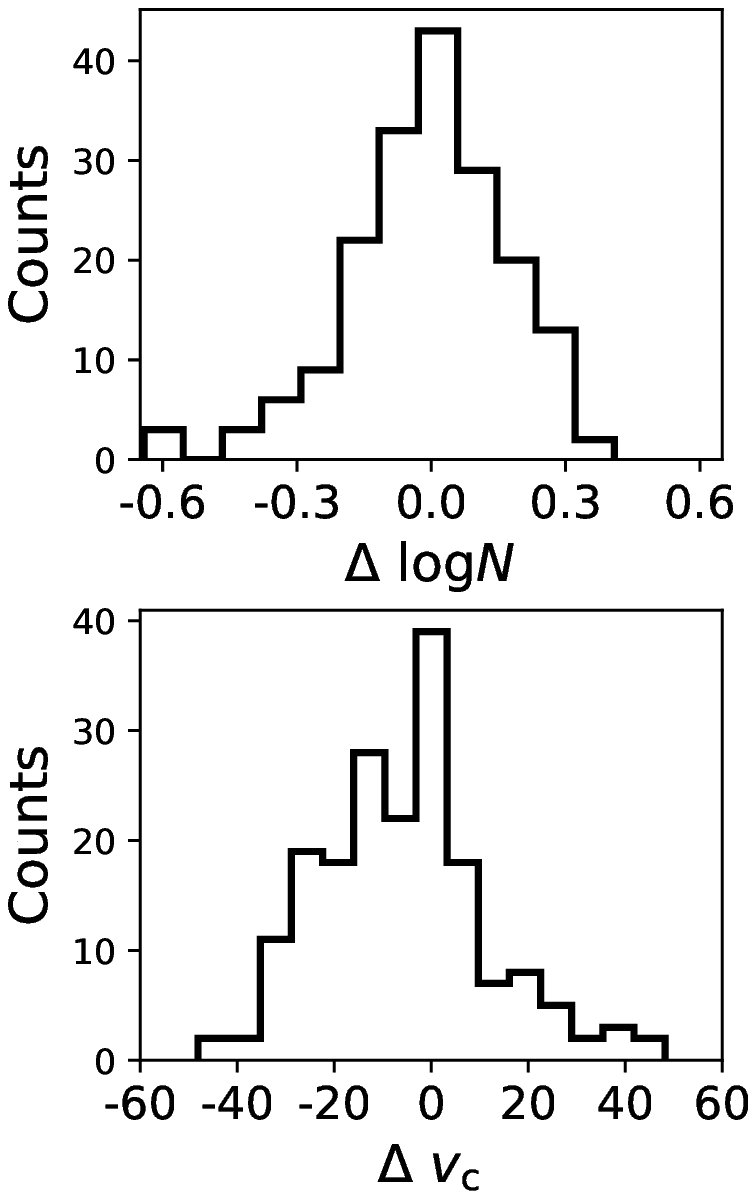}
\end{center}
\caption{The column density residual (left), the velocity shift (middle) maps and their distributions (right). For the column density residuals, both hemispheres show average values about the zero, which shows no large scale differences ($\gtrsim 90^\circ$). However, at small scale ($\approx 20^\circ$), the column density residual shows clustering (e.g., the positive structure around $l=180^\circ$ and $b=60^\circ$). Similarly, the velocity shift also shows the small scale structure, but the kinematical structure ($\approx 50^\circ$) is larger than the column density variation. These qualitative results are confirmed in more accurate estimation in Fig. \ref{angular_size}.}
\label{dN_dv_dist}
\end{figure*}

For individual sight lines, the difference between the model predictions and the observed total column density are shown in Fig. \ref{dN_dv_dist}.
Broadly, these residuals of column density are uniformly distributed over the entire sky at the large scale but shows some clustering (coherence) on scales of $\approx 20^\circ$ (with similar enhancements or deficits).
There is no significant connection between the column density enhancement or deficit and the known HVCs or IVCs detected in \ion{H}{1}.
The velocity shifts of individual sight lines (Fig. \ref{dN_dv_dist} right panels) also show clustering, but on a larger scale than the column density variation (more details are discussed in Section \ref{gas_property}).
The positive and negative shifts occur over the entire sky, which suggests that the bulk velocity field included in the kinematical model is the first-order approximation of the velocity of absorption features.

\subsection{The Gas Distribution}
\label{gas_dist}

In the fiducial model, the cloud path-length density of the warm gas clouds is about $1.3\pm0.2$ per kpc at the GC, while it is $0.6\pm 0.1$ per kpc at the Solar neighborhood.
For the Solar neighborhood, this path-length density is consistent with the nearby ($d\lesssim 0.2$ kpc) WD stellar sample, where no ISM \ion{Si}{4} has been detected.
The path-length cloud densities set upper limits of the cloud size of about 0.8 kpc at the GC and 1.5 kpc at the Solar neighborhood.
If individual clouds have sizes much larger than 1.5 kpc, the warm gas traced by \ion{Si}{4} will be smoothly continuous rather than the cloud-like variation (i.e., the intrinsic scatter traced by the patchiness parameter).
It is not clear whether there is a real difference between the GC and the Solar neighborhood, but some differences are expected because the GC has higher pressure for the warm gas, where could lead to smaller cloud sizes.
This constraint for the warm gas is consistent with our more accurate estimation of cloud size in Section \ref{gas_property} (1.3 kpc).

\begin{figure*}
\begin{center}
\includegraphics[width=0.49\textwidth]{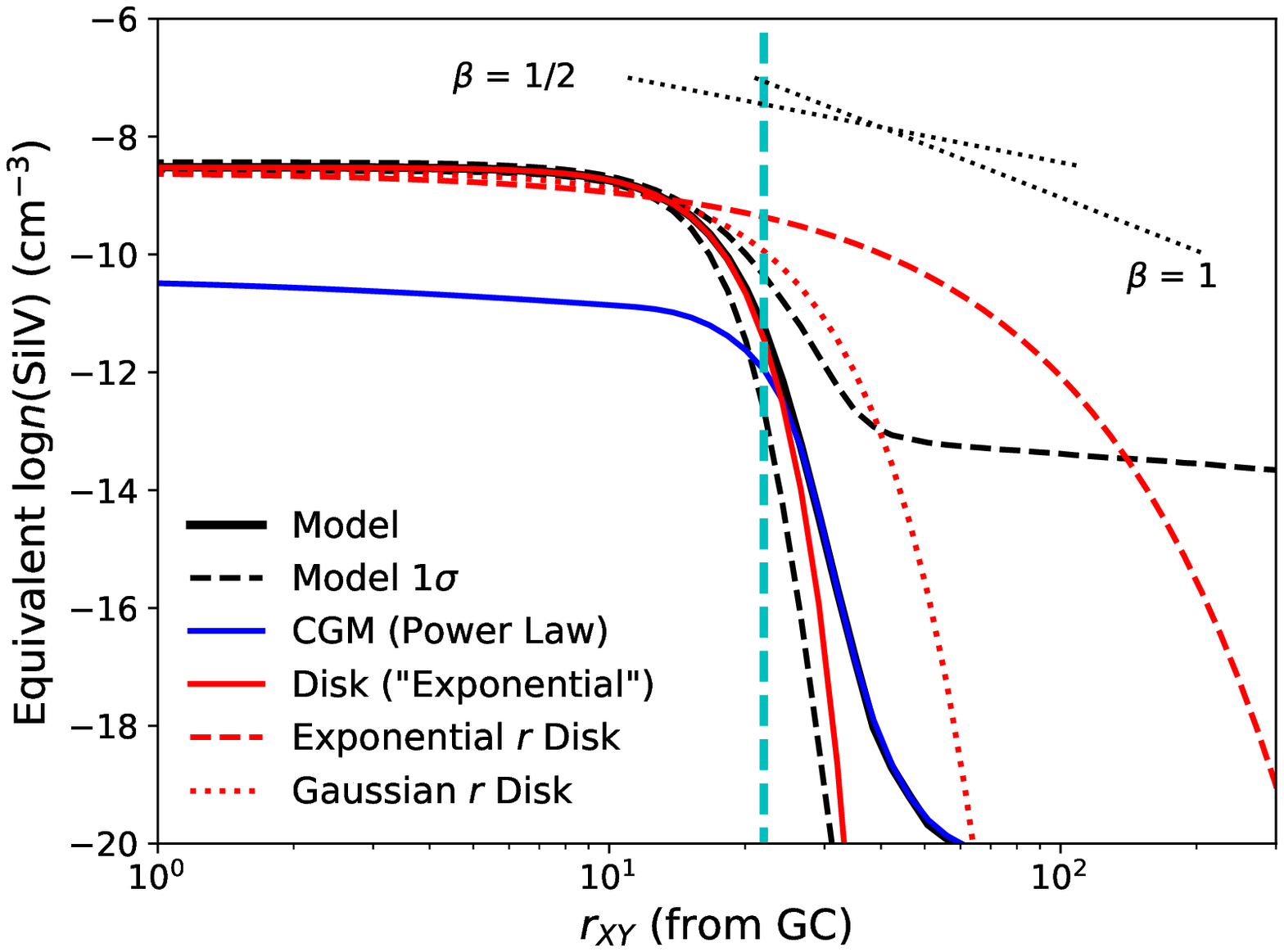}
\includegraphics[width=0.49\textwidth]{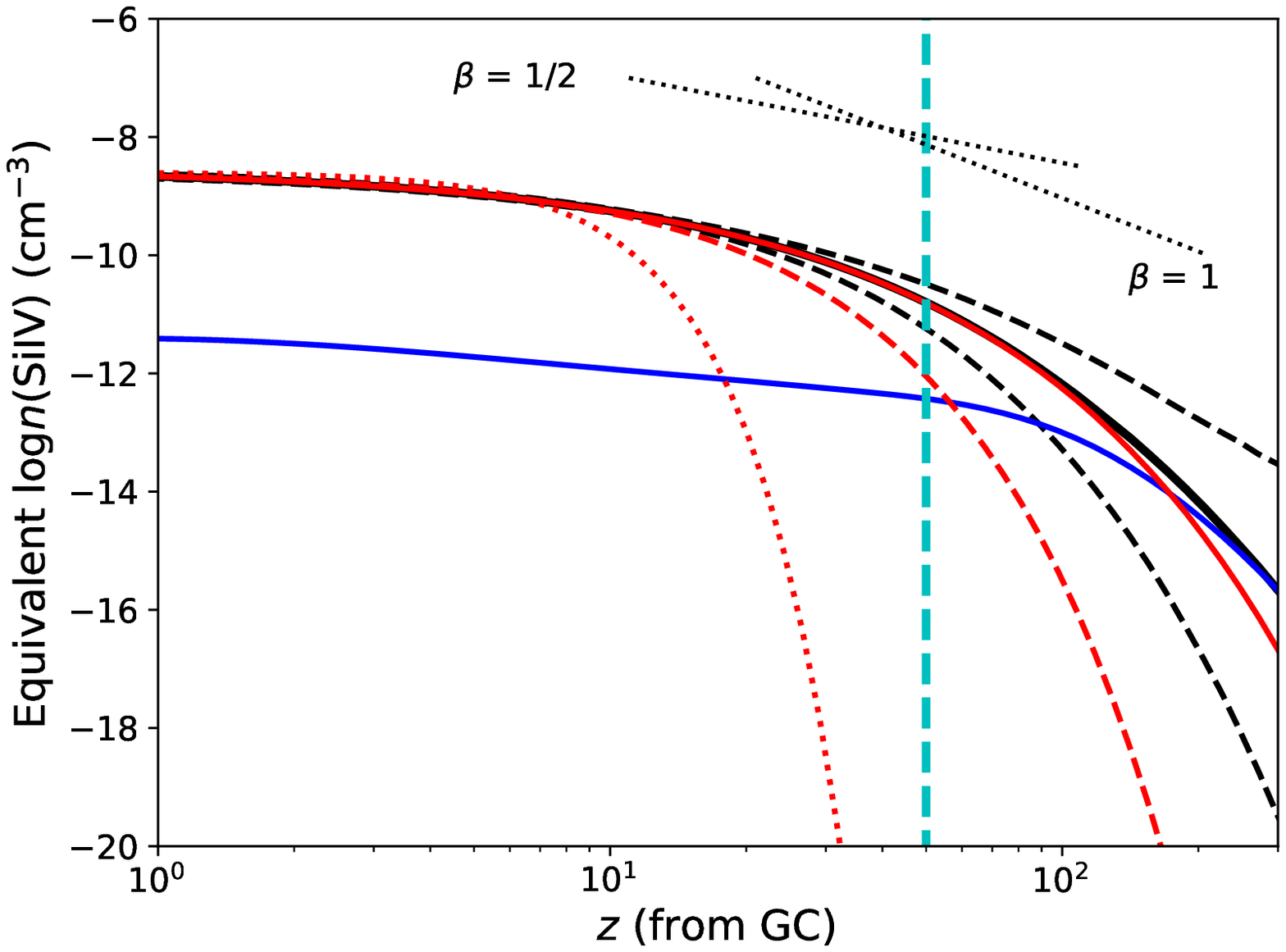}
\end{center}
\caption{The equivalent \ion{Si}{4} density distribution along the radial (left) and the $z$ (right) directions. The vertical cyan dashed lines indicate the limiting radius ($20-50$ kpc), which is set by the limiting column density. In the fiducial model, the radial direction density distribution (the red solid line) shows a sharp decay at about $10-20$ kpc compared to the exponential function (the red dashed line) and the Gaussian function (the red dotted line). Oppositely, the $z$ direction density distribution is more extended than the exponential function. The difference between the two directions suggest that the MW warm gas traced by \ion{Si}{4} is affected by feedback processes originated from the disk.}
\label{GC_rz}
\end{figure*}

We calculate the equivalent density profile (the cloud path-length density times the column density per cloud; $X N_{\rm sg}$)  for every single point in the MCMC chain and obtain the median and  the 1 $\sigma$ uncertainty (Fig. \ref{GC_rz}).
For both density distributions, the outskirts suffer from large uncertainties.
Here, we set the observation limits as $\approx 10^{-11}\cc$, which is the ratio of the limiting column density ($10^{12}\cmsq$) and the typical path length of ($30-50$ kpc).
Then, we suggest that the warm gas density distribution are well constrained within the limiting radius (although the model extend to 250 kpc): $\approx 20$ kpc in the $r_{\rm XY}$ direction and $\approx 50$ kpc in the $z$-direction.

The majority of gas contributing to the column density is within 20 kpc to the GC.
Quantitatively, we calculate the average distance of the warm gas with weights as the column density ($\int n r {\rm d} r/\int n {\rm d} r$).
In $r_{\rm XY}$ and $z$-directions, the distance are 20 and 13 kpc, respectively.
With the origin at the Sun, we could calculate the average observed gas distance.
Then, the average distance of the observed absorbing gas has distances of 3 kpc ($r_{\rm XY}$), 5 kpc ($z$ in the south), and 9 kpc ($z$ in the north).
Therefore, we suggest that the observed \ion{Si}{4} features can be considered at $\approx 5$ kpc, and we use this value as a typical distance to estimate the cloud physical size in Section \ref{gas_property}.

There is a significant difference between the $r_{\rm XY}$ and $z$ directions, where the density distribution along the $z$-direction is more extended than the $r_{\rm XY}$-direction.
For the $z$-direction density distribution, the CGM component begins to take over at about 100 kpc, although it is a minor contributor to the column density, so the statistical constraints are poor.
Within the distance range of $10-100$ kpc, the approximated power law slope is about $-1.5$ to $-2$ ($\beta \approx 0.5-0.7$).
The $r_{\rm XY}$ density distribution has a much sharper edge than the $z$-direction density distribution at about $10-20$ kpc, which is about the size of the stellar disk.

The difference of the \ion{Si}{4} density distributions between the $r_{\rm XY}$ and $z$-directions indicates that the warm gas distribution traced by \ion{Si}{4} depends on the disk orientation.
This could be explained as that the warm gas is more associated with the disk phenomena (i.e., feedback) rather than accretion from the IGM.
Theoretically, feedback processes could enrich the warm gas above or below the disk by ejecting gas and metals or providing more ionizing photons to photo-ionize \ion{Si}{4}.
The gas above and below the disk is more affected by Galactic feedback than the disk radial direction.
Therefore, it is expected that the gas dominated by feedback follows a disk-like shape, while accretion leads to more spherical geometry (e.g., \citealt{Stern:2019aa}).
The origin of the warm gas is discussed further in Section \ref{implication}.

\subsection{The Gas kinematics}
\label{gas_kinematics}
In the kinematical model, we consider the first-order approximation of the bulk velocity field as the combination of the rotation velocity and the radial velocity.
Here we mainly consider the kinematics of the major absorption features (with centroids at $\approx 0\kms$), but these features are not necessary to be low velocity (e.g., in some sky regions, the features could extend to HVCs; $|v|>100\kms$).

The fiducial model suggests that the rotation velocity of the warm gas is about $-214.8_{-3.0}^{+3.3} \kms$ at the midplane, which is comparable to the \ion{H}{1} disk (and halo; $\approx -220\kms$\citealt{Kalberla:2008aa}) and the stellar disk ($\approx -200 \kms$; \citealt{Huang:2016aa}).
This measured rotation velocity in the kinematical model has a dependence on the Solar motion, especially the $Y$-component, which is fixed to $-232\kms$ in the fiducial model.
Physically, the line shapes of the \ion{Si}{4} absorption features in the AGN sight lines suggests that the rotation velocity of warm gas is about $15-20 \kms$ smaller than the Solar rotation velocity.

Above or below the midplane, the rotation velocity is found to have a velocity gradient (i.e., lagging).
In the kinematical model, we do not employ the linear format of the lagging, which could reverse the rotation direction at $z \approx 20-30$ kpc, if the lagging is $\approx 10\kms ~\rm kpc^{-1}$. 
However, absorption line investigations on external galaxies shows that the CGM is co-rotating (the same rotation direction) with the disk \citep{Martin:2019aa}.
By adopting $v_{\rm rot}(r, \theta) = v_{\rm rot} \cos \theta$, we assume a cosine function for the velocity lagging, which never break the co-rotation between the CGM and the disk.
In this assumption, an equivalent linear velocity gradient is ${\rm d}|v|/{\rm d} |z| \approx -8 \kms~\rm kpc^{-1}$ within $|z| = 10$ kpc around the Solar system.

The radial velocity is found to behave differently in the northern and southern hemispheres.
The northern hemisphere has a radial inflow velocity of $-69.3_{-6.0}^{+6.9}\kms$ at 10 kpc, while the radial velocity in the southern hemisphere is not well constrained, with  $v_{\rm rad, N}<-12 \kms$.

\begin{figure}
\begin{center}
\includegraphics[width=0.49\textwidth]{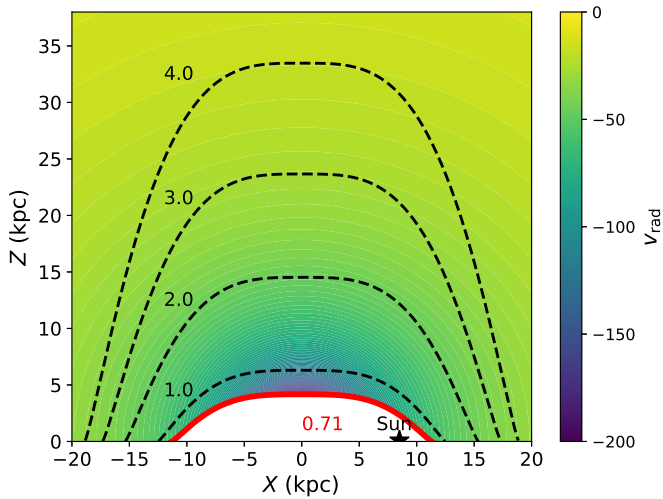}
\end{center}
\caption{The boundary of the radial velocity in the northern hemisphere. The thick red solid line is the boundary of the radial velocity, below which there is no radial velocity. This line follows the isodensity contour of the disk component of $(r_{\rm XY}/r_0)^{\alpha_{\rm rxy}} + (|z|/z_0)^{\alpha_{\rm z}}= 0.72$. The black dashed lines are also isodensity contours at different levels. Because the majority of absorbing gas is within $20-50$ kpc, the majority of radial velocity is about $30-200\kms$ in the northern hemisphere.}
\label{boundary}
\end{figure}

In the northern hemisphere, the kinematical model assumes that radial velocity depends on the radius (Eq. 16), which is approximated as $v_{\rm rad} \propto r^{3\beta - 2}$.
The $\beta$ factor is $0.21^{+0.15}_{-0.16}$, which indicates that the accretion velocity is larger in the inner region of the northern hemisphere.
The accretion velocity is $30 - 200 \kms$ between 30 kpc and the boundary of the accretion.
The boundary for the radial velocity ($R_{\rm disk}$) is given by the boundary surface:
\begin{equation}
(\frac{r_{\rm XY}}{12.35 \rm~ kpc})^{\alpha_{r_{\rm XY}}} + (\frac{z}{4.9\rm~kpc})^{\alpha_{z}} = 0.72_{-0.19}^{+0.22}.
\end{equation}
We show this boundary for the northern hemisphere in Fig. \ref{boundary}.
Because most column densities are from the warm gas within $30-50$ kpc, the accretion velocity is also dominated by the behavior in this region; the radial velocity beyond 50 kpc is unconstrained.
Because the radial format of the radial velocity is a model assumption, we do not suggest that the radial velocity is necessary to be the $r^{3\beta-2}$ format.
However, it is concluded that the inner region has higher radial velocities, because the constant radial velocity model (beyond the boundary) is significantly worse ($\Delta$BIC $>6$)

For the southern hemisphere, both the radial velocity ($v_{\rm rad}< -12 \kms$) and the boundary of the radial velocity ($R_{\rm disk} > 3.4$) are unconstrained.
The boundary and the radial velocity are degenerate to some degree for the southern hemisphere.
Ideally, the radial velocity determines the velocity shift of the gas beyond the boundary, while the boundary position determines the amount of gas that is shifted.
Thus, the total column density away from the peak of the rotation-only model is roughly proportional to the gas beyond the boundary times the velocity shift.
The absorption features in the southern hemisphere do not show shifted column densities that would constrain both the radial velocity and the boundary.

To show the effect of the radial velocity, we plot the model without radial motions for both northern and southern hemispheres (other parameters are the same as the fiducial model) in Fig. \ref{combine_ls}.
All regions in the northern hemisphere show line shapes (for both the observation and the fiducial model) with negative shifts compared to the model without the radial velocity.
These negative shifts indicate that there is accretion in the northern hemisphere.
For the southern hemisphere, the models with/without radial velocity and the observation do not have a significant differences in the line centroids for most sky regions.
However, the region C.1 shows the outflow feature as a positive shift to the model.
In this sky region, 4/5 of sight lines pass through the Fermi Bubble (FB), so expansion of the FB might be responsible for this feature.
This outflow feature is for the majority of the warm gas (e.g., the disk) in this direction, so it is different from individual HVCs as detected in \citet{Karim:2018aa}.
As a comparison, there is no sight line passing through the FB in regions C.5 and B.1.
The region B.5 has 5/9 sight lines passing through the FB, but this region does not show significant outflow for the absorption features around $0\kms$, although HVCs associated with the FB are detected in \citet{Bordoloi:2017aa}.

The velocity shifts ($v_{\rm shift}$) of individual sight lines are shown in Fig. \ref{dN_dv_dist}, which is introduced to account for the random motion of warm gas cloud along sight lines.
The random motion is not completely random, because there is clustering in both hemispheres.
In the northern hemisphere, random velocities are reduced in four regions roughly with centers at ($270^\circ$, $40^\circ$), ($180^\circ$, $55^\circ$), ($135^\circ$, $40^\circ$), ($45^\circ$, $55^\circ$).
In the southern hemisphere, these clustering regions have centers at ($225^\circ$, $-55^\circ$) and ($90^\circ$, $-40^\circ$), and the possible FB feature in C.1.
These clustering regions imply that the warm gas has kinematical structures with angular sizes of coherent $40^\circ-50^\circ$.

We exclude the possibility that these features are artifacts of how the model was constructed.
If the radial velocity is not well determined, the northern or the southern hemispheres should show the systematic positive or negative shifts in Fig. \ref{dN_dv_dist}.
However, this feature is not evident, as the average shifts in both hemispheres are close to the zero.
For the rotation velocity, the most significant features should occur near $270^\circ$ or $90^\circ$, and they should be of opposite sign: $270^\circ$ negative and $90^\circ$ positive (this means the fitting rotation velocity $|v_{\rm rot}|$ is larger), or $270^\circ$ positive and $90^\circ$ negative (the smaller fitting rotation velocity $|v_{\rm rot}|$).
If the rotation velocity is not well determined, both hemispheres should have the same signal at different Galactic latitudes (e.g., both positive or negative shifts at $270^\circ$).
These features also are not seen in the velocity shift map in Fig. \ref{dN_dv_dist}.
Therefore, we suggest that both rotation and radial velocities are well determined and the clustering patterns are due to local features rather than poor fitting of global features.
The physical origins of these features are discussed in Section \ref{implication}.

\subsection{The Gas Properties}
\label{gas_property}

In the kinematical model, we assume that the warm gas is cloud/layer-like, which introduces intrinsic scatter for the observed column density.
The fiducial model suggests that the column density of single cloud ($\log N_{\rm sg} = 12.86_{+0.05}^{-0.04}$) is slightly larger than the estimation based on the patchiness parameter ($\log N_{\rm sg} = 12.68$; Section \ref{cloud_model}) in the column density-only model (no kinematics).
This is because when applying $\log N_{\rm sg}$ to the line shape, it not only accounts for the uncertainty of the column density, but also the uncertainty of the kinematical model.

The absorption broadening velocity is $29.8_{-1.0}^{+1.3} \kms$, which is an average of all sight lines (Fig. \ref{combine_ls}).
This broadening velocity contains three major contributors: the COS instrumental broadening, the thermal broadening, and the turbulence broadening.
The resolution of COS/FUV is $12.8 \kms$ at $1400\rm~ \AA$, so the intrinsic broadening due to the warm gas is $26.9 \kms$.
This velocity is equivalent to the $b$-factor of $38 \kms$, and a full width half maximum (FWHM) of $63 \kms$.
This result is consistent with the direct measurements of from \citet{Wakker:2012aa}, which has an FWHM of $63\pm 11\kms$ for \ion{Si}{4}.

We suggest that the intrinsic broadening ($\approx 27 \kms$) is more dominated by turbulent broadening rather than thermal broadening.
It is not clear whether the observed \ion{Si}{4} has multiple components in our sample, because the {\it HST}/COS cannot resolve features with velocity separation $\lesssim 15 \kms$.
However, using higher-resolution STIS spectra ($\Delta v \approx 2 \kms$), \citet{Lehner:2011ab} detected much narrower \ion{Si}{4} features in sight lines toward disk stars, which set the constraint on the temperature and the thermal broadening of the \ion{Si}{4} gas ($\approx 10 \kms $), which is well below the intrinsic broadening of $27 \kms$.

Another fundamental property of the warm gas cloud is the cloud size.
Here we use a similar method to \citet{Werk:2019aa}, which proposed the method to estimate the cloud size based on the relationship between the angular distance and the absolute column density difference of AGN-AGN pairs.
It is equivalent to extracting the angular power spectrum of the column density variation.
When two sight lines are close enough to pass through the same cloud, the absolute difference of the column density is less at a smaller distance.
If two sight lines are distant, there is no relation between the two column density measurements in these sight lines, so the absolute column density difference is more random.

\begin{figure*}
\begin{center}
\includegraphics[width=0.99\textwidth]{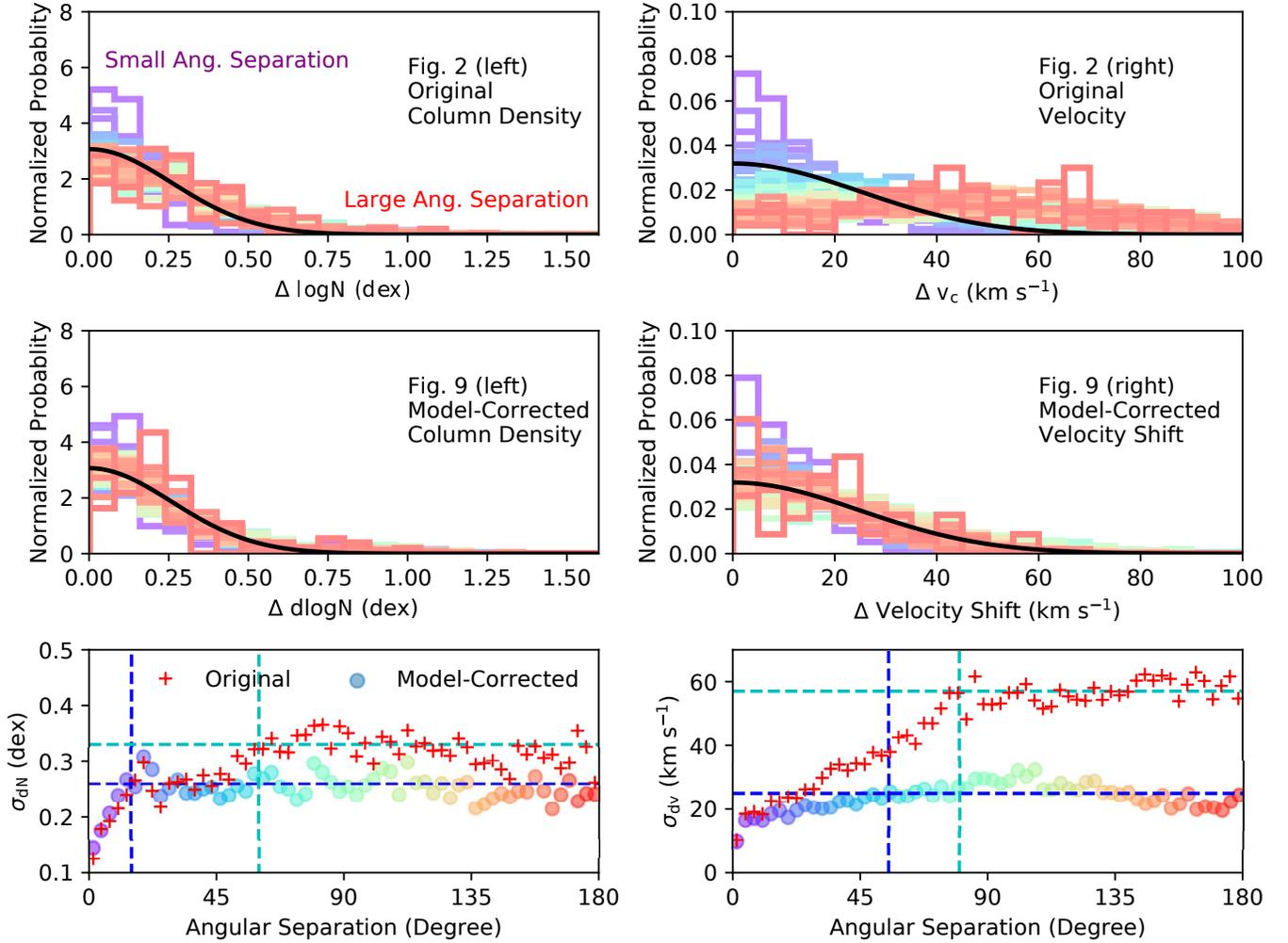}
\end{center}
\caption{Plots to show the the coherence of the column density (left panels) or the velocity variations (right panels) as a function of angular scales.
In the top two rows, the distribution of the absolute difference of column density or velocity (see the text for details). 
These distributions are color-encoded by the angular separation (in $3^\circ$ bins) from purple ($0^\circ$) to red ($180^\circ$; encoded in rainbow colors, see the lower panels for a detailed match).
The first row is for the original maps (Fig. \ref{N_v_dist}), while the second row is for the model-corrected maps (residual maps; Fig. \ref{dN_dv_dist}).
The common feature is that the purple lines (small separation) typically have higher peaks and narrower wings than the red lines (large separation).
This indicates that the smaller angular separation leads to a smaller variation of both the column density and the velocity.
The $1\sigma$ widths of the distribution are plotted in the lower panels: original data (red cross) and model-corrected data (circles colored for increasing angular separation).
The flat part of the $1\sigma$ dependence on the angular separation means that there is no correlation between two sight lines at this angular separation.
Based on both original and residual variations, we divide the angular separation behavior into three parts: the cloud variation (lower than the vertical blue lines), the global variation (between the blue and cyan lines), the random variation (larger than the cyan lines).
For the column density variation, the cloud size is about $15^\circ$, and the global variation is about $55^\circ$.
For the velocity variation, the kinematic structure is about $55 ^\circ$, and the global variation goes up to $80^\circ$.}
\label{angular_size}
\end{figure*}

We examine column density and velocity variations over the angular separation by using 17205 AGN-AGN pairs extracted from 186 sight lines.
First, for each AGN-AGN pair, we extract the angular separation and the absolute difference between the measured (original) column density (left panel of Fig. \ref{N_v_dist}).
Then, we extract the distribution of the absolute column density difference for every angular distance bin of $3^\circ$.
For each angular distance bin, the histogram of the absolute column density difference distribution (upper left panel in Fig. \ref{angular_size}) is approximated by a Gaussian function (i.e., even at a large angular separation, the largest possible difference is also 0).
Combing histograms, we found that at small angular distances (purple lines), the peaks of the distributions are higher and the corresponding wings are narrower than the distributions at large angular distance (red lines).
This becomes clearer when we extract the $1 \sigma$ width ($68\%$ percentile; $\sigma_{{\rm d}N}$) of the absolute column density difference distribution (lower left panel in Fig. \ref{angular_size}).
This value is equivalent to the power spectrum of the column density variation at a given angular separation.
The value of $\sigma_{{\rm d}N}$ keeps increasing within the angular distance of $\approx 55^\circ$, after where $\sigma_{{\rm d}N}$ has a flat part.
Within the angular distance of $\approx 55^\circ$, $\sigma_{{\rm d}N}$ is significantly smaller, which indicates there are physical correlations between sight lines within this separation.
This is consistent with the conclusion in \citet{Werk:2019aa}, who found a tight correlation for halo stars with $z$-heights of $\approx 3-10$ kpc up to 40$^\circ$, indicating an angular size of the warm gas of at least $40^\circ$.

However, the column density has a global variation due to the disk and the CGM variations (e.g., the minimum of column density in $|b|= 40^\circ-60^\circ$ and $|l|=180^\circ$; \citetalias{Qu:2019ab}).
This global variation could introduce additional correlations between column densities at large angular separations.
Therefore, we try to correct for the global variation by subtracting the fiducial model from the column density measurements.
We make similar plots using the column density residuals in the fiducial model (left panel of Fig. \ref{dN_dv_dist}) instead of the original column density.
By using the column density residuals, we exclude the global variation of column densities.
The result is shown in the middle left panel of Fig. \ref{angular_size}, which appears similar to the upper left panel.
Also, we extract the $1 \sigma$ width of the column density difference distribution (in the lower left panel of Fig. \ref{angular_size}), which shows that the correlation of the column density is within $\approx 15^\circ$ (i.e., flat part after $15^\circ$).
By comparing the non-corrected variation and the model-corrected variation, it is clear that the correlation disappears at angular scales of $\approx 15^\circ-55^\circ$.
This phenomenon means that the moderate scale correlation at $\approx 15^\circ-55^\circ$ is due to the global variation of the \ion{Si}{4} column density rather than the local variation due to individual clouds.
Then, we divide the entire angular separation into three ranges: individual clouds ($\lesssim 15^\circ$), the global variation ($\approx 15^\circ-55^\circ$; due to the large-scale warm gas distribution), the random variation ($\gtrsim 55^\circ$; no correlation).
This does not mean the correlation seen in the \citet{Werk:2019aa} sample is also due to the global variation, because these halo star sight lines have relatively smaller distances and different z-heights, which may affect the column density behaviors.

This two-point correlation of the \ion{Si}{4} gas is consistent with what has been found for HVCs, where a legacy HVC survey was done for the MW using the COS archival data \citep{Richter:2017aa}.
It is found that a strong correlation occurs within $\approx 20^\circ$, and a weak correlation is present up to $\approx 70^\circ$ \citep{Richter:2017aa}.
This consistency indicates that rotation and inflow in the northern hemisphere are the two dominant factors for the HVCs, which is the high-velocity tail of the warm gas distribution.

The 15$^\circ$ angular size can be converted into a physical size by adopting a distance.
Using the density distribution in Section \ref{gas_dist}, the mean distance of the warm gas (weighted for the column density; $\int n z {\rm d} z / \int n {\rm d} z$ or $\int n r_{\rm XY} {\rm d} r_{\rm XY} / \int n {\rm d} r_{\rm XY}$) is about 5 kpc from the Solar system.
Then, the warm gas cloud size is estimated to be 1.3 kpc.
This value is consistent with the estimation based on the cloud path-length density, which has an upper limit of $\lesssim  1.5$ kpc around the Solar system.
We can estimate the volume filling factor by combining the physical size of the warm gas and the cloud path-length  density.
At the Solar neighborhood, the volume-filling factor is about $({1.3\rm~ kpc} \times {0.6\rm~ kpc^{-1}})^3\approx 50 \%$.

These kpc-size structures are also seen in external galaxies.
H$\alpha$ and {\it BVI} imaging of NGC 891 indicates that there are $0.1-1$ kpc-size diffuse ionized gas features at $z$-heights of $1-2$ kpc \citep{Howk:2000aa}.
The \ion{Si}{4} gas has a higher temperature than the gas traced by H$\alpha$, so the smaller size of H$\alpha$ is expected.

Combining the cloud size with the single cloud column density, one can derive a \ion{Si}{4} density of $10^{12.86}\rm~cm^{-2}/1.3 ~kpc = 2 \times 10^{-9}\cc$, which is slightly higher than the average \ion{Si}{4} density around the Solar system.
The value $2 \times 10^{-9}\cc$ is the average \ion{Si}{4} in a warm gas cloud, but it is not necessary for the \ion{Si}{4} density to be uniform in the cloud.
The \ion{Si}{4}-bearing gas could be a shell surrounding a cooler core (e.g., \ion{H}{1} seen in H$\alpha$).
Then, we suggest that the estimated volume filling factor ($50\%$) of warm gas is the upper limit, because it is not clear whether the core region is also the warm gas.

For kinematical structures, we also extract the power spectra of the line centroid velocity (right panel of Fig. \ref{N_v_dist}) and the model velocity shift (right panel of Fig. \ref{dN_dv_dist}), which allow us to understand the size of the kinematical structures of the MW warm gas (Fig. \ref{angular_size}).
The difference between the non-corrected and model-corrected angular variation is much more significant than the column density.
This is because the kinematics of the gas (i.e., the rotation and the accretion in the northern hemisphere) has more significant global effects (e.g., rotation leads to opposite shifts at $l=90^{\circ}$ and $270^{\circ}$).
For the kinematical structure, the angular size is $80^\circ$ without the model correction, while it is reduced to $55^\circ$ after the model correction.
Then, the kinematical structure has a physical size of 4.8 kpc at a distance of 5 kpc.
This is larger than the column density structures by a factor of $\approx 3-4$, which indicates that every kinematical structure could contain multiple clouds (Section \ref{implication}).

\section{Discussion}
\label{discussion}
\subsection{Comparison with the QB2019 Model}
\label{comparison}

In Section \ref{previous_models}, we introduced three previous models, which are the basis of the kinematical model.
The 2D disk-CGM model proposed in \citetalias{Qu:2019ab} could fit both the stellar and the AGN samples simultaneously, showing a comparable contribution of the disk ($\log N \approx 13.0$) and the CGM component ($\log N \approx 13.2$) for \ion{Si}{4}.
This conclusion differs from the new results based on the kinematical model, where we find that the cloud path-length density of the CGM component only has an upper limit.
Therefore, the CGM column is much lower than the disk component in the kinematical model.
Here, we discuss whether there are physical differences between the \citetalias{Qu:2019ab} model and the kinematical model.
For the comparisons between three previous models, more details can be found in \citet{Zheng:2019aa} and \citetalias{Qu:2019ab}.

There are three major differences between the kinematical model and the \citetalias{Qu:2019ab} model: the inclusion of the cloud nature, the density distribution (disk and CGM), and the kinematical constraints.
The kinematical model assumes the cloud nature of the warm gas (with the path-length density and the column density of a single cloud; Section \ref{cloud_model}) instead of the ion density distribution.
This method not only predicts the column density of individual sight lines ($l$, $b$, and $d$), but also predicts the intrinsic uncertainty (based on the Poisson noise of the number of clouds).
Therefore, the cloud nature of the gas also introduces additional variation to the measured column density, which is similar to the traditional method of the patchiness parameter.
However, the patchiness method assumes a constant additional variation to all AGN sight lines, which implies a constant weight on different sight lines.
The cloud nature suggests the variation has a dependence on the path length.
For AGN sample, the path length is large ($\approx250$ kpc), which leads to similar uncertainty for these sight lines.
For the stellar sample, the uncertainty variation is large, such as from 0.4 dex to 0.2 dex from 1 kpc to 10 kpc, which is equivalent to having different weights for stellar sight lines.
These weights of the stellar sample do not affect the large scale structure in the fitting results (e.g., $r_0$ and $z_0$) significantly, because the stellar sample has typically small distances and mainly determine the midplane properties.
Therefore, we suggest that the cloud nature mainly determines the column density of individual clouds, and has little effect on the gas distribution due to the weights on the stellar sample.

The density distributions in the kinematical model are similar to the \citetalias{Qu:2019ab} model, but there are three significant differences.
First, the CGM component in \citetalias{Qu:2019ab} has a column density distribution over $l$ and $b$ that is limited by the ability to constrain the CGM radial density distribution.
This method implies an origin at the Solar system, which is impractical.
In the kinematical model, the CGM radial density distribution could be extracted based on the kinematics, so we consider the GC as the origin ($\phi$ and $\theta$ in Fig. \ref{cartoon}).
Second, the \citetalias{Qu:2019ab} CGM model only has a column density distribution, and only applies to AGN sight line predictions, which implies no CGM gas cospatial with the disk.
In the kinematical model, we also consider the core region of the CGM, which could affect the stellar sight lines.
Third, we introduce the $\alpha$ parameters in the disk model, which is a more detailed way of representing the disk extension in both direction r and z.
This variation could affect the distribution of CGM, i.e., the $z$-direction density distribution of the disk is more extended, which suppresses the CGM component.

Finally, the most important factor is the kinematics, based on which the distance to the gas could be estimated.
In \citetalias{Qu:2019ab}, we extract a density distribution for the disk component, but this is mainly based on the global variation of the AGN sample (and the midplane gas properties determined by the stellar sample).
The scale length is constrained by the variation at low Galactic latitude of AGN sight lines.
However, the distribution profile was not well constrained in \citetalias{Qu:2019ab}, in which the exponential and Gaussian profiles show similar fitting results.
The scale height is the variation over different Galactic latitudes.
Similarly, the profile in the $z$ direction was also poorly constrained.
Therefore, we suggest that the scale height and scale length in \citetalias{Qu:2019ab} are not accurately measured parameters from the column density-only sample.

With the bulk velocity field in the kinematical model, we measure the spatial density distribution of the warm gas, from the line shape at different velocities.
We find that the most of warm gas (contributing to the MW absorption features) is close to the disk ($\lesssim 20-50$ kpc) rather at large radii (i.e., 100 kpc) (Section \ref{gas_dist}).
For gas close to the disk, we find the modified disk model (with $\alpha$) adequately models the gas distribution, which sets the upper limit for the CGM.
For the CGM at large radii ($> 100$ kpc), we obtain the upper limit for the average ion density.

\subsection{The MW Warm Gas Mass and Accretion Rate}
\label{section_mass}

With the measured density distribution, we estimate the mass of the warm gas.
Combining the density distribution with kinematics, we could also estimate the accretion rate from the warm gaseous halo.

\begin{figure*}
\begin{center}
\includegraphics[width=0.49\textwidth]{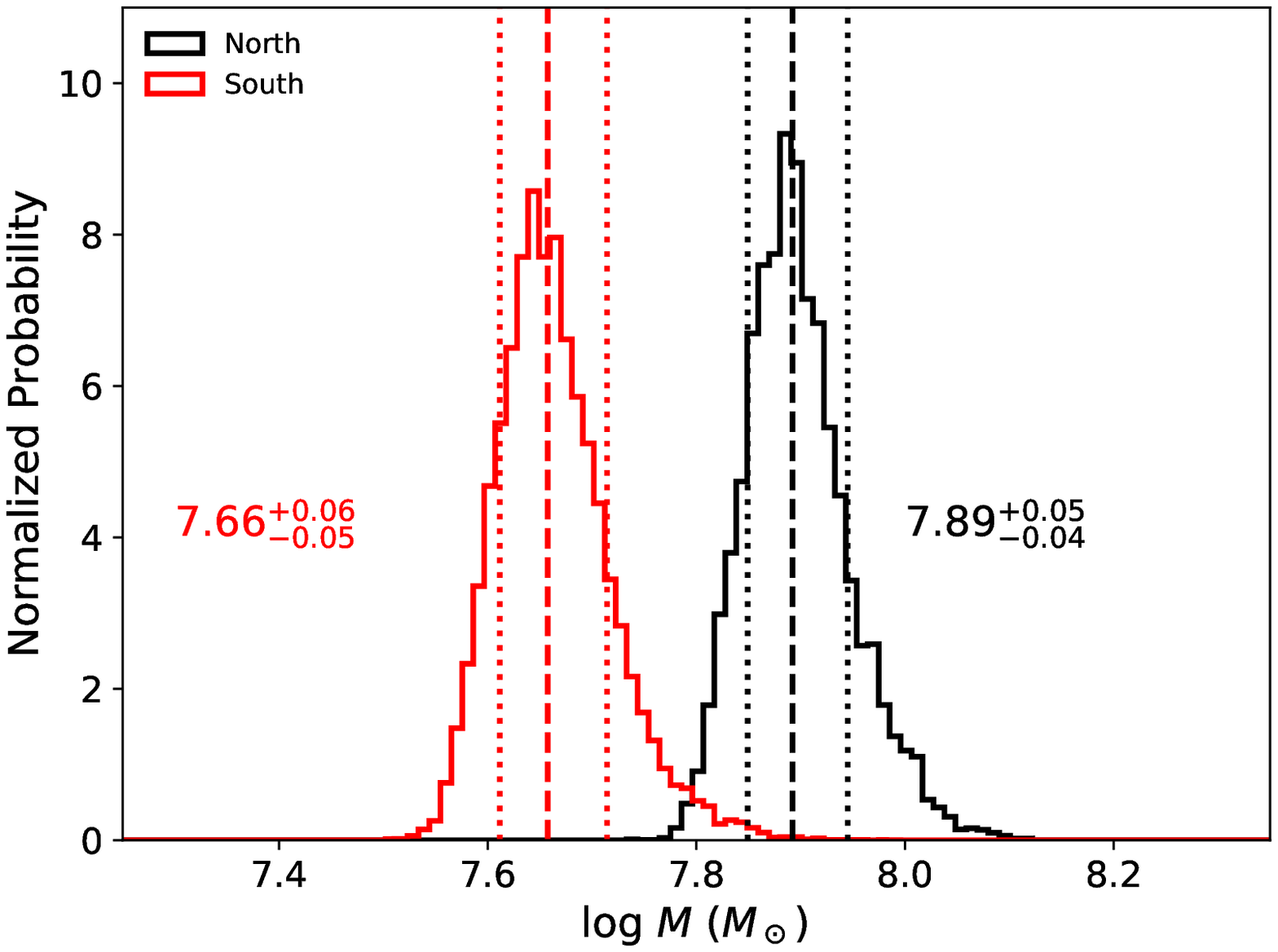}
\includegraphics[width=0.49\textwidth]{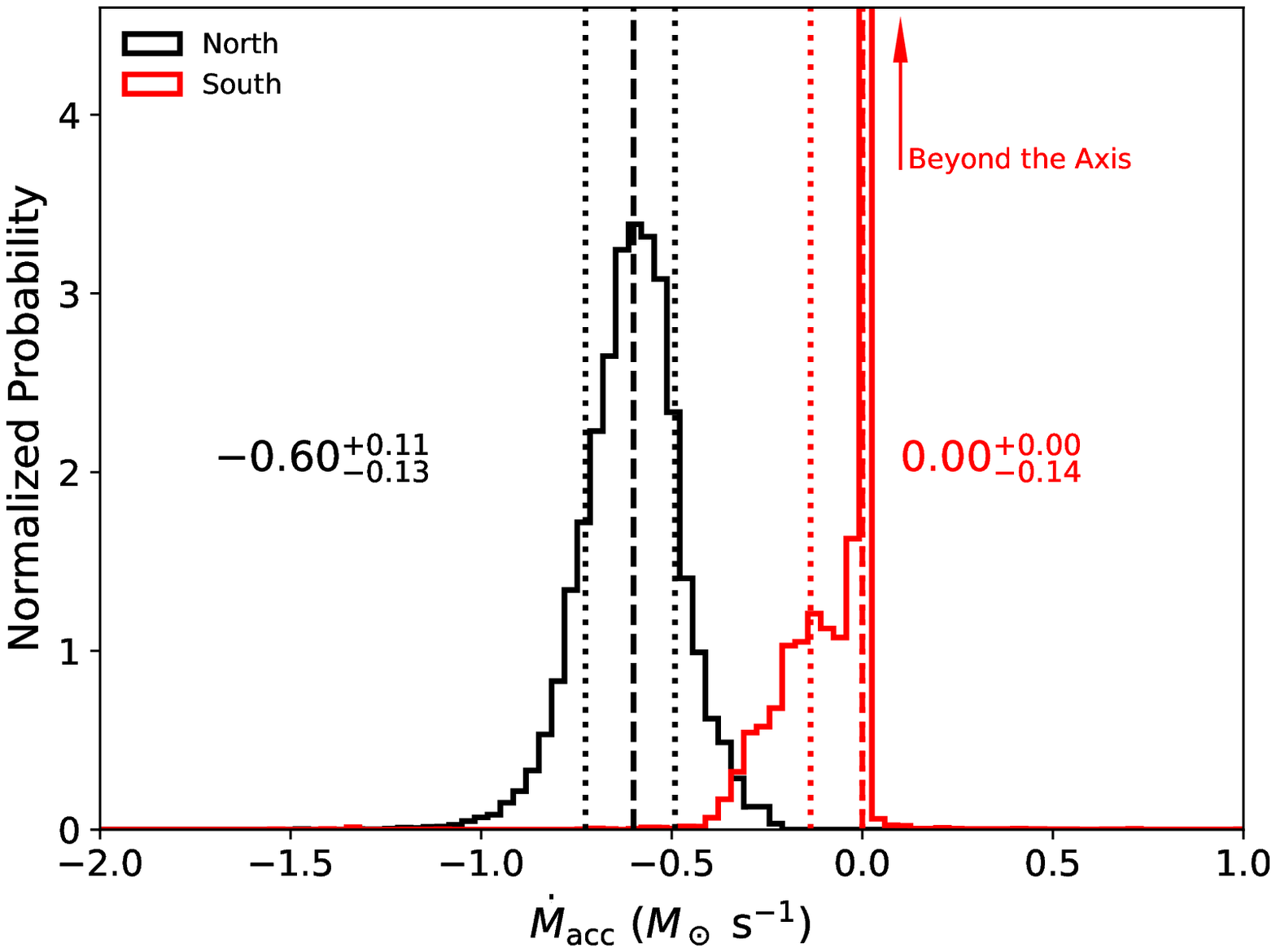}
\end{center}
\caption{The estimations of the warm gas disk mass (left panel) and the accretion rate (right panel). Vertical dashed lines and dotted lines are the median and the $1\sigma$ uncertainty. The disk component dominates the mass within 50 kpc, which has a total mass of $\log M = 8.09_{-0.04}^{+0.05}$. The CGM component might dominate the mass upto 250 kpc, which has a $3\sigma$ upper limit of $\log M < 9.1$ (no MS contribution). The accretion rate in the southern hemisphere is close to the zero, but has a $3\sigma$ upper limit of $-0.4 M_\odot~\rm s^{-1}$.}
\label{mass_acc}
\end{figure*}

\subsubsection{Mass}

In the kinematical model (Section \ref{kmodel}), we use two functions to approximate the density distribution of the warm gas: an ``exponential" disk (sharp decrease at large radii) and a power-law CGM (slow decrease at large radii).
As stated in Section \ref{gas_dist}, the disk component dominates the ion density distribution within 50 kpc, and the density distribution has a large uncertainty at $> 50$ kpc.

To estimate the mass, we first integrate the ion density distribution to obtain the total ion number of \ion{Si}{4} ($\mathcal{N}_{\rm SiIV}$).
Then, the total silicon number is $\mathcal{N}_{\rm SiIV}/f$, where $f$ is the ionization fraction, which is assumed to be 0.2 for \ion{Si}{4} (bout half of the maximum in CIE or PIE to represent the average ionization fraction; \citealt{Gnat:2007aa, Oppenheimer:2013aa}).
The hydrogen number is estimated by accounting for the metallicity (assumed to be $Z =0.5 Z_\odot$; \citealt{Bregman:2018aa}) and silicon abundance ($a = 3.24\times 10^{-5}$; \citealt{Asplund:2009aa}).
Therefore, the total hydrogen number $\mathcal{N}_{H} = \mathcal{N}_{\rm SiIV}/f/Z/a$.
Finally, the total mass is $1.3 \mathcal{N}_{H} m_{\rm H} $, where 1.3 accounts for the helium mass and $m_{\rm H}$ is the mass of the atom hydrogen.

We use two ways to report the masses.
First, we calculate the masses based on the disk and the CGM components.
To obtain the model-predicted mass (with uncertainty), we not only use the median value of the posterior distribution in Fig. \ref{params_corner}, but all models in the MCMC chain to estimate the uncertainty of the mass.
For the disk component, the mass distributions are shown in Fig. \ref{mass_acc} for both hemispheres.
The northern disk has a mass of $\log M =7.89^{+0.05}_{-0.04}$, while southern disk is $\log M =7.66^{+0.06}_{-0.05}$; the total disk component mass is about $\log M= 8.09^{+0.05}_{-0.04}$.
These masses could be scaled according to the ionization fraction and the metallicity as $-\log (f/0.2) - \log(Z/0.5Z_\odot)$.
For the CGM component, we could only obtain an upper limit.
Within 250 kpc, the $3 \sigma$ upper limit is $\log M < 9.1$ for the northern hemisphere and 8.0 for the southern hemisphere.
Combining the two hemispheres, the total CGM component has a $3 \sigma$ upper limit mass of $\log M< 9.1$ (excluding the Magellanic system; MS).

Another approach is to report the total mass within the given radii.
As shown in Fig. \ref{GC_rz}, the density distribution of the warm gas are well constrained at 30-50 kpc for both $z$ and $r$ directions.
The density distribution of warm gas is dominated by the disk component within 50 kpc, and the total mass is $\log M (r < 50 {\rm~kpc}) = 8.09_{-0.04}^{+0.05}$.
For warm gas within 250 kpc, although the column density measured from the Sun is dominated by the gas close to the disk, the mass may be dominated by the gas at large radii ($> 50$ kpc).
The mass upper limit, from combining the disk and the CGM component, is $\log M <9.1$ within 250 kpc at 3 $\sigma$(excluding the MS).
This is consistent with \citetalias{Qu:2019ab} that the warm gas in the MW is dominated by the MS, which has a mass of $\log M \approx 9.4$ \citep{Fox:2014aa}.

\subsubsection{Accretion Rate}

With the kinematical model, we determine the accretion velocity (for the northern hemisphere) and obtain the bulk accretion velocity field at different radii.
We estimate the accretion rate by combining the density distribution and the radial velocity.
As described in Section \ref{kmodel}, we set a boundary for the radial velocity ($R_{\rm disk}$), within which there is no radial velocity.
Here we calculate the total mass accretion rate at this boundary.

At the boundary surface, we integrate the product of the inflow velocity $v_{\rm rad}$ and the equivalent ion density $n = X N_{\rm sg}$:
\begin{equation}
\dot{M} = A\int n(z, r) v_{\rm rad} (r) {\rm d}S,
\end{equation}
where $S$ is the surface area of the boundary, and $A=1.3m_{\rm H}/f/Z/a$ is the conversion factor from \ion{Si}{4} ion number to total mass.
Similar to the mass estimation, we calculate the accretion rate for every model in the MCMC chain and plot the posterior distribution in Fig. \ref{mass_acc}.
For the northern hemisphere, the accretion rate is $-0.60_{-0.13}^{+0.11}~M_\odot~\rm s^{-1}$.
For the southern hemisphere, the accretion rate is estimated to be $0.00^{+0.00}_{-0.14}~M_{\odot}~\rm s^{-1}$, and the 3 sigma upper limit of accretion is $-0.4~M_\odot~\rm s^{-1}$.
The distribution of the southern hemisphere accretion rate is not a Gaussian-like distribution, but has a very high peak and a sharp edge at $0.00 ~ M_\odot~\rm s^{-1}$.
Therefore, the preferred accretion rate in the southern hemisphere is $0.00 ~M_\odot~\rm s^{-1}$, although the distribution of the accretion rate shows a negative wing.
Combining both hemispheres, the total accretion rate is $-0.64^{+0.14}_{-0.17}~M_\odot~\rm s^{-1}$.

\subsection{Physical Implications of the kinematical Model}
\label{implication}
The fitting results of the kinematical model are described in Section \ref{results}.
Here we discuss the implications of the kinematical model, focusing on the warm gas origins and the warm gas kinematics.

\subsubsection{Galactic Fountain Origin of the Warm Gas}
The warm gas in the MW has four possible origins considering the location where it is formed from the inner region to the outskirts: the ISM on the disk, the ejected material from the disk (feedback), the cooling flow in the MW gaseous halo, and the direct accretion from the IGM.
The warm gas density distribution could provide hints to distinguish between these origins.
Theoretically, the first two possibilities indicate that the warm gas distribution follows a disk shape \citep{Fielding:2017aa}, while the accretion from IGM is expected to be more spherical, because no cosmic filament is observed in local galaxies around the MW \citep{Tully:2019aa}.

As introduced in Section \ref{gas_dist}, there is a significant difference between the vertical direction (perpendicular to the disk; $z$) and the radial direction (along the midplane; $r_{\rm XY}$).
The warm gas in the $z$ direction is more extended than in the $r_{\rm XY}$ direction, so we suggest that the warm gas distribution is affected by Galactic feedback.
This effect is not only for the disk itself, but also extends to $20-50$ kpc, as shown in Fig. \ref{GC_rz}.
For the MW, more warm gas above or below the disk implies that the disk feedback processes are important to understand the warm gas formation.

The most direct explanation is that the warm gas is formed by the Galactic wind, where ejected gas could reach the virial radius (and beyond).
Galactic winds could produce a large amount of warm gas at $10-100$ kpc in the $z$ direction through radiative cooling \citep{Thompson:2016aa}.
Along the disk $r$-direction, a Galactic wind could not be launched, which leads to less warm gas.
This scenario could explain the warm gas distribution observed in the MW.
However, one issue remains for the Galactic wind model as the warm gas is mainly accreted (the northern hemisphere) or there is no systematical radial flow (the southern hemisphere).
In the Galactic wind model, Galactic wind is mainly outflowing, except for the large mass loading case (i.e., more ejected hot gas than the star formation rate; \citealt{Thompson:2016aa}).

An alternative model is the Galactic fountain, where some of the ejected material (typically warm gas) is recycled (accreted) to the disk instead of leaving the galaxy halo with the Galactic wind \citep{Bregman:1980aa, Fraternali:2017aa, Kim:2018aa}.
In the Galactic fountain scenario, the warm gas fountain is a byproduct of the hot wind, which is formed in the Galactic wind shocks.
The kpc-scale numerical simulation suggests that the warm gas in the hot wind has a periodic outflow and inflow cycles with a timescale of $\approx 50$ Myr (regulated by stellar activity; \citealt{Kim:2018aa}).
The majority of the outflow and inflow velocity is within $-50$ to $50\kms$ at any epoch.
Then, one could estimate the length scale of the size of the Galactic fountain structure to be $\approx 3$ kpc.

The theoretical predictions of Galactic fountains are consistent with our measurements of the MW warm gas.
Based on the kinematical model, inflow and outflow features are observed in addition to the bulk velocity field (i.e., rotation and inflow).
These inflow and outflow features have a velocity dispersion of $\approx 20 \kms$ ($v_{\rm rand}$).
This velocity dispersion is the projected value along sight lines, so the 3D dispersion is higher.
Assuming the random motion is isotropic, the 3D velocity dispersion is $\approx 35 \kms$, which is consistent with the velocity distribution predicted in \citet{Kim:2018aa}.

As stated in Section \ref{gas_property}, the kinematical structure typically has a size of $\approx 5$ kpc, which is $3-4$ times the cloud size (1.3 kpc; i.e., column density structures).
One kinematical structure could contain multiple clouds.
In sub-kpc-scale ($\approx 0.1$ kpc) numerical simulations, the warm gas is generated in interaction (or mixing) layers between cool gas ($\approx 10^3-10^4$ K) and hot gas ($\approx 10^6$ K; \citealt{Gnat:2010aa, Kwak:2010aa}).
Recent simulations predict the \ion{Si}{4} column density per layer is $\log N \approx 11$ \citep{Ji:2019aa} or $\log N \approx 12$ \citep{Kwak:2015aa}.
For a cloud, the sight line could pass though the mixing layer at least twice for a cool gas core or filament, so the predicted \ion{Si}{4} column density per cloud is about $\log N = 11.5$ or 12.5.
The \citet{Ji:2019aa} model has a lower number density ($10^{-2}\cc$; hence the pressure) for the cool gas than the \citet{Kwak:2015aa} model ($10^{-1}\cc$).
We suggest that the \citet{Ji:2019aa} model is more similar to \ion{Si}{4} clouds in the galaxy halo, while the \citet{Kwak:2015aa} model is more appropriate to the galaxy disk.
For the MW, the observed \ion{Si}{4} features are mainly due to warm gas in or close to the disk ($\lesssim 20$ kpc), which has the higher pressure, and is more similar to the \citet{Kwak:2015aa} model ($\log N \approx 12.0 - 12.5$).

\subsubsection{The Origins of the Net Inflow in the Northern Sky}
As a significant NS asymmetry, the northern hemisphere has a net accretion flow (also more massive; Section \ref{section_mass}), while the southern hemisphere does not.
This net accretion flow is also observed as prominent IVCs and HVCs in the northern hemisphere, which are detected by the \ion{H}{1} 21 cm line \citep{Wakker:2004aa} and nebular lines such as H$\alpha$ \citep{Haffner:2001aa}.
For the hotter gas ($\log T \approx 6$), current X-ray instruments are not sensitive to this level of accretion. 
Here, we limit the discussion to the cool and warm gas.

There are two possible origins for the inflow in the northern hemisphere: asymmetric disk activities or accretion histories.
We suggest that even though the inflow might be due to the accretion from the IGM, the warm gas is still shaped by feedback processes to account for the density distribution (Section \ref{gas_dist}).

For the disk origin, the northern hemisphere inflow might be due to a one-sided hot wind burst.
In zoom-in simulations, the hot wind is not necessarily symmetric between the two hemispheres, which could lead to different density distribution and kinematics \citep{Kim:2018aa}.
If the hot wind is temporarily blocked in one side, one only expects the recycled (accreted) gas in the other side. 
Then, an one-sided gigantic burst in the past might regulate the Galactic fountain, and lead to net accretion in the northern hemisphere.
However, there are three remaining issues for this scenario.
First, small Galactic fountain structures are supported by numerical simulations (i.e., $\approx 3$ kpc; \citealt{Kim:2018aa}).
This typical size cannot cover the entire northern hemisphere ($>10$ kpc at 5 kpc).
Second, the timescale of the Galactic fountain cycle is about $\approx 50-150$ Myr.
Then the burst should happen within the past $\lesssim 1$ Gyr to keep the systematic accretion feature, otherwise the random motion will dominate the kinematics.
Third, there is about $M \approx 3\times 10^7~M_\odot$ more warm gas in the northern hemisphere, which is comparable with the total mass of the warm gas disk of the MW.
It is almost impossible that the disk activities lifted all of the $\log M \approx 7.5$ more warm gas in the north.

Therefore, the inflow in the northern hemisphere should also reflect the accretion history of the MW.
There are two possible accretion modes onto the MW, through the sub-halo or the cosmic filament, which cannot be distinguished by our observations.
In the sub-halo scenario, the inflow in the northern hemisphere is due to a merger of a dwarf galaxy in the past.
The mass of the dwarf galaxy cannot be small, which is limited by the momentum conserved in the gas inflow.
Then, this merger may lead to observable features in the MW stellar halo \citep{Deason:2019aa}.
Another possibility is that the gas inflow is due to the continuous accretion of the IGM through a cosmic filament in a given direction.
The materials accreted from the IGM could cool down within the MW halo, which could lead to significant inflow ($v_{\rm r}\approx 100 \kms$) close to the disk \citep{Stern:2019aa}.

\subsubsection{The kinematics of IVC/HVC}
\label{IHVC}
The absorption systems of the MW are divided into three classes based on the velocity: absorption features due to the MW disk ($|v_{\rm LSR}| \lesssim 20 \kms$), IVCs ($20 \kms \lesssim |v_{\rm LSR}| \lesssim 90 \kms$), and HVCs ($90 \kms \gtrsim |v_{\rm LSR}|$).
As discussed in \citet{Wakker:1991ab}, this classification is arbitrary, and it does not consider Galactic rotation.
Therefore, \citet{Wakker:1991ab} introduced the deviation velocity ($v_{\rm DEV}$), which is the velocity difference between the observed line centroids and the predictions in a rotation-only model.
Based on the deviation velocity, \citet{Wakker:1991ab} defined IVCs ($35\kms < |v_{\rm DEV}| < 90 \kms$) and HVCs ($|v_{\rm DEV}| > 90 \kms$).
As summarized in \citet{Wakker:2004aa}, prominent \ion{H}{1} IVCs with negative velocities show up in the northern hemisphere (e.g., IV-Arch and IV-Spur).
Because the deviation velocity excludes the effect of Galactic rotation, the existence of prominent IVCs indicates that the kinematics of IVC cannot be accounted for by a rotation-only model.
Similarly, \citet{Sembach:2003aa} found that the velocity distribution of \ion{O}{6} HVCs prefers a static halo (at $|z|>3$ kpc) rather than a co-rotating halo.

However, as shown in Fig. \ref{combine_ls}, the major absorption features ($-150\kms \lesssim v_{\rm helio} \lesssim 150 \kms$ including some IVCs and HVCs) could be reproduced by our fiducial model, which contains both Galactic rotation and inflow.
There are three differences between our fiducial kinematical model and the rotation-only model used in \citet{Wakker:1991ab} and \citet{Sembach:2003aa}, which lead to the different conclusions whether the IVCs and HVCs could be reproduced in a rotation scenario.
First, the velocity lagging is important to understand the velocity of warm gas at high latitudes.
A lagging of $10 \kms\rm~kpc^{-1}$ leads to a velocity difference of $-30 \kms$ at a vertical height of 3 kpc.
After the projection along the sight lines, it is about $-20$ to $-30 \kms$ at high Galactic latitudes ($|b|\gtrsim 60^\circ$).
Second, our density distribution extends to higher $z$ and larger $r_{\rm XY}$, while previous studies typically assume a boundary of $z =3$ kpc for the warm gas.
A more extended density distribution means a larger lagging effect to account for more negative absorption features.
Third, the radial velocity (inflow) is important for the northern hemisphere.
This is of special importance for warm gas at $|l|\approx 180^\circ$, which has a negative-velocity HVC.
Therefore, we suggest that the MW rotation-inflow halo could explain most IVCs and HVCs seen in the \ion{Si}{4} absorption features.

Besides these MW halo IVCs and HVCs, there are still high-velocity features that cannot be modeled in the fiducial model.
In Fig. \ref{combine_ls}, the region C.2 shows a features at about $250 - 400 \kms$, while regions C.4, C.5, D.4, and D.5 show unaccounted for features $-400$ to $-100\kms$.
We suggest that these features could be divided into two populations.
One is the HVCs associated with the MS (mainly around LMC).
For this population, \citet{Fox:2014aa} has a detailed analyses, which is found to contribute a significant amount of the total warm gas mass in the MW halo (\citealt{Zheng:2019aa}; \citetalias{Qu:2019ab}).
Another population is the gas associated with the LG, which shows extremely high-velocity HVC at $v<-300 \kms$ towards the barycenter of the LG \citep{Richter:2017ab, Bouma:2019aa}.
The LG barycenter is located in the region C.4 in Fig. \ref{combine_ls}, where one sees the unaccounted features.
We note that the LG population may not only contain extremely HVCs but also HVCs (even low-velocity features), which might be misidentified as MW warm gas, and lead to additional variance in our kinematical model.
To distinguish better between the LG and the MW absorption features, one needs a better understanding of the LG kinematics, which is beyond the scope of this paper.
Here we suggest that the unaccounted features around the region C.4 may be associated with the LG.

\section{Summary and Conclusions}
\label{conclusion}
In this paper, we extract the line shape sample of the MW \ion{Si}{4} absorption features using the {\it HST}/COS archival data, and develop a kinematical model.
Using the kinematical model to reproduce the line shape sample, we constrain the ion density distribution of \ion{Si}{4}, the bulk velocity field (i.e., the rotation velocity and the radial velocity), and the warm gas properties (i.e., the broadening velocity).
Here, we summarize the key results:
\begin{enumerate}
\item In the kinematical model, we approximate the warm gas density distribution by two components: an exponential-like disk component of $n(r_{\rm XY}, z) = n_0 \exp(-(r_{\rm XY}/r_0)^{\alpha_{r_{\rm XY}}}) \exp(-(|z|/z_0)^{\alpha_{z}})$ and a $\beta$-model CGM component of $n(r) = n_0 (1+(r/r_{\rm c})^2)^{-3\beta/2}$.
The parameters $\alpha_{r_{\rm XY}}$ of $3.4\pm 0.8$ and $\alpha_{z}$ of $0.8\pm 0.2$ indicate that the warm gas distribution is significantly more extended in the $z$ direction (perpendicular to the disk) than the radial direction of the disk (Fig. \ref{GC_rz}).
The scale length of the \ion{Si}{4} disk is $12.5\pm0.6$ kpc, which leads to a shape decay of density at the disk edge.
Similar to \citetalias{Qu:2019ab}, we note that there is a significant NS asymmetry for the warm gas distribution.
The northern hemisphere has a larger scale height ($z_{\rm 0, N} = 6.3\pm1.5$ kpc) than the southern hemisphere ($z_{\rm 0, S} = 3.6_{-0.9}^{+1.0}$ kpc).
The CGM component only has upper limits, which indicates that the majority of the observed column density is close to the disk rather than at large radii of $>50$ kpc.

\item The warm gas in the MW is co-rotating with the stellar or \ion{H}{1} disk at a rotational velocity of $-215\pm 3 \kms$.
Above and below the disk, there is also rotation velocity gradient (lagging) of $\approx 8 \kms\rm~kpc^{-1}$ at $z = 3$ kpc, and smaller at higher $z$ heights.
This velocity gradient is important for fitting the absorption features at high Galactic latitudes.
The radial velocity shows a NS asymmetry as a significant inflow of $-69\pm7\kms$ (at 10 kpc) in the northern hemisphere, while the southern hemisphere does not show a significant net outflow or inflow.

\item The total mass of the disk component is $\log M = 8.09 - \log (Z/0.5 Z_\odot)$, which is also the dominant mass contributor in the inner 50 kpc.
At larger radii, the total mass of the warm gas in the MW halo might be dominated by the CGM component, which has an upper limit of $\log M < 9.1 - \log (Z/0.5 Z_\odot)$ at $3 \sigma$ (excluding the warm gas associated with the MS).
Combining the ion density distribution with the kinematics, we estimate the accretion rate in the northern hemisphere is $-0.60_{-0.13}^{+0.11}~ M_\odot~\rm yr^{-1}$.
For the southern hemisphere, we set a $3\sigma$ upper limit to the accretion of $-0.4 ~ M_\odot~\rm yr^{-1}$.

\item In the kinematical model, we adopt the cloud model rather than continuously smooth density distribution (Section \ref{cloud_model}).
Using the cloud model, we determine the average column density of individual clouds of $\log N({\rm SiIV}) \approx 12.6- 12.8$.
By subtracting the model from the observation, we estimate the cloud size by analyzing the angular power spectrum of the column density residual (Fig. \ref{dN_dv_dist} and Fig. \ref{angular_size}).
The angular size of the cloud is found to be $\approx 15^\circ$, which corresponds to a physical size of 1.3 kpc at 5 kpc (the average distance of the observed column density; Fig. \ref{GC_rz}).
Similarly, we determine the size of the kinematical features based on the velocity residuals, showing an angular size of $\approx 55^\circ$ and a physical size of 4.8 kpc at 5 kpc.
This indicates that every kinematical structure contains multiple clouds.

\item We suggest that most of the observed features could be explained in the Galactic fountain scenario rather than Galactic winds and continuous accretion from the IGM.
First, the warm gas observed in absorption is mainly co-rotating with the stellar disk (with the velocity gradient). 
Second, the \ion{Si}{4} density distribution shows a significant dependence on the disk-shape (not spherically distributed), which suggests the origin of \ion{Si}{4} is more associated with feedback processes rather than accretion.
Third, we see kpc-size variations of the kinematical structures for both northern and southern hemispheres, which contain both inflow and outflow (after excluding the bulk velocity field), which is consistent with recent simulations \citep{Kim:2018aa}.

\item Based on the kinematical modeling, we find that a considerable amount of IVCs and HVCs could be explained in the scenario of Galactic rotation and inflow in the northern hemisphere.
The remaining HVCs in Fig. \ref{combine_ls} might be associated with the MS and the LG.

\end{enumerate}

\acknowledgments
The authors would like to thank the anonymous referee, Eric Bell, Chris Howk, Edward Jenkins, Chang-Goo Kim, and Nicolas Lehner for kindly helps and thoughtful discussions on this work.
Z.Q. acknowledges Astropy \citep{Astropy-Collaboration:2013aa}, Emcee \citep{Foreman-Mackey:2013aa}, and HSLA \citep{Peeples:2017aa} to making various resources public, without which this work might be delayed by years. 
This work is supported by NASA through a grant HST-AR-15806.002-A from the Space Telescope Science Institute.
J.B. and E.H. would also like to thank supports from grants NNX16AF23G and 80NSSC19K1013.
\bibliographystyle{apj}
\bibliography{MissingBaryon}

\end{document}